\documentclass[preprint2]{aastex}
\usepackage{apjfonts}
\newcommand{\cotan}{\mathrm{cotan\,}} 
\newcommand{\sign}{\mathrm{sign\,}}   
\newcommand{\vect}[1]{\mbox{\boldmath{$#1$}}}   
\newcommand{\gvect}[1]{\mbox{\boldmath{$#1$}}} 
\newcommand{\scalprod}{\bullet}       
\newcommand{\sunmass}{M_\sun}         
\newcommand{\kms}{km~s$^{-1}$}
\newcommand{\HeI}{\ion{He}{1}}
\newcommand{\HeII}{\ion{He}{2}}
\newcommand{\HI}{\ion{H}{1}}
\newcommand{\HII}{\ion{H}{2}}
\newcommand{\NIII}{\ion{N}{3}}
\newcommand{\Brg}{Br$\gamma$}

\shorttitle{Stellar disks in the Galactic Center}
\shortauthors{Paumard et al.}
\begin{document}
\title{The  Two  Young  Star  Disks  in  the Central  Parsec  of  the  Galaxy:
  Properties, Dynamics and Formation\footnote{Based on observations with the
    Very  Large  Telescope  of  the European  Southern  Observatory,  Paranal,
    Chile.}}

\author{T.~Paumard\altaffilmark{1}, R.~Genzel\altaffilmark{1,2},
  F.~Martins\altaffilmark{1}, S.~Nayakshin\altaffilmark{3,4},
  A.~M.~Beloborodov\altaffilmark{5,6}, Y.~Levin\altaffilmark{7,8},
  S.~Trippe\altaffilmark{1}, F.~Eisenhauer\altaffilmark{1},
  T.~Ott\altaffilmark{1}, S.~Gillessen\altaffilmark{1},
  R.~Abuter\altaffilmark{1}, J.~Cuadra\altaffilmark{3},
  T.~Alexander\altaffilmark{9,10}, A.~Sternberg\altaffilmark{11}\\
  \textit{(paumard@mpe.mpg.de, genzel@mpe.mpg.de)}}

\altaffiltext{1}{Max-Planck Institut f\"ur extraterrestrische Physik (MPE),
  Giessenbachstra{\ss}e, 85748 Garching, Germany}
\altaffiltext{2}{Department of Physics, University of California, 366 LeConte
  Hall, Berkeley, CA 94720, USA}
\altaffiltext{3}{Max-Planck Institut f\"ur Astrophysik (MPA),
  Karl-Schwarzschild-Str. 1, 85741 Garching, Germany}
\altaffiltext{4}{Theoretical Astrophysics Group, Department of Physics \&
  Astronomy, University of Leicester, Leicester, LE1 7RH, United Kingdom}
\altaffiltext{5}{Physics Department and Columbia Astrophysics Laboratory, Columbia University, New York, USA}
\altaffiltext{6}{Astro-Space Center of Lebedev Physical Institute, 84/32 Profsoyuznaya st., Moscow, 117997, Russia}
\altaffiltext{7}{Canadian Institute for Theoretical Astrophysics, University
  of Toronto, 60 St. George Street, Toronto, Ontario, M5S 3H8, Canada}
\altaffiltext{8}{Sterrewacht Leiden, Leiden University, P.O. Box 9513,NL-2300
  RA  Leiden, The Netherlands}
\altaffiltext{9}{Faculty of Physics, Weizmann Institute of Science, Rehovot 76100, Israel}
\altaffiltext{10}{William Z. and Eda Bess Novick career development chair}
\altaffiltext{11}{School of Physics \& Astronomy, Tel Aviv University,
  P.O. Box 39040, Tel Aviv 69978, Israel}

\begin{abstract}
  We report the definite spectroscopic identification of $\simeq40$ OB
  supergiants, giants and main sequence stars in the central parsec of the
  Galaxy. Detection of their absorption lines have become possible with
  the high spatial and spectral resolution and sensitivity of the adaptive
  optics integral field spectrometer SPIFFI/SINFONI on the ESO VLT.  Several
  of these OB stars appear to be helium and nitrogen rich. Almost all of the
  $\simeq80$ massive stars now known in the central parsec (central arcsecond
  excluded) reside in one of two somewhat thick
  ($\langle|h|/R\rangle\simeq0.14$) rotating disks.  These stellar disks have
  fairly sharp inner edges ($R\simeq1\arcsec$) and surface density profiles
  that scale as $R^{-2}$. We do not detect any OB stars outside the central
  $0.5$~pc. The majority of the stars in the clockwise system appear to be on
  almost circular orbits, whereas most of those in the `counter-clockwise'
  disk appear to be on eccentric orbits.  Based on its stellar surface density
  distribution and dynamics we propose that IRS~13E is an extremely dense cluster
  ($\rho_\mathrm{core}\gtrsim3\times10^8\sunmass$~pc$^{-3}$), which has formed in the
  counter-clockwise disk.  The stellar contents of both systems are remarkably
  similar, indicating a common age of $\simeq6\pm2$~Myr.  The K-band luminosity
  function of the massive stars suggests a top-heavy mass function and limits
  the total stellar mass contained in both disks to
  $\simeq1.5\times10^4\sunmass$. Our data strongly favor in situ star
  formation from dense gas accretion disks for the two stellar disks.  This
  conclusion is very clear for the clockwise disk and highly plausible for the
  counter-clockwise system.
\end{abstract}

\section{INTRODUCTION}

The Galactic Center (GC) is a unique laboratory for studying galactic nuclei.
Given its proximity, processes in the GC can be investigated at resolutions
and detail that are not accessible in any other galactic nucleus (unless
otherwise specified we adopt a distance of 8~kpc for simplicity of comparison
to earlier work; we specifically use the most recent value
$R_0=7.62\pm0.32$~kpc by Eisenhauer et al. 2005 when the $\simeq5\%$ error
could lead to a significant bias).  The GC has many features that are thought
to occur in other nuclei (for reviews, see Genzel \& Townes 1987; Morris \&
Serabyn 1996; Mezger, Duschl \& Zylka 1996; Alexander 2005).  It contains the
densest star cluster in the Milky Way intermixed with a bright \HII\ region
(Sgr~A West or the `mini-spiral') and hot gas radiating at X-rays. These
central components are surrounded by a $\simeq1.5$~pc ring/torus of dense
molecular gas (the `circum-nuclear disk', CND). At the very center lies a very
compact radio source, Sgr~A*. The short orbital period of stars (in particular
the B star S2) in the central arcsecond around Sgr~A* show that the radio
source is a $3$--$4\times10^6\sunmass$ black hole (BH) beyond any reasonable
doubt (Sch\"odel et al. 2002; Ghez et al.  2003).  The larger Galactic Center
region contains three remarkably rich clusters of young, high mass stars: the
Quintuplet, the Arches, as well as the parsec-scale cluster around Sgr~A*
itself (Figer 2003).

In seeing-limited near-infrared images of the central region of the Galactic
Center, several bright sources dominate the $\simeq20\arcsec\times20\arcsec$
field centered on Sgr~A*. Among these, the IRS~16 cluster\footnote{The sources
  named ``GCIRS'' for Galactic Center Infrared Source are often referred to
  simply as ``IRS'' sources in the GC-centric literature.} (Becklin \&
Neugebauer 1975) is a bright source of broad \HeI~2.058~\micron\ line emission
(Hall, Kleinmann \& Scoville 1982).  IRS~16 has been since then resolved into
a cluster of about a half a dozen stars (Forrest et al. 1987, Allen, Hyland \&
Hillier 1990, Krabbe et al. 1991, 1995, Tamblyn et al. 1996, Paumard et al.
2001). These appear to be post main-sequence OB stars in a transitional phase
of high mass loss (Morris et al.  1996), between extreme O supergiants and
Wolf-Rayet (WR) stars (Allen et al. 1990; Najarro et al. 1994, 1997; Trippe et
al.  2005). They have been classified as Ofpe/WN9 stars (Allen, Hyland, \&
Hillier 1990) and have been suggested as Nitrogen-rich OB stars (OBN) stars by
Hanson et al. (1996) and as Luminous Blue Variables (LBV) candidates by
Paumard et al.  (2001).  There is no a-priori incompability between these
tentative classifications that are based on different properties.  Several
dozens of even more evolved WR stars have been observed in the same region
(e.g. Krabbe et al.  1995; Blum, Sellgren \& DePoy 1995; Paumard et al.
2001). The lack of OB stars in these earlier studies is puzzling. The question
is whether this lack is due to a true depletion or is merely a selection
effect due to veiling of the weak absorption lines in near main sequence stars
by bright nebular emission.  Adaptive optics (AO) spectroscopy of the
center-most arcsecond around Sgr~A* (mostly devoid of nebular emission) has
already revealed a dozen massive stars. These stars appear to be main sequence
late O and B stars and orbit the central black mass at distances as short as a
few light-days (Sch\"odel et al.  2003; Ghez et al. 2003, 2005; Eisenhauer et
al. 2005).

These observations show that massive star formation has occurred at or near
the Galactic Center within the last few million years. This is surprising. All
obvious routes to creating or bringing massive young stars in(to) the central
region face major obstacles. In situ star formation, transport of stars from
far out, scattering of stars on highly elliptical orbits and rejuvenation of
old stars due to stellar collisions and tidal stripping have all been proposed
and considered (for a recent review of the rapidly growing body of literature
see Alexander 2005). No explanation at this point is the obvious winner (or
loser). Perhaps the two most prominent and promising scenarios for explaining
the young massive stars outside the central cusp, at radii of 3--$10\arcsec$
from Sgr~A*, are
\begin{enumerate}
\item the `\emph{in situ, accretion disk}' scenario (Levin \& Beloborodov
  2003; Genzel et al.  2003; Goodman 2003; Milosavljevic \& Loeb 2004;
  Nayakshin \& Cuadra 2005). Here the proposal is that stars have formed near
  where they are found today, very close to the central black hole. However,
  in situ star formation is impeded by the tidal shear from the central black
  hole and surrounding dense star cluster. To overcome this shear, gas clouds
  have to be much denser ($\simeq10^{12}R_{1\arcsec}^{-3}$~cm$^{-3}$) than
  currently observed (Morris 1993).  The tidal shear can be overcome if the
  mass accretion was large enough at some point in the past -- perhaps as the
  consequence of the in-fall and cooling of a large interstellar cloud -- such
  that a gravitationally unstable (outside of a critical radius) disk was
  formed. The stars were formed directly out of the fragmenting disk;
\item the `\emph{in-spiraling star cluster}' scenario (Gerhard 2001; McMillan
  \& Portegies Zwart 2003; Portegies\linebreak Zwart et al. 2003; Kim \& Morris 2003;
  Kim, Figer, \& Morris 2004; G\"urkan \& Rasio 2005). Here the idea is that
  young stars were originally formed outside the hostile central parsec and
  only transported there later on. Individual transport of stars by two body
  relaxation and mass segregation from further out takes too long a time
  ($\simeq10^{7.5}$--$10^9$ years: Alexander 2005).  Stars in a bound, massive
  cluster can sink in much more rapidly owing to dynamical friction (Gerhard
  2001). To sink from an initial radius of a few parsec or more to a final
  radius of $\ll1$~pc within an O star lifetime (a few Myr) requires a cluster
  mass $>10^5\sunmass$.  To prevent the final tidal disruption of such a
  cluster at too large a radius {--} resulting in the deposition of its stars
  there {--} the core of the original star cluster also has to be much denser
  ($>10^7\sunmass$~pc$^{-3}$) and more compact ($\ll1$~pc) than any known
  cluster.  However, as a helpful by-product, dynamical processes in such a
  hypothetical super-dense star cluster may then lead to the formation of a
  central, intermediate mass black hole (IMBH; Portegies-Zwart \& McMillian
  2002; G\"urkan et al, 2004). Such a black hole may help to stabilize the
  cluster core against tidal disruption and lessen the high density
  requirement somewhat (Hansen \& Milosavljevic 2003).
\end{enumerate}

The `paradox of youth' in the central S-star cluster, with $>15$ apparently
normal main sequence B stars residing in tightly bound orbits in the central
light month around the central black hole, probably requires yet another
explanation (recently Ghez et al.  2003, 2005; Genzel et al. 2003; Hansen \&
Milosavljevic 2003; Gould \& Quillen 2003; Alexander \& Livio 2004; Eisenhauer
et al. 2005; Alexander 2005; Davies \& King 2005). Perhaps the most promising
route to get the B stars into the central arcsecond is a scattering process
from the reservoir of massive, young stars at $\simeq3$--$10\arcsec$ (e.g.
Alexander \& Livio 2004).

To test these proposals, the detailed properties and dynamics of the massive
stars in the central parsec must be studied. These properties include exact
stellar type, spatial distribution and 3D space velocities. For this purpose,
high-resolution imaging and spectro-imaging are required. The new adaptive
optics assisted, near-infrared integral field spectrometer on the ESO-VLT,
SPIFFI/SINFONI (Eisenhauer et al.  2003b; Bonnet et al. 2004) represents a key
new capability for addressing the issues discussed above. We report in this
paper SPIFFI/SINFONI observations in 2003, 2004 and 2005 that give important
new information on the location, dynamics and evolution of the massive, early
type stars in the central parsec.  We begin by discussing the SPIFFI/SINFONI
observations and data analysis in Sect.~\ref{sect:obs}. This is followed by
presentation of our results in Sect.~\ref{sect:res}.  In
Sect.~\ref{sect:discuss} we discuss the implications of our findings. Further
technical details are presented in the Appendices.

\section{OBSERVATIONS AND DATA ANALYSIS}
\label{sect:obs}

\subsection{Observations}

SPIFFI (Eisenhauer et al. 2003b,c) is a near-infrared integral field
spectrometer providing a 2048 pixel spectrum simultaneously for a contiguous,
$64\times32$-pixel field. Its salient features include a reflective image
slicer and a grating spectrometer with an overall detective throughput
(including pre-optics module and telescope) of $\geq30\%$. Its $2048^2$-pixel
Hawaii II detector covers the J, H and K (1.1 to 2.45~\micron) atmospheric
bands. In its 2003 version with a smaller $1024^2$-pixel detector the
spectrometer provided 1024 spectra for a $32^2$-pixel field. Spectral
resolving powers range from $R=1000$ to 4000. Three pixel scales
($12.5\times25$ square milli-arcseconds (mas), $50\times100$~mas$^2$ and
$125\times250$~mas$^2$) can be chosen on the fly. In the SINFONI ESO VLT
facility, SPIFFI is mated with the MACAO adaptive optics module (Bonnet et al.
2003) employing a 60-element wave-front-curvature sensor with avalanche
photodiodes. This mode makes it possible to perform spectroscopy at the
smallest (diffraction limited) pixel scale.

\begin{deluxetable}{lllllp{5.5cm}}
  \tablecaption{Summary of SPIFFI/SINFONI Data Sets\label{table:obs}}
  \tablehead{
    \colhead{Date} & \colhead{Band} & \colhead{Pixel} &  \colhead{2D res.} & \colhead{Mosaic size\tablenotemark{a}} & \colhead{Comments}\\
    &&mas&mas&arcsec$\times$arcsec & } \tablecolumns{6} \startdata
  2003 Apr 8--9 & K & 100 & 250 & $10\times10$ & Excellent seeing\\
  2003 Apr 8--9 & H+K & 250 & 900 & $38\times32$ & \nodata \\
  2004 Aug 18--19 & K & 100 & 220 & $10\times10$ & \nodata \\
  2005 Mar 14--23 & K & 100 & 200 & \phn$6\times16$ & Centered $\simeq15\arcsec$ N of Sgr~A*; very deep spectroscopy ($m_{K\mathrm{,lim}}\simeq16$--17.5)\\
  2005 Mar 14--23 & K & 100 & 200 & \phn$4\times20$ & Long dimension SW--NE;
  centered $\simeq15\arcsec$ NE of Sgr~A* \\
  2005 Jun 17 & K & 250 & 1000 & $40\times40$ & `Frame' completing 2003 H+K mosaic 
\enddata
\tablenotetext{a}{Unless otherwise specified, East--West $\times$ North--South.}
\end{deluxetable}
\begin{figure*}
\begin{center}\includegraphics[width=10cm]{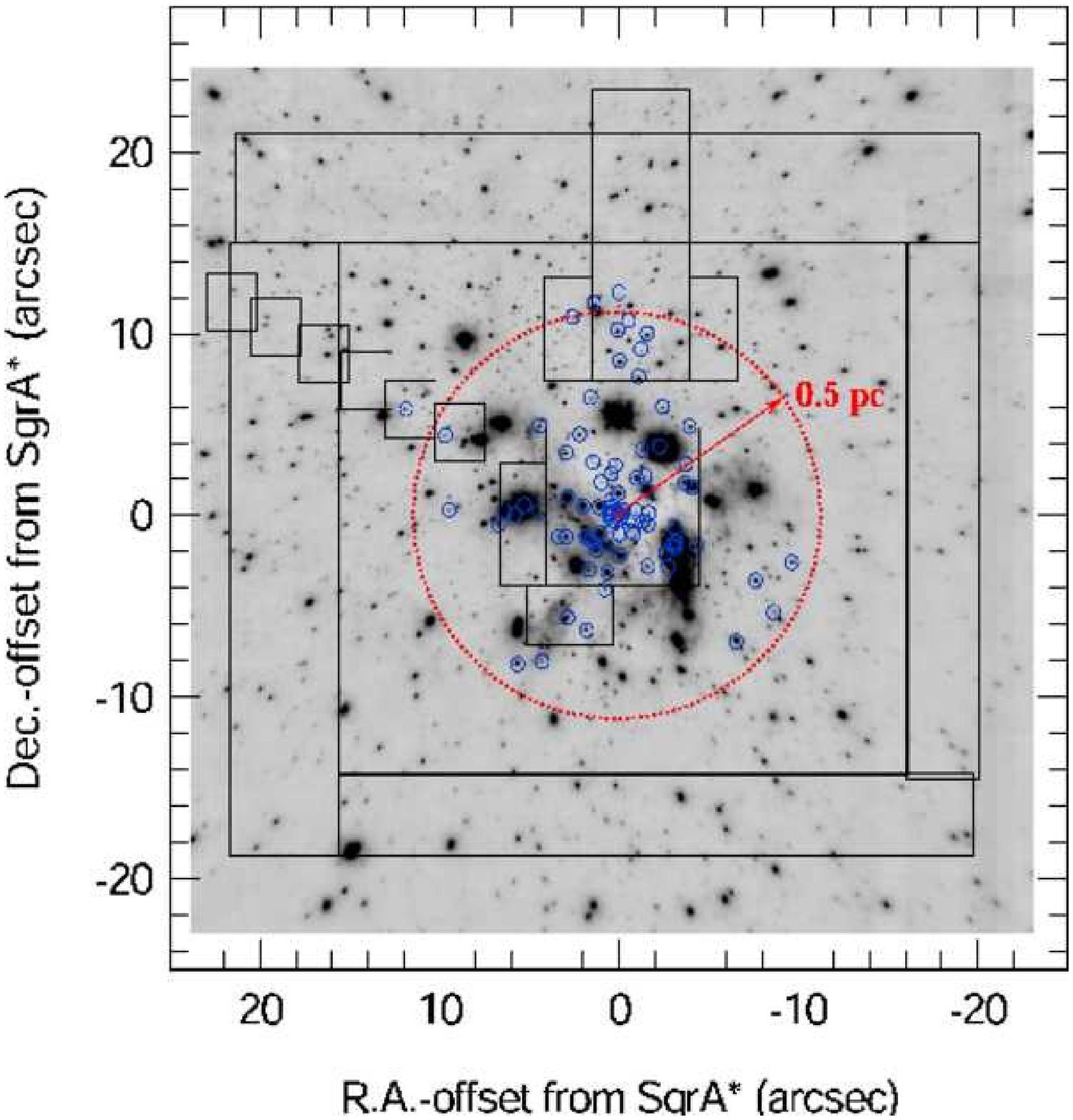}\end{center}
\caption{Outline of the various 2003--2005 SPIFFI/SINFONI H+K and K-band cubes
  (\emph{outlines}), superposed on a $\simeq100$~mas resolution, L-band NACO
  image (logarithmic scale). Small circles denote the 90 quality 1 and 2 early
  type stars (OB I--V, Ofpe/WN9, WR-stars) reported in this paper
  (Table~\ref{table:big1}).  A dotted circle denotes a 0.5~pc radius zone
  centered on Sgr~A*, within which essentially all OB stars we have found
  appear to lie.\label{fig:obs}}
\end{figure*}
Table~\ref{table:obs} lists the various data sets we have been able to obtain
with both SPIFFI as a guest instrument (in 2003) and SINFONI during its
commissioning in 2004 and guaranteed time observation (GTO) runs in 2005.
These data have been taken in the K-band with a resolution $R=4000$ (full
width at half maximum $\mathrm{FWHM}=85$~\kms) and in the H+K mode ($R=1500$,
$\mathrm{FWHM}=230$~\kms).  Preliminary results from the 2003 seeing limited
datasets were presented in Genzel et al.  (2003), Horrobin et al. (2004) and
Paumard et al. (2004a). The 2004 mosaic supersedes the 2003 K-band one.
Although the FWHM resolution of these two sets is almost identical, the AO
reduced dramatically the wings of the point spread function (PSF), rending
unnecessary the complex method for correction of the nebular emission
described in Paumard et al. (2004a).  Figure~\ref{fig:obs} shows the coverage
of these spectral cubes, superposed on a diffraction limited L$'$-band
(3.76~\micron) image taken with NACO.

We also analyzed several high-quality H- and
K$_\mathrm{s}$-band, diffraction limited images taken during
2002--2004 with NACO during our GTO astrometry imaging program to
construct very deep images of the IRS~13E and IRS~16 regions. We will
return to this when we discuss these images in Sect.~\ref{sect:irs13}.

\subsection{Spectral Identification of Early Type Stars}

Interpretation of stellar spectra in the Galactic Center is hindered by
stellar crowding and ionized interstellar gas emission, especially in \HI\
$\lambda$~2.166~\micron\ (\Brg) and \HeI\ $\lambda$~2.058~\micron. As a
result, previous seeing limited spectroscopic observations have relatively
easily detected broad emission line stars, such as WR stars and other evolved
objects (Ofpe/WN stars), but have not been successful in detecting near
main-sequence OB stars. These are characterized by relatively weak, absorption
features of \HeI\ ($\lambda$~2.058, 2.113, 2.163~\micron) and \HI\ \Brg\
(Hanson et al. 2005; the feature at 2.113~\micron\ is in fact a compound of 4
\HeI\ lines and three \NIII\ lines). The new high-resolution SINFONI data
overcome both of these issues to a considerable extent. We can now reliably
detect all OB supergiants and giants ($m_K\simeq11$--13). In less crowded
regions without strong nebular emission we are now also successful in
detecting OB main sequence stars ($m_K\simeq13$--15; see also Eisenhauer et
al. 2005).

We searched for OB stars by visual inspection of 2D continuum-subtracted line
maps in the above-mentioned lines.  \Brg\ and \HeI\ $\lambda$~2.058~\micron\
are usually intrinsically stronger than \HeI\ $\lambda$~2.113~\micron, but the
latter does not suffer from the problem of nebular emission. This line is,
therefore, a good choice for identification of OB stars, especially in the
lower resolution H+K data. The spectra of the OB candidates so identified were
then extracted with the interstellar emission removed by subtracting an
off-source spectrum (generally from a ring around each source).

For the medium plate scale ($50\times100$~mas$^2$/pixel) K band data, the
diffuse emission can in most cases be successfully removed by this
subtraction. The correction is not perfect, especially for stars embedded in
the interstellar medium, which they excite locally. In other cases the
profiles of the interstellar emission are complex and vary rapidly from
position to position (Paumard et al. 2004b). As a result there remain
considerable uncertainties around \Brg\ and \HeI\ $\lambda$~2.058~\micron\ in
the spectra extracted near/on mini-spiral streamers. In a few cases,
especially for fainter stars, the spectral identification of the stars is not
certain. We have taken these issues into account by grouping stars into three
quality codes. Stars with code 2 have high quality spectra and certain
identification. Stars with code 1 have possible uncertainties in the
extraction of some of the key spectral features.  In the case of stars with
code 0 the identification (as early type stars) is preliminary and needs to be
confirmed.  For a positive identification of a candidate \Brg\ absorption line
star we require that the stellar spectrum does not exhibit any signs for the
2.3--2.4~\micron\ CO overtone absorption bands characteristic of late type
stars.

In the larger K and H+K cubes at lower spectral and spatial resolution (the
outer square and rectangles in Fig.~\ref{fig:obs}), the nebular emission
subtraction becomes much harder. There remains only the possibility of looking
for the \HeI/\NIII\ complex at 2.113~\micron\ as explained above but the cube
in this case is filled with an unresolved background of late-type stars: the
CO break at 2.29~\micron, which is obvious in almost every pixel in the field,
cannot be used as a discriminator. All spectra are contaminated to some extent
by the numerous absorption lines that the cool stars exhibit (Wallace \&
Hinkle 1997 speak of a ``grass'' of absorption lines). We see in particular
\ion{Al}{1} ${\lambda}$~2.1099, 2.1132 and 2.1170~\micron. This makes the
identification of \HeI\ ${\lambda}$~2.1127~\micron\ uncertain or even
doubtful. While we find many emission line stars in this cube, we regard the
identification of a number of absorption line stars as reasonably safe, and
classify them with the same quality codes.

To obtain stellar identifications we compared the extracted spectra with
templates from the literature. For the wind-dominated stars and WR stars we
mainly used the atlas of Morris et al. (1996) and Figer, McLean \& Najarro
(1997). For the OB stars we referred to Hanson et al. (1996, 2005). We
compared spectra from the latter atlas individually to the Galactic Center
stars and made spectral identifications based on the strength of the \HeI,
\HI, \HeII, and \NIII\ lines, along with equivalent widths and line
profiles/widths. We computed absolute magnitudes from the observed magnitudes
and individual extinction corrections (Appendix~\ref{app:HR}).  These absolute
K magnitudes are used in our final classifications (Table~\ref{table:big1}) as
an additional constraint for placing the stars in the different luminosity
classes.
\clearpage
\pagestyle{empty}
\begin{deluxetable}{p{2cm}rrrrrrrrrrrrrrrrlcrr}
\rotate
\tablewidth{\textheight}
\tabletypesize{\scriptsize}
\tablewidth{0pt}
\tablecaption{Properties of Early Type Stars (Quality 1+ 2) in the Central
Parsec\label{table:big1}}
\tablehead{\colhead{Name} & \colhead{$p$} & \colhead{$x$} & \colhead{$y$} & \colhead{$z$} & \colhead{$\pm$} & \colhead{$m_K$} & \colhead{$v_x$} & \colhead{$\pm$} & \colhead{$v_y$} & \colhead{$\pm$} & \colhead{$v_z$} & \colhead{$\pm$} &  \colhead{$j$} & \colhead{$\pm$} &  \colhead{$e$\tablenotemark{a}} & \colhead{$\pm\tablenotemark{a}$} &  \colhead{Type} & \colhead{Q} & \colhead{$M_K$} & \colhead{$\pm$}}
\tablecolumns{21}
\startdata
E1:S2,S02 & 0.12 & 0.04 & 0.12 & \nodata & \nodata & 14.0 & 9 & 32 & 1830 & 43 & -1060 & 25 & 0.29 & 0.02 &  0.876 &  0.007 & B0-2 V & 2 & -3.9 & 0.6 \\
E2:S14,S016 & 0.14 & 0.12 & 0.07 & \nodata & \nodata & 15.7 & 2106 & 191 & 1103 & 88 & 300 & 80 & -0.04 & 0.05 &  0.939 &  0.008 & B4-9 V & 2 & -2.2 & 0.6 \\
E3:S13,S020 & 0.16 & -0.16 & 0.00 & \nodata & \nodata & 15.8 & 359 & 93 & 1483 & 52 & -280 & 50 & -0.98 & 0.03 &  0.395 &  0.032 & B4-9 V & 2 & -2.1 & 0.6 \\
E4:S1,S01 & 0.21 & -0.04 & -0.20 & \nodata & \nodata & 14.5 & 801 & 28 & -1183 & 44 & -1033 & 25 & 0.71 & 0.02 &  0.358 &  0.036 & B0-2 V & 2 & -3.4 & 0.6 \\
E5:S12,S019 & 0.23 & -0.06 & 0.26 & \nodata & \nodata & 15.5 & 255 & 79 & 1098 & 56 & 280 & 50 & -0.46 & 0.07 &  0.902 &  0.005 & B4-9 V & 2 & -2.4 & 0.6 \\
E6:S4,S03 & 0.29 & 0.26 & 0.11 & \nodata & \nodata & 14.4 & 623 & 26 & 74 & 24 & -570 & 40 & -0.28 & 0.04 & \nodata & \nodata & B0-2 V & 2 & -3.5 & 0.6 \\
E7:S08 & 0.40 & -0.30 & 0.27 & \nodata & \nodata & 15.8 & 121 & 23 & -471 & 21 & -390 & 70 & 0.54 & 0.04 & \nodata & \nodata & B4-9 V & 2 & -2.1 & 0.6 \\
E8:S5 & 0.40 & 0.36 & 0.17 & \nodata & \nodata & 15.0 & -134 & 30 & 355 & 30 & 30 & 90 & 1.00 & 0.08 & \nodata & \nodata & B4-9 V & 2 & -2.9 & 0.6 \\
E9:S9,S05 & 0.40 & 0.18 & -0.36 & \nodata & \nodata & 15.1 & 109 & 23 & -499 & 23 & 610 & 40 & -0.26 & 0.05 & \nodata & \nodata & B0-2 V & 2 & -2.8 & 0.6 \\
E10:S8,S04  & 0.45 & 0.37 & -0.26 & \nodata & \nodata & 14.5 & 536 & 45 & -569 & 41 & 15 & 30 & -0.21 & 0.05 &  0.927 &  0.019 & B0-2 V & 2 & -3.4 & 0.6 \\
E11:S6,S07 & 0.48 & 0.47 & 0.09 & \nodata & \nodata & 15.4 & 295 & 30 & -21 & 24 & 160 & 60 & -0.25 & 0.08 & \nodata & \nodata & B V & 2 & -2.5 & 0.6 \\
E12:S7,S011 & 0.53 & 0.53 & -0.05 & \nodata & \nodata & 15.2 & -225 & 22 & -93 & 23 & -20 & 150 & -0.46 & 0.10 & \nodata & \nodata & B V & 2 & -2.7 & 0.6 \\
E13 & 0.68 & 0.53 & 0.43 & \nodata & \nodata & 15.1 & 153 & 23 & -20 & 22 & -890 & 31 & -0.73 & 0.15 & \nodata & \nodata & B V & 2 & -2.8 & 0.6 \\
E14:S014 & 0.82 & -0.78 & -0.28 & \nodata & \nodata & 13.7 & 34 & 21 & -2 & 20 & -14 & 40 & 0.39 & 0.60 & \nodata & \nodata & O9.5-B2 V & 2 & -3.9 & 0.6 \\
E15:S1-3 & 0.96 & 0.43 & 0.86 & 1.57 & 0.33 & 12.2 & -518 & 21 & 115 & 22 & 68 & 40 & 0.97 & 0.04 &  0.000 &  0.151 & ? & 2 & -4.9 & 1.1 \\
E16:S015 & 0.98 & -0.94 & 0.26 & 0.22 & 0.24 & 13.6 & -262 & 23 & -374 & 26 & -424 & 70 & 0.94 & 0.06 &  0.096 &  0.173 & O9-9.5 V & 2 & -3.4 & 0.9 \\
E17 & 1.01 & -0.04 & -1.01 & -1.70 & 0.48 & 14.7 & 432 & 29 & 8 & 28 & 26 & 30 & 1.00 & 0.07 &  0.107 &  0.255 & ? & 1 & -3.0 & 0.5 \\
E18 & 1.09 & -0.66 & -0.87 & -1.96 & 0.56 & 14.1 & 313 & 28 & -144 & 28 & -364 & 40 & 0.98 & 0.08 &  0.636 &  0.229 & OB & 1 & -3.6 & 0.5 \\
E19:IRS16NW & 1.21 & 0.03 & 1.21 & \nodata & \nodata & 10.0 & 199 & 52 & 67 & 44 & -44 & 20 & -0.94 & 0.25 &  0.898 &  0.052 & Ofpe/WN9 & 2 & -7.4 & 0.4 \\
E20:IRS16C & 1.23 & 1.13 & 0.48 & 1.22 & 0.46 & 9.7 & -342 & 50 & 302 & 44 & 125 & 30 & 0.90 & 0.10 &  0.075 &  0.279 & Ofpe/WN9 & 2 & -6.7 & 0.4 \\
E21 & 1.31 & -0.85 & -1.00 & -1.91 & 0.45 & 13.8 & 397 & 28 & -65 & 27 & -24 & 30 & 0.86 & 0.07 &  0.349 &  0.159 & OB I? & 2 & -3.9 & 0.7 \\
E22 & 1.40 & -1.36 & -0.31 & -0.90 & 0.33 & 12.8 & 157 & 29 & -277 & 27 & -434 & 50 & 0.96 & 0.08 &  0.204 &  0.219 & O8-9.5 III/I & 2 & -4.2 & 0.6 \\
E23:IRS16SW & 1.43 & 1.05 & -0.98 & -1.46 & 0.51 & 9.9 & 261 & 47 & 90 & 43 & 320 & 40 & 0.88 & 0.16 &  0.410 &  0.190 & Ofpe/WN9 & 2 & -6.5 & 0.4 \\
E24 & 1.68 & -1.67 & 0.14 & -0.21 & 0.38 & 13.1 & 62 & 29 & -206 & 28 & -344 & 50 & 0.93 & 0.13 &  0.413 &  0.177 & O9-9.5 III? & 2 & -4.5 & 0.5 \\
E25 & 1.72 & -1.64 & -0.50 & -1.15 & 0.66 & 12.7 & 273 & 28 & -81 & 28 & -224 & 50 & 0.55 & 0.10 & \nodata & \nodata & O8.5-9.5 I? & 2 & -4.2 & 0.5 \\
E26:IRS16SSW & 1.75 & 0.72 & -1.60 & \nodata & \nodata & 11.5 & 118 & 28 & -207 & 29 & 206 & 30 & 0.10 & 0.12 & \nodata & \nodata & O8-9.5 I & 2 & -5.5 & 0.5 \\
E27:IRS16CC & 2.08 & 2.01 & 0.54 & 1.35 & 0.54 & 10.4 & -85 & 44 & 219 & 45 & 241 & 25 & 0.99 & 0.19 &  0.478 &  0.137 & O9.5-B0.5 I & 2 & -6.7 & 1.1 \\
E28:IRS16SSE2 & 2.08 & 1.45 & -1.49 & -2.13 & 0.54 & 12.4 & 292 & 28 & 120 & 27 & 286 & 20 & 0.93 & 0.09 &  0.228 &  0.241 & B0-0.5 I & 2 & -4.4 & 0.4 \\
E29 & 2.08 & 0.99 & 1.83 & 3.21 & 0.57 & 13.7 & -254 & 28 & 67 & 27 & -94 & 50 & 0.97 & 0.11 &  0.490 &  0.170 & O9-B0 & 2 & -3.5 & 0.8 \\
E30:IRS16SSE1 & 2.09 & 1.59 & -1.36 & -1.85 & 0.57 & 12.2 & 291 & 29 & 116 & 28 & 216 & 20 & 0.88 & 0.09 &  0.057 &  0.212 & O8.5-9.5 I & 2 & -4.9 & 0.5 \\
E31:IRS29N & 2.14 & -1.60 & 1.41 & \nodata & \nodata & 10.0 & 130 & 50 & -119 & 45 & -190 & 90 & 0.02 & 0.27 & \nodata & \nodata & WC9 & 2 & -7.7 & 0.4 \\
E32:MPE+1.6-6.8(16SE1) & 2.18 & 1.85 & -1.15 & -1.52 & 0.63 & 10.9 & 184 & 43 & 124 & 44 & 366 & 70 & 0.91 & 0.20 &  0.260 &  0.347 & WC8/9 & 2 & -5.9 & 0.5 \\
E33:IRS33N & 2.19 & -0.06 & -2.19 & \nodata & \nodata & 11.1 & 85 & 30 & -212 & 40 & 68 & 20 & 0.40 & 0.16 & \nodata & \nodata & B0.5-1 I & 2 & -5.8 & 0.4 \\
E34:MPE+1.0-7.4(16S) & 2.26 & 1.27 & -1.88 & -2.70 & 0.61 & 10.7 & 301 & 47 & 1 & 43 & 100 & 20 & 0.83 & 0.15 &  0.100 &  0.287 & B0.5-1 I & 2 & -6.1 & 0.6 \\
E35:IRS29NE1 & 2.28 & -0.99 & 2.06 & 2.99 & 0.60 & 11.7 & -370 & 51 & 25 & 43 & -100 & 70 & 0.87 & 0.13 &  0.140 &  0.306 & WC8/9 & 2 & -6.0 & 0.4 \\
E36 & 2.34 & 0.45 & 2.29 & 3.52 & 0.56 & 12.5 & -317 & 29 & 85 & 27 & 41 & 20 & 1.00 & 0.09 &  0.192 &  0.121 & O9-B0 I? & 2 & -5.3 & 0.4 \\
E37 & 2.62 & -1.47 & 2.17 & \nodata & \nodata & 14.8 & 130 & 29 & -143 & 28 & -114 & 30 & -0.14 & 0.15 & \nodata & \nodata & O8-9 I? & 2 & -3.3 & 0.5 \\
E38 & 2.76 & 0.19 & 2.76 & 3.47 & 0.54 & 13.1 & -342 & 46 & 85 & 43 & 36 & 20 & 0.98 & 0.13 &  0.279 &  0.166 & O8-9 III/I & 2 & -5.0 & 0.5 \\
E39:IRS16NE & 3.05 & 2.87 & 1.03 & \nodata & \nodata & 8.9 & 104 & 49 & -379 & 47 & -10 & 20 & -1.00 & 0.12 &  0.000 &  0.234 & Ofpe/WN9 & 2 & -7.5 & 0.6 \\
E40:IRS16SE2 & 3.17 & 2.94 & -1.19 & -1.20 & 0.71 & 12.0 & 107 & 28 & 181 & 29 & 327 & 100 & 0.99 & 0.14 &  0.206 &  0.288 & WN5/6 & 2 & -4.5 & 0.4 \\
E41:IRS33E & 3.19 & 0.65 & -3.12 & -3.57 & 0.50 & 10.1 & 182 & 47 & -9 & 42 & 170 & 20 & 0.97 & 0.26 &  0.630 &  0.186 & Ofpe/WN9 & 2 & -6.3 & 0.4 \\
E42 & 3.20 & -3.13 & -0.66 & \nodata & \nodata & 14.6 & -52 & 28 & 257 & 28 & 40 & 40 & -1.00 & 0.10 &  0.358 &  0.218 & B V/III & 2 & -2.7 & 0.6 \\
E43 & 3.21 & -1.60 & -2.79 & -3.57 & 0.50 & 12.2 & 227 & 29 & 1 & 28 & -114 & 50 & 0.87 & 0.13 &  0.214 &  0.297 & O8.5-9.5 I & 2 & -4.7 & 0.4 \\
E44 & 3.29 & 1.45 & 2.95 & 3.63 & 0.42 & 13.8 & -259 & 29 & 53 & 27 & -114 & 40 & 0.97 & 0.11 &  0.508 &  0.164 & O9-B0 II/I? & 2 & -4.1 & 0.5 \\
E45 & 3.33 & -2.61 & -2.08 & \nodata & \nodata & 12.5 & 175 & 27 & 106 & 28 & 63 & 30 & 0.13 & 0.13 & \nodata & \nodata & O9-B0 I & 2 & -4.3 & 0.4 \\
E46:IRS13E1 & 3.37 & -2.94 & -1.64 & \nodata & \nodata & 10.7 & -201 & 45 & -50 & 42 & 71 & 20 & -0.26 & 0.21 & \nodata & \nodata & B0-1 I & 2 & -5.6 & 0.3 \\
E47 & 3.41 & 1.67 & -2.97 & \nodata & \nodata & 12.5 & -49 & 25 & 150 & 25 & 91 & 30 & 0.20 & 0.16 & \nodata & \nodata & B0-3 I & 2 & -4.4 & 0.4 \\
E48:IRS13E4 & 3.50 & -3.19 & -1.42 & \nodata & \nodata & 11.7 & -316 & 29 & 76 & 29 & 56 & 70 & -0.61 & 0.09 &  0.809 &  0.058 & WC9 & 2 & -4.7 & 0.4 \\
E49:IRS13E3\tablenotemark{b} & 3.53 & -3.19 & -1.51 & \nodata & \nodata & 13.0 & -157 & 29 & 118 & 30 & 87 & 20 & -0.88 & 0.15 &  0.725 &  0.098 & ? & 2 & -5.2 & 0.3 \\
E50:IRS16SE3 & 3.54 & 3.35 & -1.16 & -0.93 & 0.76 & 11.9 & 7 & 29 & 201 & 27 & 281 & 20 & 0.96 & 0.14 &  0.319 &  0.224 & O8.5-9.5 I & 2 & -4.7 & 0.6 \\
E51:IRS13E2 & 3.59 & -3.14 & -1.74 & \nodata & \nodata & 10.8 & -303 & 44 & 68 & 46 & 40 & 40 & -0.66 & 0.15 &  0.749 &  0.099 & WN 8 & 2 & -5.6 & 0.4 \\
E52 & 3.84 & -1.26 & 3.62 & \nodata & \nodata & 13.3 & 214 & 28 & 214 & 26 & -167 & 20 & -0.90 & 0.09 &  0.378 &  0.183 & O8-9 III & 2 & -4.8 & 0.4 \\
E53 & 3.95 & -2.76 & -2.83 & \nodata & \nodata & 12.4 & -65 & 25 & -154 & 25 & 29 & 20 & 0.36 & 0.15 & \nodata & \nodata & B0-1 I & 2 & -4.7 & 0.6 \\
E54:IRS34E & 4.08 & -3.67 & 1.80 & 2.12 & 0.88 & 12.6 & -221 & 28 & -131 & 27 & -154 & 25 & 0.84 & 0.10 &  0.171 &  0.215 & O9-9.5 I & 2 & -5.0 & 0.6 \\
E55 & 4.14 & 0.77 & -4.06 & \nodata & \nodata & 12.5 & -65 & 29 & -159 & 27 & 76 & 20 & -0.54 & 0.17 & \nodata & \nodata & B0-1 I? & 2 & -4.3 & 0.4 \\
E56:IRS34W & 4.35 & -4.05 & 1.59 & 1.55 & 0.89 & 11.4 & -79 & 28 & -166 & 27 & -290 & 30 & 1.00 & 0.15 &  0.217 &  0.354 & Ofpe/WN9 & 2 & -5.8 & 0.5 \\
E57 & 4.43 & 4.42 & 0.25 & 1.48 & 0.96 & 13.5 & -109 & 28 & 114 & 27 & 196 & 40 & 0.76 & 0.17 &  0.343 &  0.260 & O7-9 III? & 2 & -3.0 & 0.5 \\
E58:IRS3E & 4.48 & -2.26 & 3.80 & \nodata & \nodata & 15.0 & \nodata & \nodata & \nodata & \nodata & 107 & 100 & \nodata & \nodata & \nodata & \nodata & WC5/6 & 1 & -2.9 & 0.6 \\
E59:[PMM2001] B9\tablenotemark{c} & 4.54 & 2.94 & 3.46 & \nodata & \nodata & 13.0 & 250 & 28 & 32 & 26 & -150 & 100 & -0.67 & 0.11 &  0.794 &  0.078 & WC9 & 2 & -4.0 & 0.6 \\
E60 & 4.66 & -4.36 & -1.65 & \nodata & \nodata & 12.4 & -210 & 27 & 127 & 27 & 330 & 80 & -0.79 & 0.11 &  1.046 &  0.311 & WN7? & 2 & -4.5 & 0.6 \\
E61:IRS34NW & 4.69 & -3.73 & 2.85 & 3.08 & 0.81 & 12.8 & -225 & 28 & -112 & 27 & -150 & 30 & 0.90 & 0.11 &  0.000 &  0.230 & WN7 & 2 & -4.6 & 0.6 \\
E62 & 4.99 & 2.18 & 4.48 & \nodata & \nodata & 11.5 & 229 & 42 & -66 & 43 & -134 & 40 & -0.99 & 0.18 &  0.325 &  0.229 & B0-3 I & 2 & -6.5 & 0.5 \\
E63:IRS1W & 5.30 & 5.27 & 0.57 & \nodata & \nodata & 9.6 & -108 & 44 & 209 & 55 & 35 & 20 & 0.93 & 0.23 &  0.410 &  0.304 & Be? & 1 & -7.1 & 0.3 \\
E64 & 5.81 & 5.81 & 0.05 & \nodata & \nodata & 12.4 & -20 & 30 & 170 & 25 & 40 & 25 & 0.99 & 0.15 &  0.572 &  0.151 & O9.5-B2II & 2 & -4.4 & 0.4 \\
E65:IRS9W & 6.30 & 2.85 & -5.62 & \nodata & \nodata & 12.1 & 167 & 29 & 135 & 27 & 140 & 50 & 0.98 & 0.13 &  0.665 &  0.242 & WN8 & 2 & -4.1 & 0.4 \\
E66:IRS7SW & 6.32 & -3.95 & 4.93 & \nodata & \nodata & 12.0 & -5 & 27 & -108 & 26 & -350 & 50 & 0.66 & 0.25 &  1.261 &  0.216 & WN8 & 2 & -4.4 & 0.5 \\
E67:IRS1E & 6.38 & 6.37 & 0.23 & \nodata & \nodata & 11.2 & -107 & 43 & 136 & 49 & 8 & 20 & 0.81 & 0.28 &  0.701 &  0.232 & B1-3 I & 2 & -5.6 & 0.4 \\
E68:IRS7W & 6.47 & -2.45 & 5.99 & \nodata & \nodata & 13.1 & 185 & 29 & 36 & 28 & -305 & 100 & -0.98 & 0.15 &  0.155 &  0.583 & WC9 & 2 & -4.6 & 0.4 \\
E69 & 6.58 & 1.81 & -6.32 & \nodata & \nodata & 11.1 & 202 & 29 & 91 & 28 & 153 & 50 & 0.99 & 0.13 &  0.791 &  0.359 & ? & 1 & -5.5 & 0.6 \\
E70:IRS7E2(ESE) & 6.64 & 4.41 & 4.97 & \nodata & \nodata & 12.9 & 203 & 28 & -7 & 26 & -80 & 100 & -0.77 & 0.13 &  0.714 &  0.104 & Ofpe/WN9 & 2 & -4.1 & 0.5 \\
E71 & 6.68 & 1.59 & 6.49 & \nodata & \nodata & 14.1 & -148 & 30 & 189 & 29 & -300 & 150 & 0.79 & 0.13 &  0.730 &  0.284 & WC8/9 ? & 1 & -3.8 & 0.6 \\
E72 & 6.73 & 6.71 & -0.50 & \nodata & \nodata & 13.6 & 65 & 28 & 100 & 28 & 86 & 100 & 0.87 & 0.24 &  0.555 &  0.243 & WC9? & 2 & -3.0 & 0.3 \\
E73 & 7.73 & -1.08 & 7.65 & \nodata & \nodata & 11.5 & -160 & 50 & 22 & 50 & -92 & 40 & 0.96 & 0.31 &  0.373 &  0.353 & O9-B I & 2 & -5.1 & 0.3 \\
E74:AFNW & 8.42 & -7.63 & -3.57 & \nodata & \nodata & 11.7 & -67 & 28 & -92 & 28 & 70 & 70 & 0.48 & 0.25 &  0.932 &  0.055 & WN8 & 2 & -4.5 & 0.4 \\
E75 & 8.53 & -0.02 & 8.53 & \nodata & \nodata & 11.0 & -35 & 45 & 226 & 40 & -138 & 40 & 0.15 & 0.20 &  0.727 &  0.405 & O9-B I & 2 & -5.8 & 0.5 \\
E76:IRS9SW & 9.10 & 4.28 & -8.03 & \nodata & \nodata & 13.1 & 108 & 49 & 8 & 45 & 180 & 80 & 0.91 & 0.45 &  0.521 &  0.374 & WC9 & 2 & -3.4 & 0.3 \\
E77 & 9.23 & -1.23 & 9.15 & \nodata & \nodata & 13.6 & \nodata & \nodata & \nodata & \nodata & -155 & 50 & \nodata & \nodata & \nodata & \nodata & O9-B0 V & 2 & -3.3 & 0.6 \\
E78:[PMM2001] B1\tablenotemark{c} & 9.47 & 9.46 & 0.31 & \nodata & \nodata & 13.0 & -161 & 46 & -142 & 55 & -230 & 100 & -0.64 & 0.26 &  0.781 &  0.216 & WC9 & 2 & -3.4 & 0.6 \\
E79:AF & 9.51 & -6.54 & -6.91 & \nodata & \nodata & 10.8 & 68 & 36 & 50 & 36 & 160 & 30 & 0.18 & 0.43 &  0.991 &  0.016 & Ofpe/WN9 & 2 & -5.7 & 0.8 \\
E80:IRS9SE & 9.93 & 5.65 & -8.17 & \nodata & \nodata & 11.7 & -2 & 36 & -131 & 36 & 130 & 100 & -0.58 & 0.28 &  0.766 &  0.181 & WC9 & 2 & -5.2 & 0.6 \\
E81:AFNWNW & 9.97 & -9.63 & -2.58 & \nodata & \nodata & 12.6 & 87 & 31 & -9 & 38 & 30 & 70 & 0.36 & 0.43 &  0.873 &  0.115 & WN7 & 2 & -4.9 & 0.9 \\
E82:Blum & 10.14 & -8.63 & -5.33 & \nodata & \nodata & 13.0 & -53 & 34 & 249 & 46 & -70 & 70 & -0.94 & 0.17 &  0.646 &  0.467 & WC8/9 & 2 & -3.7 & 0.8 \\
E83:IRS15SW & 10.15 & -1.58 & 10.02 & \nodata & \nodata & 12.0 & -55 & 39 & -32 & 38 & -180 & 70 & 0.93 & 0.62 &  0.863 &  0.135 & WN8/WC9 & 2 & -5.5 & 0.4 \\
E84 & 10.24 & 0.08 & 10.24 & \nodata & \nodata & 11.3 & -119 & 42 & 74 & 42 & -250 & 40 & 0.85 & 0.30 & \nodata & \nodata & O9-B I & 2 & -6.2 & 0.5 \\
E85 & 10.63 & 9.68 & 4.39 & \nodata & \nodata & 12.8 & \nodata & \nodata & \nodata & \nodata & -150 & 40 & \nodata & \nodata & \nodata & \nodata & OB  & 2 & -3.7 & 0.4 \\
E86 & 10.71 & -0.53 & 10.72 & \nodata & \nodata & 15.0 & 93 & 39 & 73 & 40 & -205 & 50 & -0.82 & 0.32 &  0.684 &  0.434 & OB V ? & 2 & -1.6 & 0.4 \\
E87 & 11.25 & 2.58 & 10.94 & \nodata & \nodata & 13.7 & -88 & 39 & -82 & 37 & -120 & 30 & 0.56 & 0.32 &  0.933 &  0.072 & B V/III & 2 & -3.5 & 0.8 \\
E88:IRS15NE & 11.76 & 1.38 & 11.68 & \nodata & \nodata & 11.8 & -8 & 39 & 103 & 46 & -65 & 40 & 0.19 & 0.37 &  0.877 &  0.114 & WN8/9 & 2 & -5.5 & 0.5 \\
E89 & 12.27 & 0.00 & 12.27 & \nodata & \nodata & 14.5 & 108 & 40 & 17 & 35 & -100 & 40 & -0.99 & 0.37 &  0.360 &  0.314 & B1-3  V & 2 & -2.9 & 0.6 \\
E90 & 13.24 & 11.87 & 5.86 & \nodata & \nodata & 12.1 & \nodata & \nodata & \nodata & \nodata & -190 & 40 & \nodata & \nodata & \nodata & \nodata & O9-B1 I? & 2 & -4.5 & 0.4 \\
\enddata
\tablenotetext{a}{for stars E1--E10, we quote Eisenhauer et al. (2005).}%
\tablenotetext{b}{cluster core (multiple object).}%
\tablenotetext{c}{designation from Paumard et al. (2001).}
\tablecomments{Each ``$\pm$'' column gives the 1${\sigma}$ uncertainty on the
  column directly to the left. These columns give in order: name(s) of the
  star; projected distance to Sgr~A*; 3D position ($z$ is derived by
  Beloborodov et al. 2006, submitted); apparent K magnitude; 3D velocity;
  sky-projected angular momentum $j$ (eq.~\ref{eq:j}); eccentricity (see
  Appendix~\ref{app:MC}); stellar type; quality (2=highest, 1=good); absolute
  K magnitude. $M_K$ and $e$ (exceot for E1--E10) assume
  $R_0=7.62\pm0.32$~kpc.  $p$, $x$, $y$, and $z$ are in (equivalent)
  arcseconds. All velocities are in \kms, assuming $R_0=8$~kpc for $v_x$ and
  $v_y$.}
\end{deluxetable}
\clearpage
\pagestyle{plain}

\subsection{Determination of Velocities}

For the emission line stars we deduced stellar radial velocities using a
variety of techniques. We fitted simple Gaussian profiles wherever possible
and averaged values obtained from different lines, giving larger weight to
single transitions. For lines with P~Cyg profiles we fitted a combination of
an emission and an absorption line. We also constructed template spectra for
well identified WR, Ofpe/WN9 and LBV profiles, either from the Galactic Center
stars themselves, or from the literature (Figer et al. 1997). Velocities were
then obtained from cross-correlation. For those stars for which velocities are
available in the literature (Genzel et al. 2000, 2003; Paumard et al. 2001;
Najarro et al. 1994, 1997), we averaged our results with the earlier values.
In the analysis of Najarro et al. (1994, 1997) the stellar velocity was a fit
parameter in an overall radiative transport, stellar atmosphere model of the
line profiles. Overall we find that for the wind-dominated stars the accuracy
of velocity determinations is dominated by the large velocity widths and
complex line profiles. In a few cases where we have line profiles over a
number of years we find some evidence for variability in the line profiles.
The 1${\sigma}$ uncertainties of the velocities are typically
$\pm50$--100~km/s.

The situation is much more straightforward for the new OB supergiants,
giants and main-sequence stars. In this case we are dealing mostly
with optically thin absorption lines of well determined transitions and
with simple line profiles. An exception is the vicinity of HI
Br${\gamma}$ for the OB supergiants and giants. In these cases the line
profiles clearly show evidence for \HeI\ 7--4, which is a complex of 7
transitions between $-80$ and $-1000$~\kms\ blueward of \HI\ 7--4
(Najarro et al. 1994). The relative strengths of \HeI\ 7--4 and \HI\
7--4 is abundance and atmosphere dependent and needs to be treated as
a free parameter. As a result of this fairly simple situation, the
1${\sigma}$ accuracy of velocity determinations is mainly limited by
signal-to-noise ratio and line width and can be as good as
$\pm20$~\kms.

\section{RESULTS}
\label{sect:res}
\subsection{OB Stars Are Finally Detected}

\begin{deluxetable}{lrrrrrrrrrrrrl}
  \tablecaption{Properties of Candidate Early Type Stars (Quality 0) in the
    Central Parsec.\label{table:big2}}
\tabletypesize{\small}
  \tablehead{\colhead{Name} &
    \colhead{$p$} & \colhead{$x$} & \colhead{$y$} & \colhead{$m_K$} &
    \colhead{$v_x$} & \colhead{$\pm$} & \colhead{$v_y$} & \colhead{$\pm$} &
    \colhead{$v_z$} & \colhead{$\pm$} & \colhead{$j$} & \colhead{$\pm$} &
    \colhead{Type}}
  \startdata
  S1-1 & 1.01 & 1.01 & 0.02 & 13.2 & 223 & 22 & 73 & 22 & ? & ? & 0.29 & 0.09 & E?\\
  ~ & 1.05 & -0.31 & -1.00 & 16.0 & -348 & 28 & -341 & 27 & ? & ? & -0.47 & 0.06 & E?\\
  ~ & 1.05 & 0.79 & -0.69 & 12.5 & 429 & 29 & 137 & 29 & ? & ? & 0.86 & 0.06 & E?\\
  ~ & 1.12 & -0.97 & 0.56 & 15.6 & -40 & 28 & -88 & 28 & ? & ? & 1.00 & 0.29 & E?\\
  ~ & 1.47 & -0.55 & -1.37 & 15.5 & -16 & 28 & -28 & 28 & ? & ? & -0.14 & 0.87 & E?\\
  ~ & 1.65 & 0.37 & -1.61 & 13.8 & 281 & 29 & -131 & 30 & 217 & 60 & 0.79 & 0.09 & OBIII?\\
  ~ & 2.34 & 2.32 & -0.26 & 12.9 & -30 & 28 & 227 & 28 & 49 & 20 & 0.97 & 0.12 & OB??\\
  ~ & 4.85 & -4.11 & -2.58 & 16.2 & -53 & 28 & 154 & 31 & -32 & 71 & -0.97 & 0.18 & OBIII\\
  IRS 7SE2 & 4.95 & 3.06 & 3.89 & 13.7 & 42 & 28 & -71 & 28 & -85 & 100 & -0.93 & 0.34 & WC\\
  ~ & 5.08 & -4.86 & 1.47 & 16.3 & 107 & 29 & -19 & 29 & 88 & 71 & -0.12 & 0.26 & OBIII\\
  & 5.80 & 3.20 & -4.84 & 12.6 & -84 & 28 & -134 & 27 & 3 & 70 & -0.91 & 0.18 & E?\\
  ~ & 6.26 & 1.54 & -6.07 & 15.8 & 42 & 29 & 102 & 29 & 128 & 50 & 0.60 & 0.27 & OBIII?\\
  ~ & 6.38 & 6.33 & 0.81 & 14.9 & -163 & 29 & 44 & 28 & 208 & 54 & 0.38 & 0.17 & OBIII?\\
  ~ & 7.11 & 6.96 & 1.43 & 15.5 & -104 & 30 & -42 & 30 & -12 & 71 & -0.18 & 0.27 & OBIII\\
  ~ & 7.45 & -4.07 & -6.24 & 15.3 & 31 & 30 & 91 & 33 & 83 & 50 & -0.25 & 0.32 & OBIII?\\
  ~ & 7.87 & -7.76 & 1.36 & 10.6 & 162 & 53 & 152 & 49 & 148 & 50 & -0.80 & 0.22 & OBIII?\\
  ~ & 8.19 & -4.81 & 6.63 & 15.8 & 54 & 32 & 28 & 33 & 229 & 51 & -0.99 & 0.53 & OBIII \\
  ~ & 9.73 & -6.32 & -7.39 & 12.6 & 52 & 39 & -95 & 37 & 108 & 51 & 0.93 &
  0.35 & OBIII?
\enddata
\end{deluxetable}
Our observations have led to the firm detection of 29 OB supergiants
(luminosity class I+II), as well as 12 OB stars of luminosity class III and V.
They are listed in Table~\ref{table:big1} as quality 1 and 2. In addition we
have 18 OB candidates whose identifications we regard as
tentative (quality 0: Table~\ref{table:big2}). Those additional stars need to
be confirmed. All these detections refer to the region outside the central
cusp, with projected radius $p_\mathrm{SgrA*}\geq0.8\arcsec$.  Eisenhauer et
al. (2005) have already reported 70~mas resolution SINFONI observations of
this central cusp, with the detection of more than a dozen main sequence B
stars, in addition to the late O9/B0 main sequence star S2 (S02) detected
earlier by Ghez et al. (2003). For completeness, these S-stars are listed as
the first 14 entries in Table~\ref{table:big1}.

\begin{figure}
\plottwo{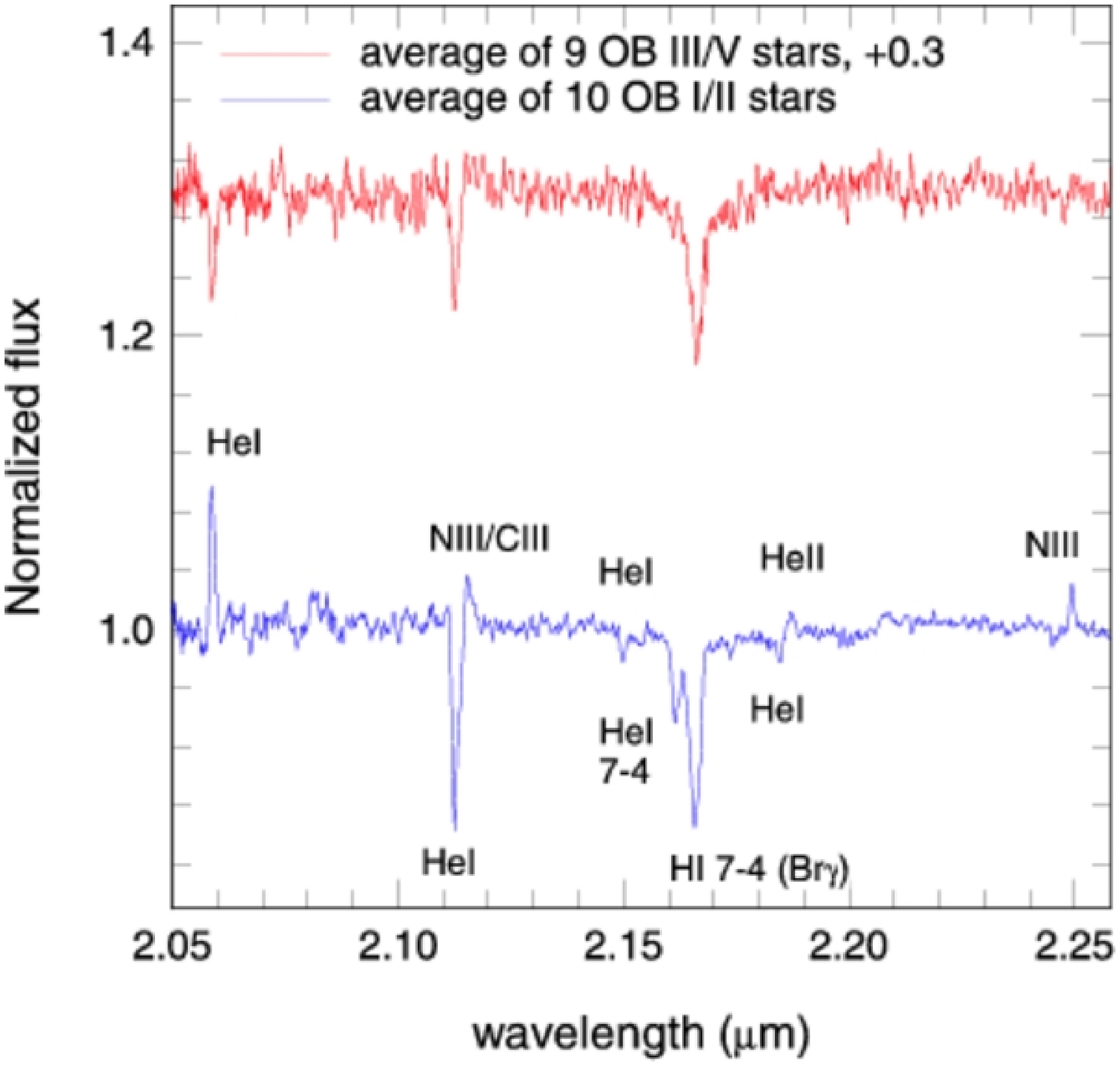}{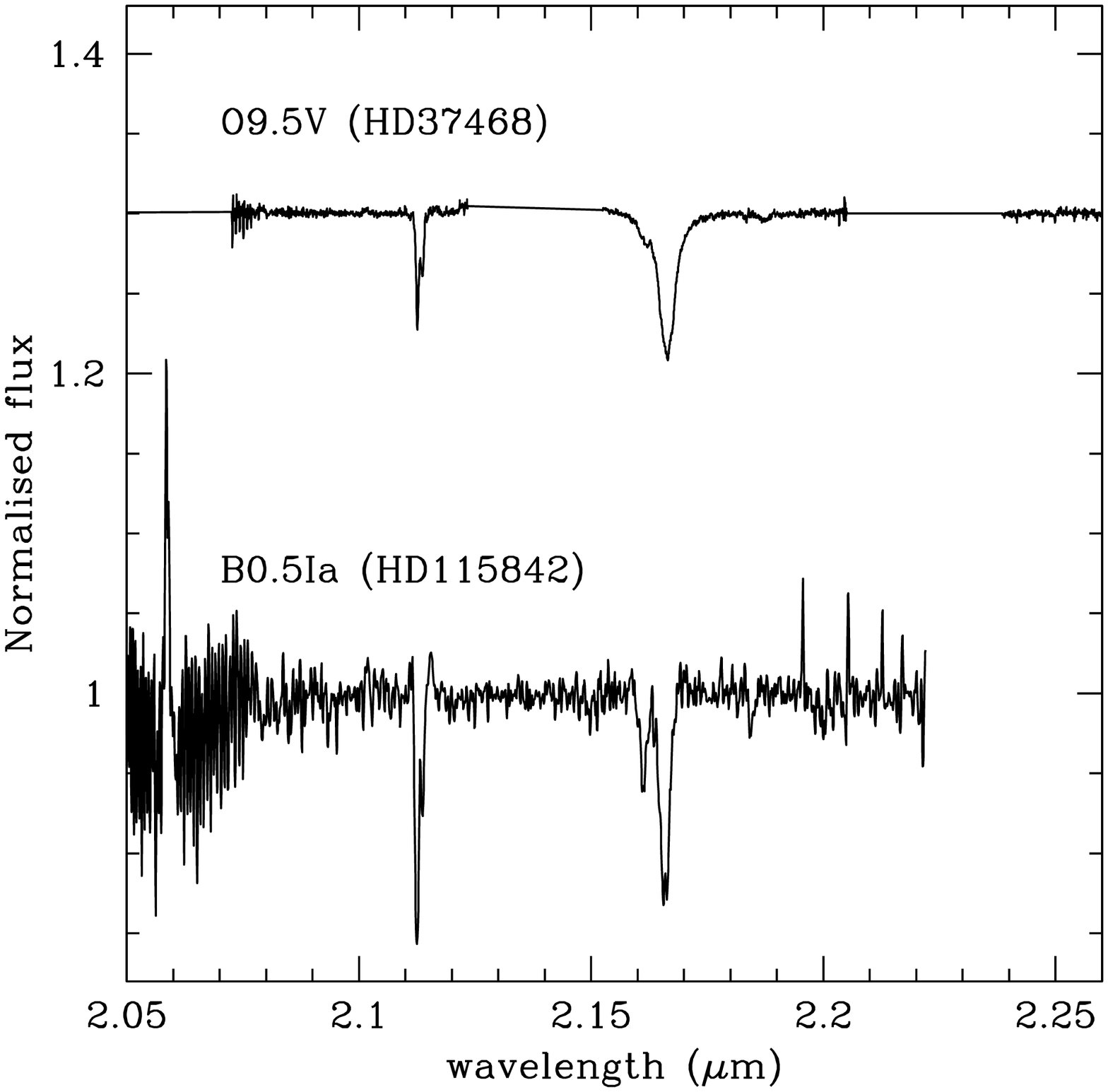}
\caption{\emph{Left:} Co-added, normalized and baseline subtracted spectra of
  the 10 best OB Iab--II supergiants (\emph{bottom}) and the 9 best O III--V
  (\emph{top}) stars in our sample.  For each star we determined the velocity
  and then shifted all stars to a common rest frame. The various transitions
  are labeled. For the top spectrum a constant of 0.18 was added.
  \emph{Right:} templates from Hanson et al. (2005) are overplotted on our
  spectra.\label{fig:avgspec}}
\end{figure}
Figure~\ref{fig:avgspec} shows the co-added spectra of the 10 best OB I stars
and the 9 best O III/V stars. A comparison with the atlas of Hanson et al.
(2005) shows that these two sample average spectra are very similar to their
respective solar neighborhood templates. After more than a decade of search,
our data finally reveal the missing OB population in the Galactic Center. It
is clear that the non-detection of these stars in earlier studies was merely
an instrumental effect.

In addition to the 41 new OB stars outside the central 0.85\arcsec, we
identify 30 post main-sequence blue supergiants and Wolf-Rayet stars, adding
several stars to the sample already known from previous work (Genzel et al.
1996, 2000, 2003; Paumard et al. 2001). Of these, we classify 17 as Ofpe/WN9
and late nitrogen rich WRs (WNL=WN7--9) stars, and 12 as carbon rich WRs (WC)
stars. There is one early WN (WNE, WN5/6) star, IRS~16SE2. The Northern Arm
bow-shock star IRS~1W, with \Brg\ in emission and \HeI\
$\lambda$~2.058~\micron\ in absorption as only features, is perhaps a Be star
(see Paumard et al. 2004a for spectrum and detailed discussion). There is also
one additional tentative WC candidate (IRS~7SE2).

Tanner et al. (2006) classify the 6 bright Ofpe/WN9 (narrow emission line
stars and LBV candidates from Paumard et al. 2001) as B stars on the basis
that they do not detect \NIII~$\lambda$~2.116~\micron\ emission in these
stars. They also do not detect the \HeI\ complex at 2.113~\micron\ for the
variable star and best LBV candidate IRS~34W (Trippe et al. 2005), in contrast
to the other 5. However we clearly detect these features although the \HeI\
feature is mostly in emission (with some P~Cyg absorption) in IRS~34W (see
spectra in Paumard et al. 2004b and Trippe et al. 2005).

Hanson et al. (1996) have suggested that these very bright stars in the IRS~16
cluster might be OBN stars. OBN stars are particular kinds of O and B stars
which show unusually strong N lines in the optical. They are also known to
show unusually strong He lines, and to be particularly bright because of a
lower atmospheric opacity (Langer 1992).  Indeed, the strengths of the
2.115~\micron\ \NIII\ compound and the 2.163~\micron\ \HeI\ 7--4 absorption,
relative to \Brg, in the average spectrum of the newly detected supergiants in
Fig.~\ref{fig:avgspec} suggests that many of the luminous OB stars in the
central parsec are nitrogen and helium rich. An extreme case is IRS~16CC where
the \HeI\ absorption line at 2.163~\micron\ is almost as deep as the \Brg\ 
line. The only stars of the Hanson et al. (1996, 2005) atlases to show
comparable depth in \HeI\ $\lambda$~2.163~\micron\ are HD~191781 and
HD~123008, two ON9.7~Iab stars.  Detailed modeling of our new stars is ongoing
(Martins et al. 2006).  Preliminary results seem to confirm a He enrichment
(He/H$\simeq0.3\pm0.1$) for the `average' star. This is still compatible with
standard evolutionary models with rotation, though. Based on simple
morphological arguments, the stronger absorption in \HeI\ 
$\lambda$~2.163~\micron\ in IRS~16CC and IRS~16SSE2 may indicate an even
larger helium enrichment, which could be in conflict with theoretical
predictions. More work is definitely needed to draw any reliable conclusion.

All these evolved stars are included in Tables~\ref{table:big1} and
\ref{table:big2}. In total, Table~\ref{table:big1} lists 90 certain detections
of early type stars.  Table~\ref{table:big2} has an additional 14 further
candidates. More than 100 early-type stars have now been detected in the
nuclear star cluster, and this number is expected to grow in the next years.

\subsection{Dynamics of the Young Stars}

\subsubsection{Two Disks of Early Type Stars}
\label{sect:disks}

Genzel et al. (1996) were the first to note that the twenty or so bright `HeI'
emission line stars between $p=1\arcsec$ and 12\arcsec\ known at that time
exhibit a coherent rotation pattern in their radial velocities. Stars north of
the center are blue-shifted while stars south of the center are red-shifted.
This pattern is opposite to Galactic rotation.  Genzel et al.  (2000) and
Paumard et al. (2001) confirmed and extended these findings.  Adding proper
motions to the radial velocities allowed a more constrained analysis (Genzel
et al. 2000, 2003; Levin \& Beloborodov 2003).  In the end, Genzel et al.
(2003) considered 26 stars with 3D velocity.  They used $j$, the normalized
angular momentum with respect to the line of sight, to demonstrate the
existence of two coherent star systems (Appendix~\ref{app:disks},
eq.~\ref{eq:j}) on near tangential orbits in projection ($|j|\simeq1$), one
rotating clockwise ($j\simeq+1$), the other counter-clockwise ($j\simeq-1$).
Using a $\chi^2$ argument proposed by Levin \& Beloborodov (2003;
eq.~\ref{eq:chi2}), they show that both systems fit disk solutions. 12--14
stars form the clockwise system. It is rather thin and its midplane has an
inclination of $i=120\degr\pm7\degr$ with respect to the plane of the sky and
a half-line of ascending (=receding) nodes at $\Omega=120\degr\pm15\degr$ east
of north (the actual numbers quoted in Genzel et al. 2003 are different
because of different conventions; a detailed definition of those used in the
present paper is given in Appendix~\ref{app:orbelms}).  The corresponding
normal vector is $\vect{n}=(n_x,n_y,n_z)=(-0.43,-0.75, +0.50)\pm(0.23, 0.17,
0.11)$.  This system is the one found earlier by Levin \& Beloborodov (2003).
The second, counter-clockwise system in Genzel et al.  (2003) is new, counts
10--12 stars, is thicker, and has $i=40\degr\pm15\degr$ and
$\Omega=160\degr\pm15\degr$ ($\vect n = (-0.6, -0.22, -0.77)\pm(0.25, 0.23,
0.17)$). The two systems are at large angles relative to each other
($87\degr\pm36\degr$). Tanner et al. (2006), adding 7 radial velocities and
improving on others, also fit disk solutions on their data (10 stars in the
clockwise system, 5 in the counter-clockwise). They find disk solutions in
good agreement with their predecessors: $\vect n=(-0.42, -0.65,
+0.76)\pm(0.05,0.03,0.06)$ and $(-0.23, -0.08, -0.97)$ $\pm0.13$ (after
normalization). From the rather high reduced $\chi^2$ they get, they conclude
that the disks must be somewhat thicker than previously thought, although they
make no quantitative statement.

\begin{figure}
\plotone{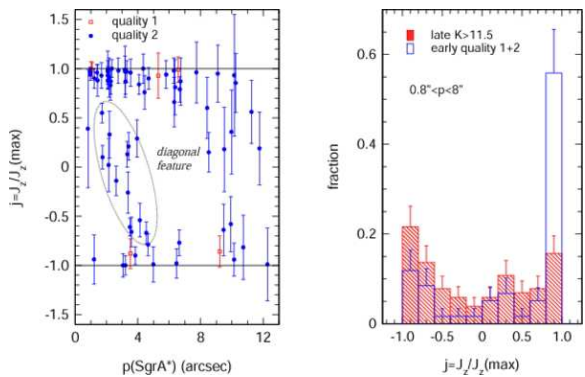}
\caption{\emph{Left}: Distribution of projected and normalized angular
  momentum on the sky $j=J_z/J_{z, \mathrm{max}}$ (eq.~\ref{eq:j}) for the early
  type stars as a function of projected separation $p$ from Sgr~A* (for
  $p>0.8\arcsec$). In this diagram stars on projected tangential, clockwise
  orbits are at $j\simeq+1$ while stars on tangential, counter-clockwise
  orbits are at $j\simeq-1$. Stars at $j\simeq0$ are on projected radial
  orbits. Filled circles denote the highest quality (2) spectroscopic stars
  and open squares denote moderate quality (1) stars. A diagonal feature (DF)
  is apparent on this diagram.  It seems statistically significant, but we
  have no physical unique interpretation to suggest.  \emph{Right}: histogram of the
  $j$ distribution of the quality 1--2 spectroscopic stars integrated over
  $0.8\arcsec<p<8\arcsec$ ($N=59$), compared to a sample of $N=102$ $m_K>11.5$
  spectroscopic late type stars that should be close(r) to relaxation.
  \label{fig:j}}
\end{figure}
Our new data increase the number of stars and the quality of the velocity
measurements very substantially. In Fig.~\ref{fig:j} we plot the same $j$ vs.
$p$ (the projected distance from Sgr~A*) diagram as in Genzel et al. (2003),
adding our new stars. We exclude stars in the central `S'-cluster
($p\leq0.8\arcsec$) that appear to be on randomly oriented, elliptical orbits
(Ghez et al. 2005; Eisenhauer et al.  2005). For $p\geq8\arcsec$ the velocities
are smaller and proper motion uncertainties increase. As a result the typical
uncertainty in $j$ increases to $\pm0.3$--0.5 and a detailed analysis is not
possible. For this reason we consider in the right panel of Fig.~\ref{fig:j}
the histogram of $j$ values for the 59 quality 1+2 stars for the range
$0.8\arcsec\leq p\leq8\arcsec$. We compare the distribution for the early type
stars to the distribution of 102 late type stars with $m_K>11.5$ in the same
range, which serve as a template for a relaxed distribution. 81\% of the early
type stars move on near-tangential orbits ($|j|>0.6$).  This is to be compared
to 59\% for the late type stars. Early type stars clearly are preferentially
on tangential orbits.

The clockwise system (CWS) at $j\simeq+1$ is particularly striking and now
contains 36 (40) quality 1+2 stars with $|j|\geq0.6$ and $p\leq8\arcsec$
(14\arcsec). The counter-clockwise system (CCWS) is less well populated with
12 (17) stars with the same quality criteria and limits as above. In fact
compared to the late type distribution in the right hand inset of
Fig.~\ref{fig:j} the enhancement in counter-clockwise stars at $|j|\geq0.6$
would not appear statistically significant.  We show below, however, that the
counter-clockwise stars do indeed lie in a common plane, just as the clockwise
stars and in contrast to the late type stars selected with the same criteria.
There are $\simeq10$ stars at $p<8\arcsec$ with small projected angular
momentum. Of these, several have large error bars in $j$ and could still be
part of the two tangential systems.  However, it is interesting to note that
these $\simeq$10 stars with $0.8\arcsec<p<8\arcsec$ and $|j|<0.6$ lie with a
fairly small scatter around a diagonal line that runs from $(p=0\arcsec,
j=0.6)$ to $(p=6\arcsec, j=0)$. Therefore this group of stars seems to have
some statistical significance, although its physical meaning is not yet clear.
We will later refer to these stars as the diagonal feature (DF) stars, as they
share other noteworthy characteristics.

\begin{figure*}
\plotone{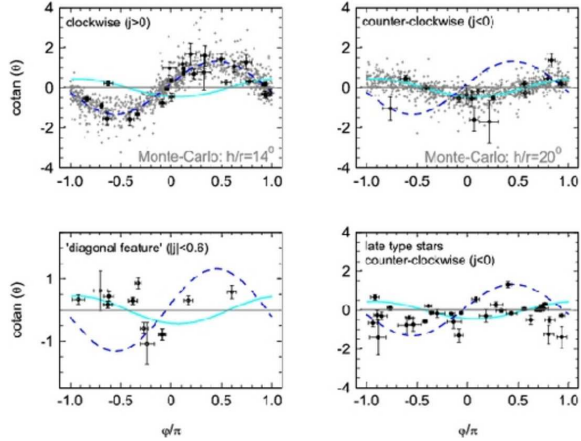}
\caption{Location of various stellar populations in the
  $\varphi$--$\cotan\theta$ plane described in Sect.~\ref{sect:disks} and
  Appendix~\ref{app:disks} (eq.  \ref{eq:cotan}): \emph{upper-left}: clockwise
  early-type stars ($j>0$); \emph{upper-right}: counter-clockwise early type
  stars ($j<-0$); \emph{lower-left}: diagonal feature stars (see
  Fig.~\ref{fig:j}: $|j|<0.6$, $p<8\arcsec$); \emph{lower-right}:
  counter-clockwise late type stars. Thick dots are quality 2 stars with
  $>3.5\sigma$ (average) detections of the 3 space velocities and (except for
  the DF stars) determination of $j$. Thin open circles are stars that have
  somewhat relaxed significance criteria ($2\sigma$).  The thick dashed line
  is the best fitting clockwise disk ($i=124\degr$, $\Omega=99\degr$) and the
  thick continuous line is the best fitting counter-clockwise disk
  ($i=24\degr$, $\Omega=167\degr$). The small crosses denote Monte-Carlo
  simulations ($\simeq10^{3}$ stars) of two disks with these parameters and a
  Gaussian distribution of opening angles ($\sigma=14\degr$ and $20\degr$
  respectively).\label{fig:cotan}}
\end{figure*}
The $\chi^2$ approach used by Levin \& Beloborodov (2003) and Genzel et al.
(2003) has the drawback that it is somewhat indirect. In the following we will
use a different way of looking for disk structures in our stellar data. This
new method is easy to visualize and can demonstrate the existence of a disk
independently from the determination of its parameters. We show in
Appendix~\ref{app:disks} (eq.~\ref{eq:cotan}) that, in the plane spanned by
$\varphi$ and $\cotan\theta$ ($\varphi$ and $\theta$ being the spherical
coordinates of the velocity vectors), stars located in a planar structure must
exhibit a telltale cosine pattern.  Figure~\ref{fig:cotan} shows the results
if stars are coarsely separated into CWS, CCWS and DF stars. Beside the value
of $j$ ($j<0$ for CCWS, $j>0$ for CWS, and $|j|<0.6$ for DF stars), the only
criteria we used was the quality of the data for selection (or rejection), as
determined from the average significance of the 3 space velocities
$(\sum_{k=x,y,z}v_k/\sigma_{vk})/3$ and the significance in the determination
of $\sign j$ ($j/\sigma_j$), and the projected distance $p$ from Sgr~A*. As
velocities decrease with $p$ and proper motion uncertainties increase with
$p$, the quality of the velocity and $j$ determinations decrease with $p$. We
find that quality 2 stars with velocity and $j$ significances $>3.5\sigma$
give by far the best determination of $\varphi$ and $\theta$. This essentially
selects stars at $|j|>0.6$ and $p<7\arcsec$ into the two tangential systems.
The DF is defined only by $|j|<0.6$ and $p<8\arcsec$.  In addition we also
considered more relaxed selection criteria ($\geq 2\sigma$).  Our main finding
is that there definitely are \emph{two well defined planar structures}, at
large angles with respect to each other ($115\degr\pm7\degr$), in the early
type star data. One at $i=127\degr\pm2\degr$ ($1\sigma$),
$\Omega=99\degr\pm2\degr$ ($\vect n=(-0.12,-0.79,0.60)\pm0.03$) fits all the
clockwise stars with our quality criteria but one, which is indeed a DF star
(\ref{fig:j}).  Again, this is the clockwise disk already found by Levin \&
Beloborodov (2003).

The second structure at $i=24\degr\pm4\degr$, $\Omega=167\degr\pm7\degr$
($\vect n=(-0.40,-0.09,-0.91)\pm(0.07,0.06,0.03)$) fits the all the
counter-clockwise stars (a few appear as outlyers but they have large error
bars). This second plane is coincident with the second plane identified by
Genzel et al.  (2003).  Remarkably it also fits very well 8 of the 11 DF
stars.  4 of the DF stars are compatible with both disks, 1 fits the CWS much
better, and 1 (IRS~16SSW) appears to fit neither. Several of the 5 DF stars
that fit best the CCWS have $j>0$, and therefore seem to counter-rotate in the
disk in which they fit best, but $j<0$ is not excluded by more than
$\simeq1.5\sigma$.

\begin{deluxetable}{lrrrrrrrrrrl}
\tablecaption{Planar Structures in the Galactic Center\label{table:planes}}
\tablewidth{0pt}
\tablehead{\colhead{Name} & \colhead{$\Omega$} & \colhead{$\pm$} &
  \colhead{$i$} & \colhead{$\pm$} & \colhead{$n_x$} & \colhead{$\pm$} &
  \colhead{$n_y$} & \colhead{$\pm$} & \colhead{$n_z$} & \colhead{$\pm$} &
  \colhead{Ref.}}
\tablecolumns{11}
\startdata
CWS    &  99\phd\phn&  2\phd\phn& 127 &  2 & -0.12 & 0.03 & -0.79 & 0.03 & +0.60 & 0.03 & \\
CCWS   & 167\phd\phn&  7\phd\phn&  24 &  4 & -0.40 & 0.07 & -0.09 & 0.06 & -0.91 & 0.03 & \\
Galaxy &  31.4      &  0.1      &  90 &  0 & +0.85 & 1e-3 & -0.52 & 1e-3 & +0.00 & 0.00 & 1 \\
Northern Arm\tablenotemark{a} &
          15\phd\phn& 15\phd\phn&  50 & 30 & +0.74 & 0.34 & -0.20 & 0.28 & -0.64 & 0.40 & 2 \\
Bar\tablenotemark{b} &
         115\phd\phn&  \nodata & 76&\nodata& -0.41&\nodata& -0.88&\nodata& -0.24&\nodata& 3 \\
CND    &  25\phd\phn&  \nodata & 70&\nodata& +0.85&\nodata& -0.40&\nodata& -0.34&\nodata& 4 \\
\enddata
\tablecomments{The parameters listed here are defined in
  Appendix~\ref{app:conventions}. They give the orientation of the given disks
  as two angles, and as one normal vector. Values from other papers have been
  translated into our conventions.}
\tablenotetext{a}{Northern Arm of the mini-spiral; values are approximate
  averages for the best defined part ($\mathrm{index}<30$) on Fig.~6 in (2).}
\tablenotetext{b}{Bar of mini-spiral.}
\tablerefs{(1) Reid \& Brunthaler (2004); (2) Paumard et al. (2004b); (3)
  Liszt (2003); (4) Jackson et al. (1993).}
\end{deluxetable}
\begin{figure}
\plotone{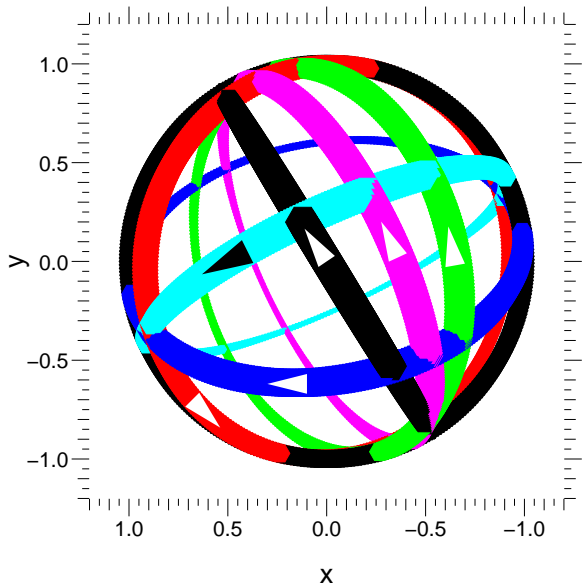}
\caption{Various planar structures in the Galactic Center
  (Table~\ref{table:planes}): the Galaxy and sky
  (\emph{black}), the clockwise (\emph{blue}) and counter-clockwise
  (\emph{red}) stellar systems, the Northern Arm (\emph{green}) and Bar
  (\emph{cyan}) components of the ionized mini-spiral, and the circum-nuclear
  disk (CND) of molecular gas (\emph{magenta}).  Each plane is represented by
  one ring. The thickness of the ring figures the proximity to the observer.
  Arrows indicate the direction of rotation.
  \label{fig:planes}}
\end{figure}
Most of the bright stars in the so-called `IRS~16' complex a few arcseconds
east and south-east of Sgr~A* are part of the clockwise system. This includes
IRS~16C and IRS~16SW. Several stars in the IRS~13E complex, as well as
IRS~16NE and IRS~16NW are part of the CCWS. In Table~\ref{table:planes}, we
list the various planar structures in the Galactic Center, including the
clockwise and counter-clockwise systems.  Figure~\ref{fig:planes} illustrates
the relative orientation of these planes. Although the CND and the Northern
Arm of Sgr~A West are at relatively low inclination to the plane of the
Galaxy, it is clear that the two stellar disks, the Bar of Sgr~A West and the
Galactic plane are all quite different from each other.

Basically the same results are obtained from the alternative and independent
method of Levin \& Beloborodov (2003; eq. \ref{eq:chi2}). For the CWS, this
method gives $i=124\degr\pm2\degr$ and $\Omega=100\degr\pm3\degr$ and for the
CCWS $i=30\degr\pm4\degr$ and $\Omega=167\degr\pm9\degr$.

Beloborodov et al. (2006, submitted) make a detailed analysis of the innermost
region of the CWS ($p<5\arcsec$) using the ``orbital roulette'' method (Beloborodov \& Levin 2004), which allows them to derive an independent estimate of the mass of Sgr~A*. This part of the CWS appears quite thin:
$10\pm4\degr$.  The best fit plane is then $\vect n=(-0.14,-0.86,+0.50)$. In
Table~\ref{table:big1}, we list the $z$ coordinate they derive for this
solution.

\subsubsection{The Disks Have Moderate Geometric Thickness}
\label{sect:disks:thickness}
The best fit $\chi^2_\mathrm{r}$ values range between 2.3 and 3.1 for both disk
systems, and for both equations \ref{eq:cotan} and \ref{eq:chi2}. The disks
are very well defined but the data require a finite thickness. If lower
quality stars are added for the fitting the resulting $\chi^2_\mathrm{r}$
increases to values above 4. We interpret this effect as likely being caused
by additional systematic uncertainties in velocity determinations, especially
for stars at $p>6\arcsec$ and for stars with poorer or broader spectral
features.

We have carried out Monte-Carlo simulations to determine the geometric
thickness of the disks. We computed the location of $\simeq10^3$ stars each in
the ${\varphi}$--$\cotan\theta$ plane, assuming a normal distribution for the
orbital inclinations to the system's midplanes. We have varied the $\sigma_i$
width of these distributions and also taken into account the errors in the
velocity determinations. We then compared the resulting model disks with the
data to determine the best fit distributions (Fig.~\ref{fig:cotan}).
For the clockwise set the best fitting value is $\sigma_i=14\degr\pm4\degr$
($\langle|h|/R\rangle=0.12\pm0.03$). For the best 11 counter-clockwise stars, the best fit
thickness is $19\degr\pm10\degr$ ($\langle|h|/R\rangle=0.16\pm0.06$).  \emph{The two
  stellar disks have significant but moderate geometric thickness.}

\subsubsection{Steep Radial Density Profile and Inner Cutoff}

We have estimated the 3D position of each star by assuming it is on the
corresponding system's midplane or, in other words, under a a very thin disk
model assumption. The thickness of the disk introduces an error. The projected
position on the CWS midplane of a CWS star at the average elevation is offset
by $\simeq0.1R$ perpendicular to the line of node ($0.07R$ for the CCWS). This
effect can be in either direction though, depending on whether the star is in
front or behind the midplane. Therefore, on average, this effect does not
introduce a significant bias.

\begin{figure}
  \plotone{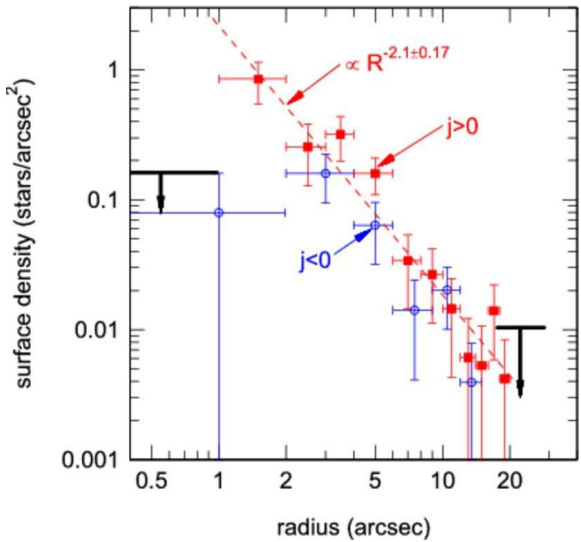}
  \caption{Surface density of stars in the clockwise (\emph{filled squares})
    and counter-clockwise (\emph{open circles}) disk systems, as a function of
    3D radius from Sgr~A* and derived in the framework of a `thin disk' model.
    The surface density distribution of the clockwise stars can be well fit by
    an extended disk with a sharp inner edge at $R\simeq1\arcsec$ and a
    surface density decreasing outward proportional to $R^{-2}$.  The
    counter-clockwise system more resembles a ring at $R\simeq4\arcsec$ (or
    has a large inner edge of about $2\arcsec$) but with the same outward
    surface density falloff as the clockwise system. No OB stars are seen
    outside about $p\simeq13\arcsec$ (Fig.~\ref{fig:obs}) despite the deep
    continuous coverage in the northern strip and deep fields.\label{fig:r2}}
\end{figure}
We have then calculated the surface density distribution as a function of true
radius for stars with quality 2 and with $v$ and $j$ significance $>1.5\sigma$
in the clockwise and counter-clockwise systems.  Figure~\ref{fig:r2} shows the
results. The clockwise disk has a well defined power-law, surface density
distribution with an index of $\simeq-2$.  There is a very well delineated
\emph{inner cutoff}, at $p\simeq R\simeq1\arcsec$. It is striking that this
inner cutoff appears to coincide almost exactly with the outer radius of the
central S-cluster of randomly oriented B-stars (Eisenhauer et al. 2005). The
CCWS has a larger inner cutoff ($\geq2\arcsec$) such that it appears to be
more like a ring centered at $R\simeq4\arcsec$, but this inner radius comes
down to roughly the same as that of the CWS when including the DF stars
($0.8\arcsec<p<8\arcsec$ and $|j|<0.6$).  Outside of this inner edge, however,
its surface density distribution is very much the same as that of the
clockwise system. The larger number of stars of the CWS as compared to the
CCWS (excluding DF, factor $\simeq$2.5) thus is mostly the result of the
former extending much further inward than the latter.

In addition to this, Fig.~\ref{fig:obs} shows already very clearly that the
early type stars all reside in the central $p\simeq13\arcsec$ ($\simeq0.5$~pc)
region despite our searches over a much larger region, and in several
directions. This non-detection of OB stars is a quite robust result.  In
particular, in the strip stretching from $y\simeq7\arcsec$ to $24\arcsec$
directly north of Sgr~A*, the effective magnitude limit (for significant
detection of spectroscopic features in early type stars) in this field is
$\simeq$15.5 across the entire field ($\simeq$110 square arcseconds) and $>17$
in two deep subsections of about 20 square arcseconds (Maness et al. 2005, in
preparation: Fig.~\ref{fig:obs}). This region is close to the visible AO
reference star so that the achieved Strehl ratio was high throughout the
observations. Judging from the extinction map of Scoville et al. (2003) the
excess K-extinction due to local dust is $<1$~mag throughout this region. This
puts a 1${\sigma}$ upper limit of $\simeq10^{-2}$ per arcsec$^2$ for O and
early B stars outside the cutoff radius of $p\simeq13\arcsec$. Over the
$>200$~arcsec$^2$ region outside $p\simeq13\arcsec$ covered by the shallower
(spectroscopic K limiting magnitude $\simeq13$) but wider large scale mosaics
a similar limit is deduced. This upper-limit is consistent with, and
strengthens, the $R^{-2}$ density profile extrapolated to $p>13\arcsec$.

\subsubsection{Isotropic Azimuthal Structure}

\begin{figure}
  \plotone{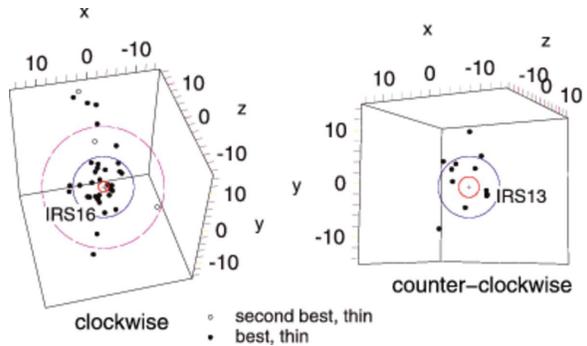}
  \caption{The azimuthal distribution of the stars in the clockwise ($j>0$,
    \emph{left inset}) and the counter-clockwise ($j<0$, \emph{right inset})
    systems. Filled circles denote the positions of the best stars
    ($>1.5\sigma$ significance of z-determination) in the `thin disk' model.
    Open circles denote the second best stars. The \emph{red circles} denote
    the inner edges ($1\arcsec$ for the CWS, $2\arcsec$ for the CCWS). The
    \emph{blue} and \emph{purple circles} have radii of $6\arcsec$ and
    $12\arcsec$ respectively.\label{fig:azimuthal}}
\end{figure}
Figure~\ref{fig:azimuthal} shows the azimuthal distribution of the stars in
the clockwise and counter-clockwise systems when viewed from the pole of the
two systems under the model assumption of very thin disks.  Again, the radial
distribution discussed in the last section, including the $\simeq1\arcsec$
sharp inner edge of the CWS and the tendency of stars in the CCWS to be at
large radii, perhaps in a ring-like shape, are apparent. The graphs also
clearly show that the azimuthal stellar distribution is azimuthally symmetric
to within the still fairly limited statistics of the data.  Apparent local
`concentrations' exist but because of small numbers they are all consistent with
statistical fluctuations.

This conclusion is somewhat in contrast to the work of Lu et al. (2005) who
have argued that the concentration of 5 stars near IRS~16SW
($\simeq1.5\arcsec$E, $1.5\arcsec$S of Sgr~A*) may be the surviving core of a
compact cluster. Their main argument is that the local 2D velocity dispersion
is at a global minimum on this group. While a group of 4--6 stars
corresponding to this `IRS~16 comoving group' is clearly visible in the left
inset of Fig.~\ref{fig:azimuthal}, to the left and down from center, there are
similar such groupings elsewhere in the disk, of similar (low) statistical
significance. However, what makes IRS~16 different from other locations in the
field is the presence of a `hole' in the CCWS at the same projected location
(this hole also has very low statistical significance). Therefore, Lu et al.
(2005) probably happen to be measuring the true velocity dispersion within the
CWS at the location of IRS~16, whereas elsewhere, their measurement must
include stars from the CCWS, and therefore naturally be higher.

Another interesting grouping of about 3 to 5 early type stars in a region of
less than $1\arcsec$ is IRS~13E, $\simeq4\arcsec$ south-west of Sgr~A*, in the
CCWS. This group has attracted recent interest, owing to
the proposal by Maillard et al. (2004) that it may be stabilized against tidal
disruption by an IMBH. We will return to this region
in a separate section (\ref{sect:irs13}).

\subsubsection{Circular and Non-Circular Motions in the Disks}
\label{sect:disks:e}

\begin{figure}
  \plottwo{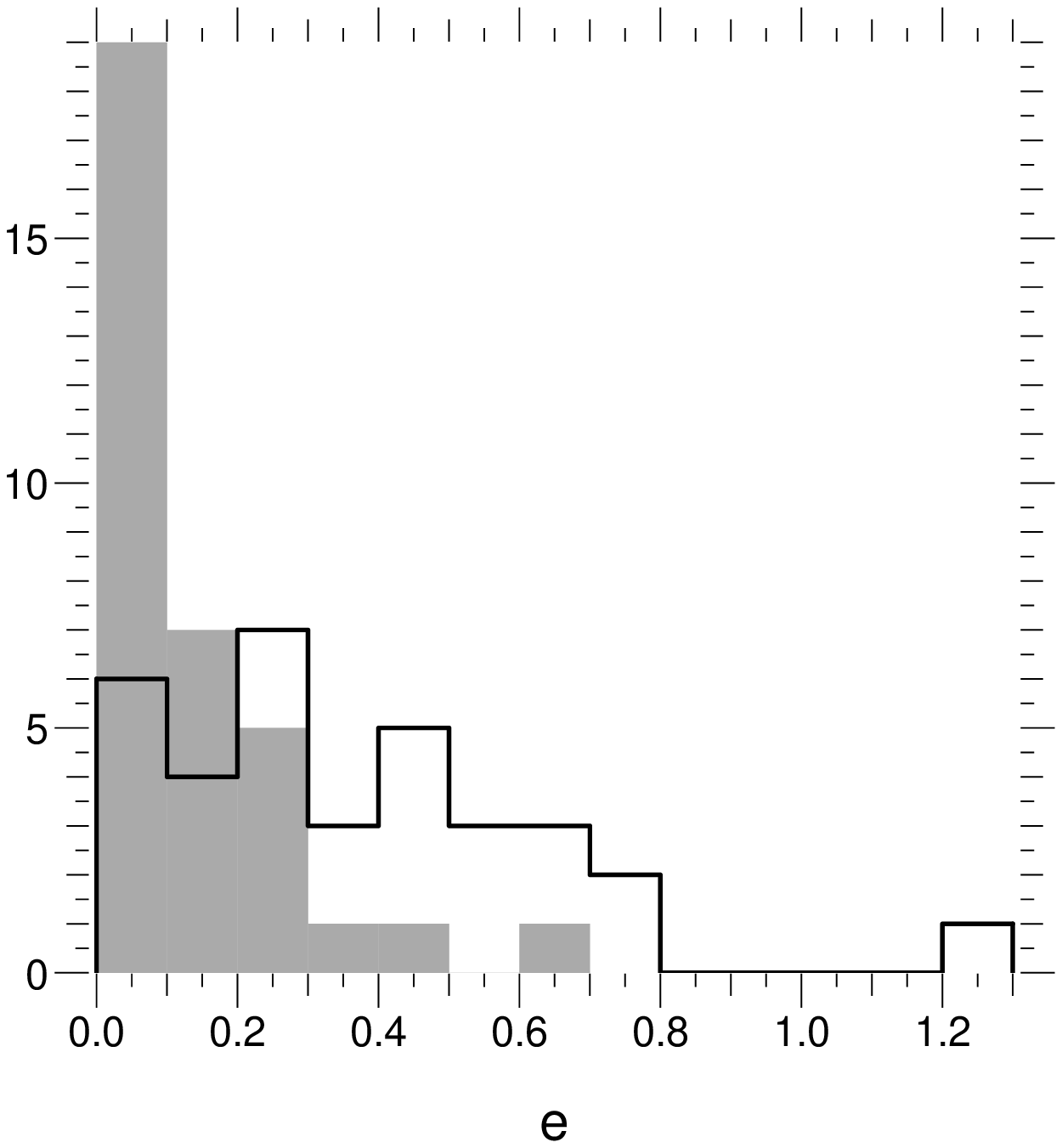}{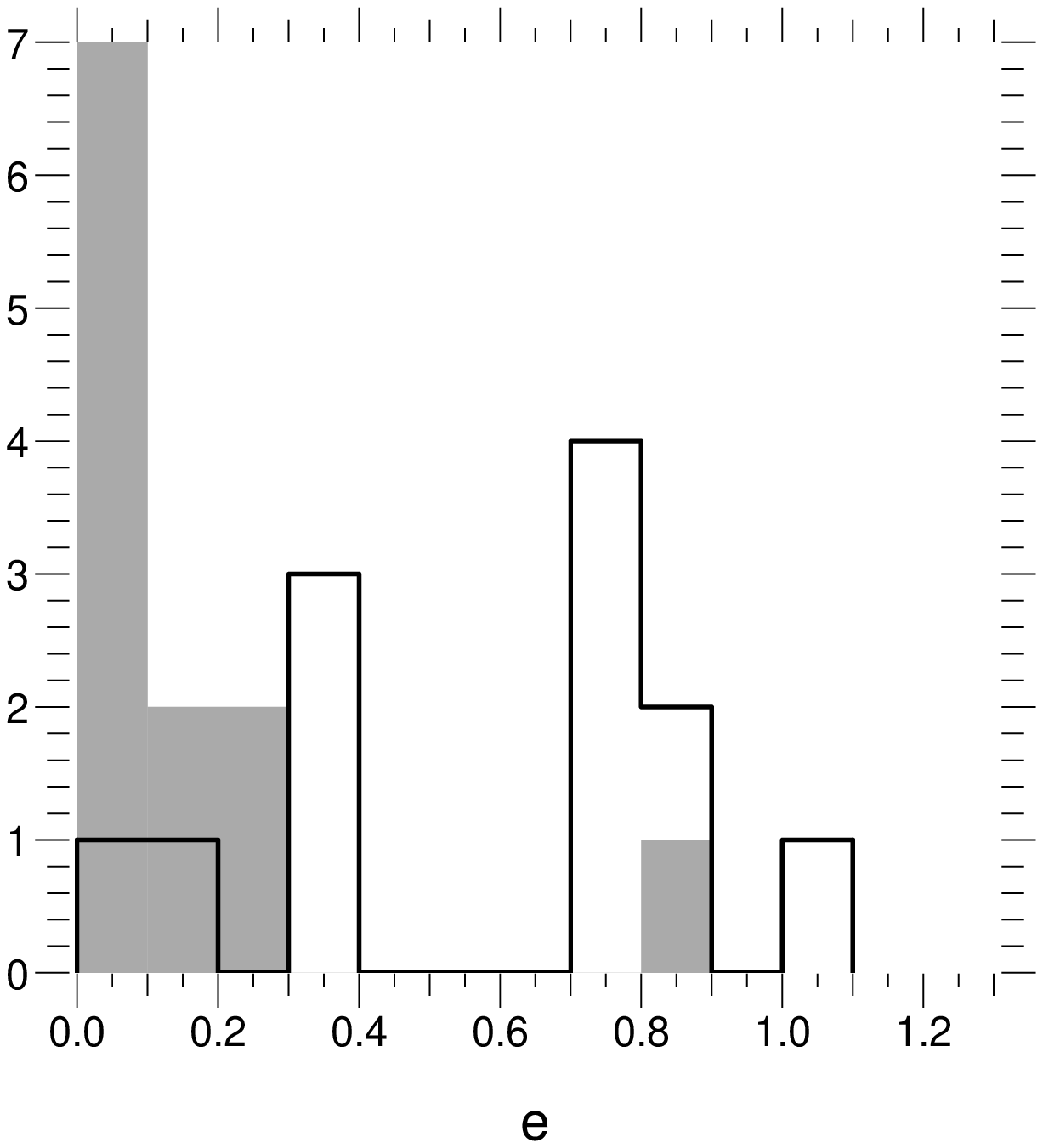}
  \caption{Eccentricity distributions in the CWS (left) and CCWS (right)
    disks, for both the real data sets (\emph{outlines}; all stars with $\vect
    v$ and $j$ significance $>2\sigma$) and artificial data sets for which all
    stars are assumed to be on circular orbits (\emph{filled histogram}).\label{fig:ehisto}}
\end{figure}
In this section we estimate the eccentricities of the stellar orbits in each
disk. This eccentricity is straightforward to derive once the line-of-sight
position ($z$) of a star is known (assuming the potential well and the distance to the GC, $R_0$,
 are known). We have used Monte-Carlo simulations to derive
estimates of the eccentricity $e$ of each star, under the assumption that it
belongs to one of the two disks. The method we used is described and discussed
(including its limitations) in Appendix~\ref{app:MC}. We produced artificial
data sets under various assumptions to validate the method. The individual
estimates we derive are listed in Table~\ref{table:big1}, and their histograms
for the CWS and CCWS are shown on Fig.~\ref{fig:ehisto}.
The completely independent method used by Beloborodov et al. (2006, submitted)
also allows them to derive stellar eccentricities. The two methods agree well
on the individual values.

These histograms show that the distributions of eccentricities of the two
systems are quite different.  The CWS is made of stars on low to medium
eccentricities ($\lesssim0.5$). For this system, we find only slightly higher
eccentricities than what we would derive for a system with the same geometry,
on which all the stars would be on circular orbits. The distribution peaks
around $e=0.2$, which is expected given the thickness of the disk
($\tan14\degr\simeq0.25$). \emph{The CWS is essentially in low eccentricity,
  close to circular motion}.

On the other hand, the CCWS contains a few low-eccentricity stars, including a
peak around $e=\tan19\degr\simeq0.35$, but is dominated by a high-eccentricity
($e\simeq0.8$) population.  Three of these stars belong to the IRS~13E complex
(Sect.~\ref{sect:irs13}), but even when counting these three stars as a single
dynamical entity, the conclusion remains that \emph{the CCWS is essentially in
  non-circular motion}. The same work performed for the DF stars shows that if
these stars live on either of the disks, then most of their eccentricities
have to be quite high (very close to 1).  These high eccentricity stars are
definitely bound to Sgr~A*.

\subsection{IRS~13E: a dissolving star cluster and/or the site of an
  intermediate mass black hole?}\label{sect:irs13}

The compact, bright source IRS~13E, $3\arcsec$ west and $1.5\arcsec$ south of Sgr~A*, merits
special discussion. IRS~13E has several properties that tend to make it unique
in the central cluster. It 
\begin{enumerate} 
\item is associated with a bright peak of dust and ionized gas emission in the
  mini-spiral, at the edge of the so-called `mini-cavity' (e.g.  Clenet et al.
  2004; Paumard et al. 2004b);
\item harbors at least $3$ bright stars within a radius of $0.25\arcsec$
  (Paumard et al. 2001);
\item is associated with a point-like X-Ray source (Baganoff et al.  2003;
  Muno et al. 2005);
\item is associated with a compact centimeter radio source (Zhao \& Goss
  1998);
\item is dynamically coherent in that its bright early type stars participate
  in a similar 3D space motion (Maillard et al.  2004; Sch\"odel et al. 2005).
\end{enumerate}
Maillard et al. concluded that IRS~13E consists of at least 7 stars, out of
which 4 have highly correlated sky velocities. From the radial velocity
difference between two of the sources, they inferred an enclosed mass in
excess of $10^3$ solar masses to bind the cluster. They argued that such a
mass could not be accounted for by (lower mass) stars in the cluster and that
the cluster might contain an intermediate mass black hole (IMBH) and be the
remaining core of an in-spiraling star cluster disrupted in the central parsec
and stabilized by this IMBH. On the basis of somewhat higher resolution images
and more accurate proper motions Sch\"odel et al.  (2005), using 4 sources,
concluded that IRS~13E could either be a local concentration of stars in the
counter-clockwise disk or a dissolving cluster core. By setting a lower limit
to the mass of a stabilizing intermediate mass black hole of $10^4\sunmass$
they felt that the presence of such an object is quite unlikely for two
reasons. First, such a massive black hole is not easy to form as a result of
core collapse even in a very massive ($10^6\sunmass$) in-spiraling cluster, as
core collapse typically creates a central concentration no more massive than
$\simeq10^{-3}$ of the original cluster mass (Portegies Zwart \& McMillan
2002). Second, an intermediate mass black hole with a mass in excess of
$10^4\sunmass$ would be inconsistent with the 2~\kms\ proper motion velocity
limit perpendicular to the Galactic plane deduced by the VLBA observations of
Reid \& Brunthaler (2004).

\subsubsection{IRS~13E is not a background fluctuation}
\label{sect:irs13:density}

\begin{figure*}
  \begin{center}\includegraphics[width=10cm]{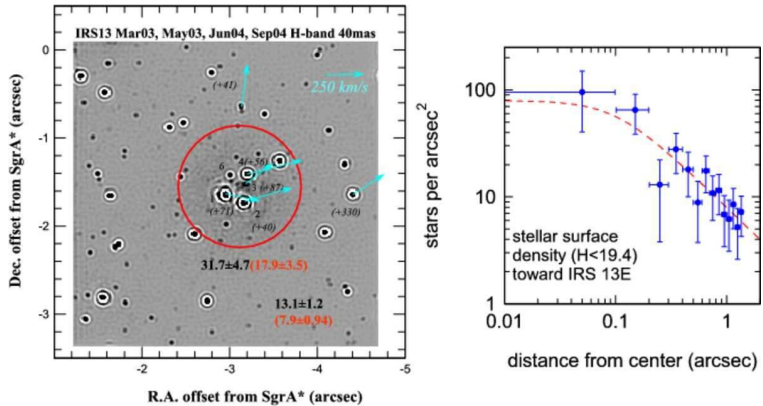}\end{center}
  \caption{\emph{Left}: Sensitive H-band image of the IRS~13E cluster region
    constructed from a combination of 4 NACO H-band data sets. The brightest
    stars have $m_K\simeq10.7$ and the image has a dynamical range of
    $\simeq9$ mags. The central circle centered on IRS~13E has a radius of
    $0.68\arcsec$. The main members of the IRS~13 group are labeled 1{\dots}6
    (nomenclature of Maillard et al.  2004), with radial velocities given in
    parenthesis. Two other members of the counter-clockwise system are in the
    image and are also marked by their radial velocities. Arrows indicate
    direction and magnitude of proper motion (calibration in the upper right
    corner). The numbers next to the circle are the surface densities of stars
    detected within it with the STARFINDER (Diolaiti et al. 2000) algorithm to
    $m_H=20.4$ and (in parenthesis) $m_H=19.4$ ($m_K\simeq18.5$ and 17.5),
    along with their 1${\sigma}$ uncertainties. For comparison, we list the
    surface densities (and uncertainties) to the same limits for the rest of
    the image shown.  \emph{Right}: surface density of stars to $m_H=19.4$, as
    a function of separation from the center of IRS~13E at $x=-3.12$,
    $y=-1.57$. The dotted curve is fit to a cluster with density profile
    $\propto1/(r^2+r_\mathrm{core}^2)$ and core radius
    $r_\mathrm{core}=0.17\arcsec$.\label{fig:irs13}}
\end{figure*}
In the following we will first use a statistical approach to determine through
stellar counts whether the IRS~13E group can be a chance association of stars
in projection. Figure~\ref{fig:irs13} shows a very deep H-band image we have
constructed from a combination of four high quality NACO images in 2003 (March
and May) and 2004 (June and September). Each of these images was individually
`cleaned' from the `dirty' AO PSF with the Lucy-Richardson algorithm and then
re-convolved with a 40~mas FWHM Gaussian.  For this purpose we constructed a
template PSF from isolated bright stars across the $\simeq3.5\arcsec$ field.
The co-added image was then again deconvolved with a Wiener filter with a PSF
constructed from fainter isolated stars immediately around IRS~13E.  The final
image shown in Fig.~\ref{fig:irs13} has a dynamical range of 9~mag and the
faintest significant stars have equivalent K-magnitudes of $\simeq19.5$.  We
then used STARFINDER (Diolaiti et al. 2000) to find and determine the
photometry of all stars in the field. We computed the surface densities for a
circular aperture of radius $0.68\arcsec$ centered on and encompassing the
core of IRS~13E, as well as for the rest of the $3.2\arcsec\times3.2\arcsec$
region shown in the figure. Fig.~\ref{fig:irs13} lists the surface densities
(and their ${1\sigma}$ uncertainties) to an H-magnitude limit of 20.4
($m_K\simeq18.5$). In parentheses we also give the same results for the more
conservative limit of $m_H\simeq19.4$. The stellar surface density in the
central aperture ($31.7\pm4.7$ ($17.9\pm3.5$) stars per arcsec$^{2}$) is 2.3
times greater than in the surrounding region ($13.1\pm1.2$ ($7.9\pm0.94$)
stars per arcsec$^{2}$).

Evaluating the significance of this result requires the careful use of Poisson
statistics given our prior knowledge. This computation is done in
Appendix~\ref{app:irs13}. We come to the conclusion that IRS~13E is very
unlikely a background fluctuation (a quite conservative upper limit on the
likelihood that this is the case is 0.2\%). We thus concur with Maillard et al.
(2004) that IRS~13E is very probably a local overdensity of stars in the CCWS.
We further note that the surface densities given above for the center of
IRS~13E are higher than anywhere but in the central cusp within $0.7\arcsec$
of Sgr~A* (Fig.~7 in Sch\"odel et al. 2006, in prep.). There, the surface
density to $m_H=19.4$ is $32.3\pm4.7$ stars per square arcsecond, 1.8 times
the value toward IRS~13E.  Because of the crowding it is not possible to
estimate an $m_H<20.4$ surface density toward Sgr~A*.  Toward the center of
the IRS~16 region, for instance, the surface densities to $m_H=20.4$ (19.4)
are $11.4\pm1.3$ ($9.3\pm1.2$) stars per square arcsecond, rather similar to
the average background density surrounding IRS~13E.

\subsubsection{The IRS~13E cluster is on an
  eccentric orbit} 

\begin{figure}
\plotone{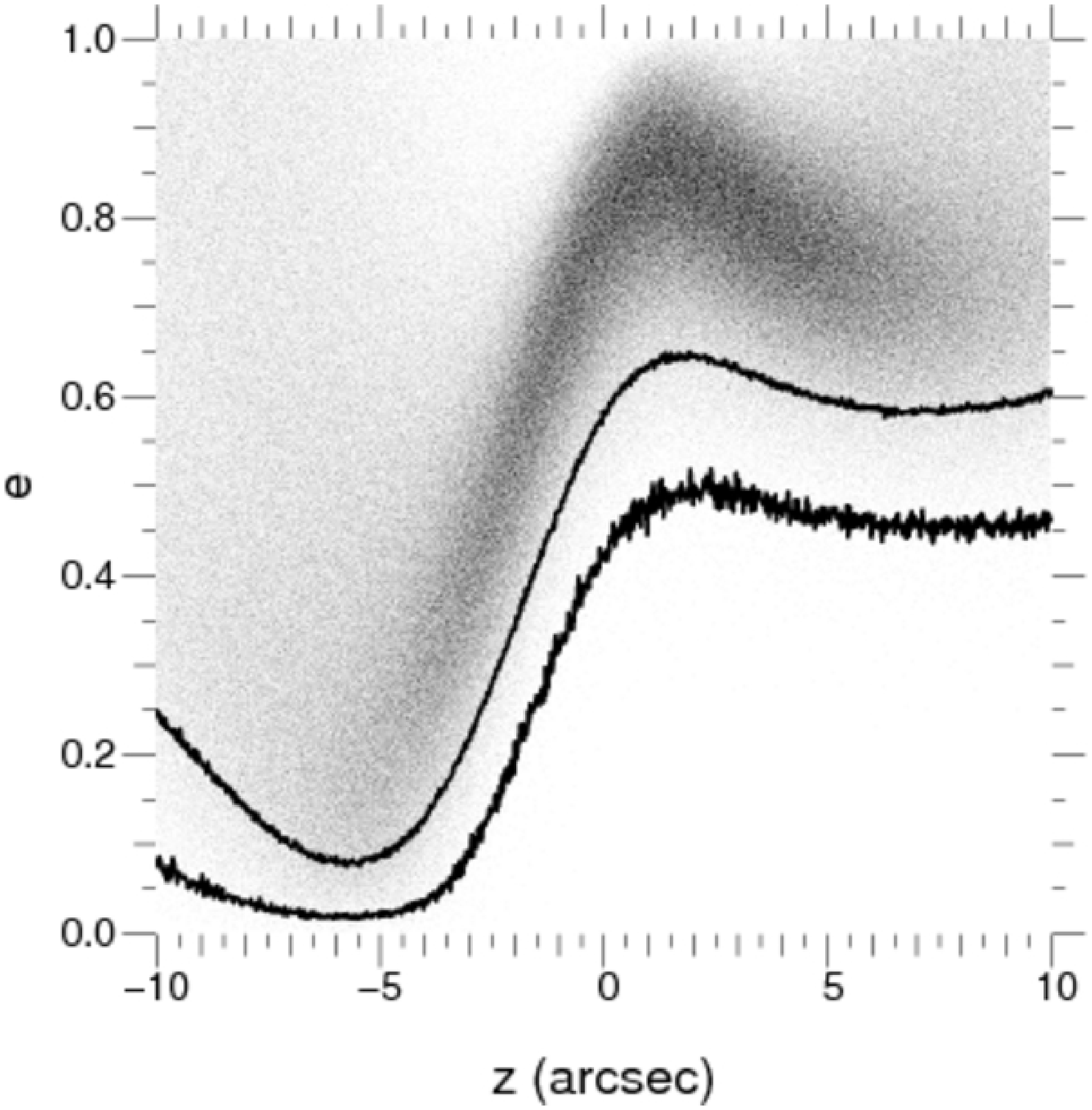}
\caption{Monte-Carlo simulation of the eccentricity
  of the IRS~13E group motion. Given the known distribution of probability for
  the accessible parameters ($R_0$, $M_\mathrm{SgrA*}$, $x$, $y$, $v_x$, $v_y$,
  $v_z$), we have computed the distribution of probability for $e$
  as a function of $z$ (shading). The two curves
  trace the 2 and $3\sigma$ lower limits on $e$. The map is not corrected for
  the low number depletion issue (Appendix~\ref{app:MC}). The data would be
  compatible with circular motion if $z\simeq-5\arcsec$.
  \label{fig:irs13:MC}}
\end{figure}
We now return to the remarkable deviation from circular motion of IRS~13E as
estimated from the Monte-Carlo simulations in the last section
(\ref{sect:disks:e}).  In the context of the cluster scenario it is now
possible to average the space motions of the 4 bright stars of IRS~13E and
obtain a more precise estimate for the motion of the cluster:
$(x,y)=(-3.12\arcsec\pm0.10\arcsec,-1.57\arcsec{\pm}0.12\arcsec)$, $(v_x, v_y,
v_z)=(-253\pm62, 57\pm48, 98\pm21)$~\kms. We have also some more information
on the line-of-sight position of the cluster. First, three of the four bright
stars in IRS~13E (13E2, 3 and 4) are formally in the counter-clockwise system
(CCWS). 13E1 is formally a DF star, and is also compatible with the CCWS (at
high eccentricity, though). In the framework of this disk, IRS~13E is at
$z\simeq1.5\arcsec$. Second, Paumard et al. (2004) argue that IRS~13E is
physically close to the `Bar' component of the ionized mini-spiral, which they
demonstrate is somewhat behind Sgr~A*. Furthermore Liszt (2003) proposes a
model of the Bar as a ring orbiting Sgr~A* at 0.3{--}0.8~pc
($7.7\arcsec${--}$20\arcsec$), which would put IRS~13E at
$z=6.8\arcsec${--}20\arcsec. It seems therefore rather safe to assume that
IRS~13E is at positive $z$. A distance of $z\simeq7\arcsec$ be in reasonable
agreement with both constraints (proximity to the Bar and to the CCWS), with
an orbial inclination to the CCWS midplane $i_\mathrm{CCW}\simeq1.8$--4.2
($1\sigma$) times the CCWS half opening angle.  The same Monte-Carlo approach
as in Sect.~\ref{sect:disks:e} allows us to estimate the eccentricity of the
orbital motion of the cluster around Sgr~A* as a function of $z$
(Fig.~\ref{fig:irs13:MC}). Since we have just shown that IRS~13E lies on the
right-hand side of this diagram, we can conclude that it is on a fairly
eccentric orbit, with $e\gtrsim0.5$.

\subsubsection{Does IRS~13E Contain an IMBH?}
\label{sect:irs13:IMBH}

We now ask the question of whether this cluster can be stable without an IMBH
at its core. The two elements to check are the tidal forces from Sgr~A* and
the internal velocity dispersion. The distribution of stellar surface density
as a function of distance from the center of IRS~13E reveals a very small
effective core radius (at which the surface density has fallen to half its
central value), $r_\mathrm{core}\simeq0.17\arcsec$ (0.0066~pc or 1,400~AU,
Fig.~\ref{fig:irs13}). After subtraction of the background, we count 12 stars
to $m_H=19.4$ within $\simeq0.35\arcsec$ ($\simeq$2 core radii). For the very
simplified assumption that these stars sample the mass function to
$\simeq5.5\sunmass$ (ZAMS for $m_H\simeq19.4$ at $A_K\simeq3.2$) and adopting
a flat mass function with $\mathrm dN\simeq m^{-1.35}\mathrm dm$
(Sect.~\ref{sect:IMF}), the inferred stellar mass within that radius is about
$350\pm100\sunmass$.  For such a flat mass function the difference between
that mass and the mass extrapolated to $1\sunmass$ amounts to about 30 solar
masses only. On the other hand, stellar crowding is an issue. Our star counts
are unlikely to be complete in the very center of IRS~13E and in the vicinity
of its brightest stars. Thus our mass estimate is obviously a lower limit to
the total stellar mass associated with the cluster, and the derived core
radius is probably an upper limit.  The core density of IRS~13E
($>3\times10^8\sunmass$~pc$^{-3}$) is higher than in any other known cluster,
except the cusp around Sgr~A*.

IRS~13E is currently at $\gtrsim4\arcsec$ from Sgr~A*. At this distance, the
tidal (`Hill') radius for such a mass is
\begin{equation}
  r_{\mathrm{Hill,IRS13E}}=\left(\frac{M_*}{M_\mathrm{BH}}\right)^{1/3}R=0.13\arcsec M_{* 400}^{1/3}
\end{equation}
where $M_*$ is the mass of the star cluster, $M_\mathrm{BH}$ the mass of Sgr~A*,
and $M_{* 400}=M_*/400\sunmass$.  The Hill radius is in remarkably close
agreement with the observed core radius.  However, the distance to take into
account is not the current one, but the periapsis distance. This parameter is
poorly constrained from our Monte-Carlo simulations. It appears that if the
cluster were would be on the CCWS midplane, then the periapsis
would be fairly small and the required cluster mass high
($\gtrsim10^4\sunmass$).  On the other hand, requiring IRS~13E to be
(currently) close the the Bar as discussed above allows the periapsis distance
to be easily above $4\arcsec$. In this case, the inferred stellar
concentration toward IRS~13E may thus be stable against tidal forces, or at
least relatively long-lived, without the need for an IMBH.

It is interesting to note that, with the same constraints on the current
line-of-sight position as above, we can estimate the date of the last
periapsis passage of IRS~13E to be $\simeq400$--$1000$~yr ago. It is therefore
possible that the past event of AGN activity of Sgr~A* 300--400~year ago
(Revnivtsev et al. 2004) was linked with the passage of IRS~13E at its
periapsis.

The main argument in Maillard et al. (2004) and Sch\"odel et al. (2005) in favor of high
cluster masses ($10^3$ and $10^4\sunmass$ respectively) came from the velocity
dispersion inside the cluster.  We still have 3D velocities only for four
`stars' in the cluster, one of which, IRS~13E3, is not even a single object
but the red core of the cluster, resolved as two sources (3A and 3B) in
Maillard et al., and that we see as no less than seven source in the present
work.  This source should therefore be discarded from velocity dispersion
analyses, and we remain with only 3 stars. Our study shows that even though
IRS~13E constitutes an over-density in the central parsec, more than one star
out of three in the aperture do belong to the background population rather
than to the compact cluster.  For this reason, any measurement of the velocity
dispersion should be considered with caution. In particular, IRS~13E1, which
drives the high value found by Sch\"odel et al. (2005), is formally a DF star rather
than a CCWS star and has (when studied independently) a higher eccentricity
than the other 13E stars.  It is possible that this star is not bound to the
cluster. We thus take the conservative position that the evidence for a
central dark mass in IRS~13E from the proper motion data presently is not
strong.

\subsubsection{Formation Scenario}

Although the IRS~13E over-density and the ring-like structure of the CCWS
centered on the radius of IRS~13E may appear to favor a dissolving cluster
scenario, it is also compatible with the idea of a star cluster formation in
situ, within an accretion disk or dispersion ring. If IRS~13E has grown from
gravitational instability in the original counter-clockwise gas disk the
maximum `isolation mass' that could have collapsed to the present cluster is
approximately the mass contained within the annulus of radial thickness
$2~r_\mathrm{Hill}$,
\begin{equation}
  M_\mathrm{isolation}=A\,M_\mathrm{disk}^{3/2} M_\mathrm{BH}^{-1/2}\;,
\end{equation}
where A is between 4 and $\simeq$30 (Lissauer 1987; Milosavljevic \& Loeb
2004). Taking a disk mass of $5000\sunmass$ (Sect.~\ref{sect:IMF}) and
$M_\mathrm{BH}=4\times10^{6}\sunmass$ this isolation mass is at least
$700\sunmass$. Our estimated stellar mass thus is also consistent with the
concept of local cluster formation within the counter-clockwise disk, in
agreement with the proposal by Milosavljevic \& Loeb (2004).

Overall, it appears that the various pieces of evidence argue for a cluster
mass of order $10^3\sunmass$, and that this mass can consist in stars, without
a black hole.  There is some indication, but no firm evidence, for a higher
mass. In particular, if IRS~13E is a concentration in the CCWS, we would
expect it to be close to the CCWS midplane. This would require a mass
$\gtrsim10^4\sunmass$. Further progress will require more proper motions and
radial velocities for the individual faint stars in the cluster.

\subsection{Stellar Content: the Disks Are Coeval And $\simeq$6~Myr Old}

\label{sect:disks:age} Wolf-Rayet (WR) stars of different sub-types appear at
different ages. The number ratios of these sub-types in a coeval population of
stars thus give information on the properties of the star forming event
leading to the observed population, in particular its age (Mas-Hesse \& Kunth
1991, Vacca \& Conti 1992, Schaerer et al. 1997). The evolution of massive
stars and the presence of a WR phase are controlled by stellar winds, which
depend on metallicity. In principle the number ratios of WR to O stars can
thus also trace the metal content. However, the strong effects of rotation on
the evolution of massive stars modify the duration of the different phases of
massive star evolution, especially for the WR phases (Meynet \& Maeder 2003;
Maeder \& Meynet 2004). Rotation varies from star to star, and the predicted
number ratios are affected by this natural spread.

\begin{deluxetable}{lrrrrrr}
  \tablecaption{Numbers and Fractions of Early Type Sub-Classes in the Two
    Disks\label{table:typeindisks}}
  \tablehead{& \multicolumn{3}{c}{clockwise} & \multicolumn{3}{c}{counter-clockwise} \\
    \colhead{Type} & \colhead{Number} & \colhead{fraction} &
    \colhead{uncertainty} & \colhead{number} & \colhead{fraction} &
    \colhead{uncertainty}} \tablecolumns{7} \startdata
  OB I/II & 18 & 0.36 & 0.08 & 3 & 0.15 & 0.09\\
  OB III/V & 13 & 0.26 & 0.07 & 6 & 0.30 & 0.12\\
  Ofpe/WNL & 12 & 0.24 & 0.07 & 5 & 0.25 & 0.11\\
  WNE & 1 & 0.02 & 0.02 & 0 & 0.00 & 0.00\\
  WC & 6 & 0.12 & 0.05 & 6 & 0.30 & 0.12\\
  ~ & ~ & ~ & ~ & ~ & ~ & ~\\
  Sum & 50 & ~ & ~ & 20 & ~ & ~\\
\enddata
\end{deluxetable}
\begin{deluxetable}{lrrr}
  \tablecaption{Ratio of Different Sub-Types\label{table:typeratio}}
  \tablehead{\colhead{Type 1/ Type 2} & \colhead{Clockwise} & \colhead{Counter-Clockwise} & \colhead{All}}
  \startdata
WR/O & 0.61 & 1.22 & 0.75\\
WR/(WR+O) & 0.38 & 0.55 & 0.43\\
WNL/(WR+O) & 0.24 & 0.25 & 0.24\\
WNE/(WR+O) & 0.02 & 0.00 & 0.01\\
WC/(WR+O) & 0.12 & 0.30 & 0.17\\
WC/WN & 0.46 & 1.2 & 0.67 \\
\enddata
\end{deluxetable}
Tables~\ref{table:typeindisks} and \ref{table:typeratio} list the numbers and
relative fractions of the different sub-types of early type stars in the two
stellar disks. Compared to observations in other star forming regions, the
WR/O star fraction in the Galactic Center disks is remarkably high. This is
partly a selection effect: our number counts are more complete for the
supergiants and WN stars than for the dwarfs, WC and WO stars.  To first
order, the fraction of different types of post main-sequence supergiants and
WR stars is strikingly similar in the two disks. The fraction of OB (I--V)
stars appears to be somewhat higher in the clockwise system, which may also
have a marginally greater fraction of OB supergiants. The striking resemblance
in content of massive stars strongly suggests that the two stellar disks are
basically coeval (Genzel et al. 2003).

\subsubsection{Population Synthesis}

From these ratios, we can attempt to estimate the age and star formation
history of the two disks. In order to investigate the physical properties of
the stellar population more quantitatively, we computed population synthesis
models for a determination of the expected number ratios under various
conditions (star formation history, metallicity, initial mass function).
Technical details on the method, based on the synthesis code developed by
Schaerer \& Vacca (1998), are given in Appendix~\ref{app:popsynth}.

\begin{figure}
\plotone{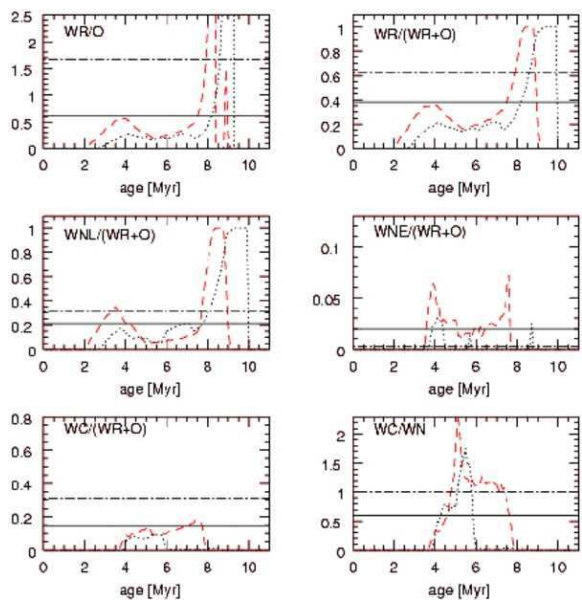}
  \caption{Number ratios of different types of massive stars as a function of
    time under the assumption of a single burst of star formation. The adopted
    initial mass function has a slope of 2.35, a lower mass limit of
    $1\sunmass$ and an upper mass limit of $100\sunmass$. We have used Geneva
    evolutionary tracks with rotation for solar (dotted line) and twice solar
    metallicity (dashed line). The horizontal solid lines correspond to the
    observed values and are upper limits. The second peak in the WR/O ratio is
    due to massive stars experiencing a blue loop.\label{fig:popsynth}}
\end{figure}
The constant star formation case can easily be ruled out since none of the
predicted number ratio matches the data. In particular, we do not detect any
O3--6 stars.  A burst of star formation is clearly preferred.
Figure~\ref{fig:popsynth} shows the results for such a scenario and a Salpeter
initial mass function (IMF). We see that the data are compatible with an age
ranging between 4 and 9~Myr. In fact, most of the ratios point to an age of
7--8~Myr except WC/WN which is also compatible with younger ages
($\simeq$4--5~Myr). All ratios involving O stars are certainly upper limits,
and as such indicate that the deduced age of 7--8~Myr is an upper limit.

The number ratios presented in Fig.~\ref{fig:popsynth} are usually higher when
Z is larger. However, in most of these diagrams we see that the age derived
from the solar metallicity case is similar to the one derived from the twice
solar metallicity case. The only exception is the ratio WC/(WR+O) for which
the $Z_\sun$ model predicts values lower than what we observe. This may be an
indication that Z in the central cluster is slightly super solar, but given
the uncertainty in the current observed number ratios of WR to O stars, this
needs to be confirmed by much more robust analyses.

Choosing a burst of star formation with finite duration has the effect
of shifting the time scale by 2 Myr but does not change significantly
the shape of the function giving the number ratios as a function of
time. We believe that a duration of ${\leq}$2 Myr is quite consistent
with our data. Longer bursts would also create large numbers of red
supergiants. Only three such supergiants (IRS7, IRS19 and IRS22, Blum,
Sellgren \& DePoy 1996) are observed in the central parsec.

As for the IMF, adopting a flatter one increases the strength of the first
``bump'' around 4 Myr observed in the evolution of the number ratios
(Fig.~\ref{fig:popsynth}), but does not strongly modify the ratios at later
epochs. For a top-heavy IMF the number ratios of the CWS are consistent with a
burst of star formation $\simeq$ 4~Myr ago (the solution at $\simeq6$--7~Myr
being still valid).

\subsubsection{Hertzsprung-Russell diagram}

We have also modeled the ages of the luminosity class I--V OB stars directly
by placing them on an infrared (IR) Hertzsprung-Russell diagram and comparing
the data with isochrones. This requires the knowledge of both an absolute
luminosity (or magnitude) and effective temperatures for all stars
(Appendix~\ref{app:HR}).  We restricted ourselves to the OB stars since for
the evolved massive stars (Ofpe/WN9 and WR stars), no calibration of effective
temperature as a function of spectral type exists (mainly due to the strong
effect of winds on the stellar properties of such objects). We excluded from
our analysis those stars for which the spectral classification is uncertain.
We also excluded the S stars near the central black hole.

\begin{figure}
  \plotone{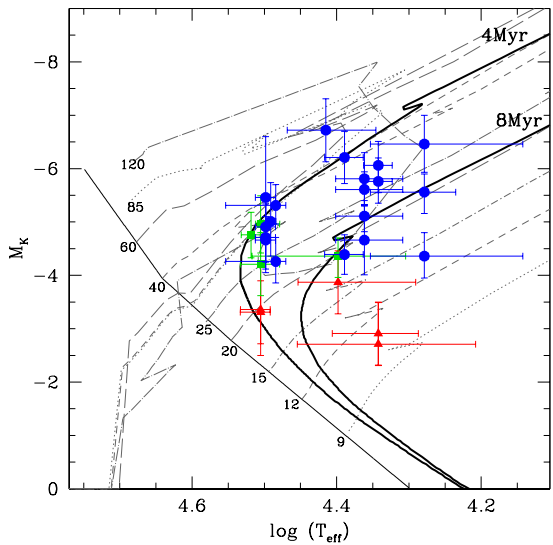}
  \caption{Location of the OB supergiants (\emph{blue dots}), giants
    (\emph{green squares}) and dwarfs (\emph{red triangles}) in an IR
    Hertzprung-Russell diagram. The numbers indicate the initial mass of the
    star on the tracks starting from the ZAMS (thin solid line). The two thick
    solid lines are two isochrones for 4 and 8~Myr. The tracks (\emph{various
      dotted/dashed lines}) are based on Geneva models without rotation
    (Lejeune \& Shaerer 2001).\label{fig:HR}}
\end{figure}
The results are shown in Fig.~\ref{fig:HR}. Overplotted are Geneva
evolutionary tracks taken from the database of Lejeune \& Schaerer (2001).
Isochrones for 4 and 8~Myr are also shown. We see that most of the stars are
located between these two isochrones. In particular, there are no stars on the
left side of the bend at $\log T_\mathrm{eff}\simeq4.5$ in the 4~Myr isochrones,
showing that the stellar population is older. There are a few outliers --
though with large error bars -- cooler than the 8~Myr isochrone, but an older
age is not likely in view of the presence of numerous WR stars not included in
this diagram for reasons highlighted above. The present age estimate confirms
the result of the previous section. The OB population in the Galactic Center
is 4 to 8~Myr old. The fairly large age uncertainties are inherent to our
methods of assigning $M_K$ and $T_\mathrm{eff}$ (Appendix~\ref{app:HR}).  The
fact that two different studies give the same result is rather convincing and
reassuring.

In summary the stellar disks in the Galactic Center have formed about
$6\pm2$~Myr ago. They are coeval to within about 1~Myr. The burst duration did
not exceed about 2~Myr. The present results thus are in excellent agreement
with the earlier findings of Tamblyn \& Rieke (1993) and Krabbe et al. (1995).
Krabbe et al. invoked a decaying burst of star formation 7~Myr ago to explain
the observed stellar population and its ionizing properties.

\subsection{Flat Mass Function and Total Mass of the Stellar Disks}
\label{sect:IMF}

\begin{figure}
\plotone{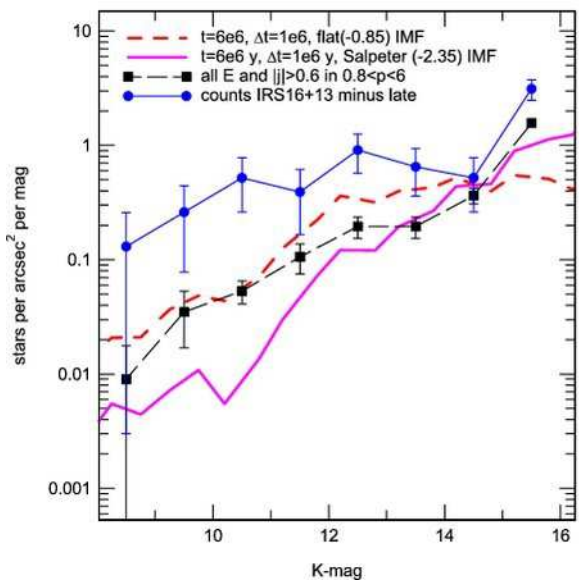}
\caption{K-band luminosity function. \emph{Filled circles} and \emph{squares} denote two
  ways of estimating the K-band luminosity function of the early type stars.
  The filled circles denote counts from the best high resolution H/K-band NACO
  images in the IRS~16 and IRS~13 regions where spectroscopically identified
  late type stars have been eliminated. The filled squares denote all stars
  with $|j|>0.6$ and $0.8\arcsec<p<6\arcsec$
  that are not spectroscopically identified as
  late type stars in our proper motion/spectroscopic catalogue (Ott et al.
  2005, in preparation). The thick continuous curve represents the KLF of a
  $t=6$~Myr old cluster forming stars over ${\Delta}t=1$~Myr burst with Salpeter
  initial mass function ($\mathrm dN/\mathrm dm\simeq m^{-2.35}$), normalized
  to the average of the observed counts in the $m_K=14$--15 bin. The long-dashed
  thick curve denotes the KLF of a burst with the same age and duration as
  above, but with much flatter IMF
  ($\mathrm dN/\mathrm dm\simeq m^{-0.85}$).\label{fig:KLF}}
\end{figure}
In this section we use the stellar number counts as a function of K-magnitude
to constrain the (initial) mass function of the young stars in the central
parsec. In Fig.~\ref{fig:KLF} we show two independent methods for estimating
the K-band luminosity function (KLF) of the stars in the disks. Obviously the
key issues are the screening against late type, background interloper stars
and the determination of the number of early type stars at the fainter levels
($m_K>14$--15) where the spectroscopy is not yet possible or incomplete.
Filled circles in Fig.~\ref{fig:KLF} denote the KLF constructed from the deep
counts from NACO H/K$_\mathrm{s}$-band data in the IRS~16 region (size
$2.5\arcsec\times2.5\arcsec$) and IRS~13E cluster (size $0.7\arcsec$, see
Fig.~\ref{fig:irs13}) and assuming that extinction is constant. We eliminated
from those counts 11 stars which are spectroscopic late type stars but
otherwise assume that the non-early type background toward these regions can
be neglected. The steep rise in the counts probably indicates that this
assumption is violated at $m_K>15$. Our alternative approach (filled squares)
was to compute counts for the entire region $0.8\arcsec\leq p\leq6\arcsec$ and
require that the detected stars either are spectroscopic early type stars or
(for the fainter stars) have $|j|>0.6$. This method thus suppresses the
background by taking advantage of the tangential motion of most of the early
type stars, in contrast to the late type stars (Fig.~\ref{fig:j}). Again, the
steep rise of the counts at $m_K>15$ probably signifies a dramatic increase in
the interloper background contribution. We note that for both methods the
average `spectroscopic completeness limit' is 13.5 to 14. The different slopes
of the two observed luminosity functions in the spectroscopically `safe'
region at $m_K<13$ is possibly a result of a decrease in average stellar
luminosity with $p$. The two methods more or less span the uncertainty box on
the number of faint and bright stars to include in the luminosity function of
the young stellar population. For comparison, the thick continuous and dashed
curves show two model luminosity functions with different mass functions. Both
curves are based on the population synthesis code STARS (Sternberg 1998;
Sternberg, Hoffmann \& Pauldrach 2003) with solar metallicity Geneva tracks
and assume a burst model of age 6~Myr and duration 1~Myr compatible with the
results of the last section. The continuous curve is for a standard
0.7--$120\sunmass$ Salpeter IMF ($\mathrm dN/\mathrm dm\simeq m^{-2.35}$),
while the dashed curve is for a much flatter mass function ($\mathrm
dN/\mathrm dm\simeq m^{-0.85}$). It is obvious that a Salpeter mass function
cannot fit the very flat counts toward IRS~16/13E and does only marginally so
for $m_K>12$. Both observed luminosity functions show a large excess above the
Salpeter model for $m_K<11$.  However, this range is dominated by very bright
evolved stars that may not be properly accounted for by the Geneva tracks.
Putting the largest weight on the more reliable range $11<m_K<14$, which
includes the OB dwarfs and giants, we conclude that the data require a
\emph{mass function flatter than a Salpeter function by $\simeq$1 to 1.5 dex}.

With this result we can now estimate empirically the total stellar mass in the
two stellar disks. As before, we exclude from the following discussion the
cluster of B-stars in the central cusp around Sgr~A*.  There are in total
$\simeq53$ and 20 observed OB+WR stars (quality 1+2) in, respectively, the
clockwise and counter-clockwise systems.  Recent stellar evolution models with
rotation by Meynet \& Maeder (2003) and Maeder \& Meynet (2004) indicate that
only stars more massive than $\simeq20$--$25\sunmass$ will go through the WR
phase we observe in the Galactic Center. Likewise the faintest OB stars we are
able to detect in the GC must have zero age main sequence masses in excess of
about $20\sunmass$. We thus assume that the observed OB and WR stars represent
the (initial) mass range of 20--$120\sunmass$. Adopting an initial mass
function with a slope of $\Gamma=-1.35$ and -0.85 ($\mathrm dN/\mathrm
dm=m^\Gamma$) required to fit the K-band luminosity function, the total mass
in $\geq20\sunmass$ stars is 2700--$3100\sunmass$ for the CWS and
1000--$1200\sunmass$ for the CCWS (the corresponding numbers for a Salpeter
($\Gamma=-2.35$) mass function would be 2100 and $790\sunmass$). Extrapolating
from the number of $\geq20\sunmass$ stars to the entire mass range above
$\simeq1\sunmass$ yields a total stellar mass of 3800--$3500\sunmass$ for the
CWS and 1420--$1320\sunmass$ for the CCWS.  Given that $\Gamma>-2$, these mass
estimates are relatively insensitive to the choice of lower mass cutoff and
slope. The corresponding numbers for a Salpeter mass function would be $10^4$
(CWS) and $3900\sunmass$ (CCWS).  Obviously these estimates are lower limits
since we almost certainly have not detected all the $M\geq20\sunmass$ stars
yet. We can establish a rough upper limit to the numbers of missing O stars by
counting all stars ($0.8\arcsec<p<13\arcsec$) in our proper motion list (Ott
et al. 2006, in preparation) that are on near tangential, clockwise orbits
($j>0.6$.) In Fig.~\ref{fig:j} we showed that only $24\%$ of late type stars
but $60\%$ of the early type stars reside in this range. Given that there are
approximately the same number of late and early type stars ($\simeq100$ each)
to $m_K\simeq13.9$, equivalent to an O8--9~V dwarf, the above criterion may
select a sample with $\simeq71\%$ early type stars.  In our proper motion
sample there are 35 such stars that are not spectroscopic early or late type
stars. Hence an upper limit to the correction factor for missing early type
stars with $M\geq20\sunmass$ in each system is about $(53 +0.71\times35)/53=
1.5$.  We conclude that the total stellar content in the clockwise and
counter-clockwise systems is $\leq6000$ and $2200\sunmass$.

These estimates (or limits) are fully consistent with two other and completely
independent lines of argumentation. In the first Nayakshin et al. (2005)
consider the precession and warping that is induced by one of the stellar
disks on the other, and vice versa. They study the evolution of the apparent
thickness of the disks with time, as a function of their masses. They require
the thickness to be as observed (Sect.~\ref{sect:disks:thickness}) at
$t=4$~Myr. Depending on whether the original configuration was two thin disks
in circular motion, or whether the counter-clockwise disk was initially an
in-spiraling cluster with significant radial motion, Nayakshin et al. find
masses of between 5000 and $15,000\sunmass$ for the CWS and between 5000 and
$10,000\sunmass$ for the CCWS. The 4~Myr age assumed in their work is somewhat
smaller than what we deduced in Sect.~\ref{sect:disks:age}.  These masses thus
are strict upper limits.

Nayakshin \& Sunyaev (2005) have\linebreak pointed out another argument for a flat
initial mass function and for stellar masses $<10^4\sunmass$ owing to the
weakness of extended X-ray emission in the central parsec. Nayakshin \&
Sunyaev argue that the bright X-ray emission of young T-Tauri stars is a
handle for probing the low mass end of the mass function.  For a $<10$~Myr old
stellar component and a Salpeter mass function the predicted diffuse X-ray
flux in the central parsec exceeds the observed value of Baganoff et al.
(2003) and Muno et al.  (2005) by a factor of 20 to 100.

From these various arguments it is safe to conclude that the mass function of
the young stars in the Galactic Center is indeed flat and that the total
stellar mass contained in the clockwise disk does not exceed
$\simeq10^4\sunmass$. The corresponding limit for the counter-clockwise system
is probably $\simeq5000\sunmass$. Our deep observations of the northern field
(3.3.3 and Maness et al. 2005, in preparation) also strongly suggest that
there is no `reservoir' of B stars tidally stripped from the in-spiraling
cluster at larger radii that would point to many missing lower mass stars. We
thus believe that the flat mass function we, Nayakshin et al. (2005) and
Nayakshin \& Sunyaev (2005) deduce approximates an initial mass function,
unaltered by dynamical effects.

\section{Discussion}
\label{sect:discuss}

\subsection{Origin of the Stellar Disks}

What have we learned from the new observations and their analysis? Let
us summarize the key conclusions:
\begin{enumerate}
\item the large majority of the early type stars in the central parsec
  (excepting those in the central cusp around Sgr~A*) reside in one of two,
  well defined rotating disks. These disks are oriented at large angles with
  respect to each other, and to the Galactic plane. One rotates clockwise in
  projection, the other one counter-clockwise. The clockwise system (CWS) has
  2.5 times as many massive stars as the counter-clockwise system
  (CCWS)\label{point:2disks};
\item the disks have a well defined inner radius (edge). The more populated
  clockwise disk has an inner edge at $p=1\arcsec$, just outside the central B-star
  cluster in the Sgr~A* cusp. Its azimuthal distribution is isotropic. The
  inner edge of the counter-clockwise disk is further out. This system thus
  resembles more a ring\label{point:inner edge};
\item Both disks have an outer density profile scaling as
  $\Sigma\simeq R^{-2}$. No O/WR-stars (in all directions)
  and lower mass B-stars (toward the north) are present outside the central
  $p=13\arcsec$ region\label{point:r2};
\item the disks have a finite but moderate thickness,
  $\sigma_i=14\degr\pm4\degr$ ($\langle|h|/R\rangle0.12\pm0.03$) for the CWS and
  $19\degr\pm10\degr$ ($\langle|h|/R\rangle=0.16\pm0.06$) for the CCWS;\label{point:h/R};
\item almost all of the stars in the CWS are on low eccentricity,
  close to circular orbits\label{point:eCW};
\item on the contrary, most stars on the CCWS appear to be on eccentric orbits\label{point:eCCW};
\item the IRS~13E group of early type stars appears to be the center of a
  larger local concentration of stars with ${\geq}$12 (K${\leq}$17.5) members
  and total stellar mass of order $10^3\sunmass$. It appears to have a large
  orbital eccentricity ($e>0.5$). This concentration represents the second
  largest stellar density in the central parsec, second only to the central
  Sgr~A* cusp, and is among the highest known core densities
  ($>3\times10^8\sunmass$~pc$^{-3}$). It is unlikely a projection effect. IRS~13E
  is compact enough that it may be long-lived in the tidal field of the
  central black hole, even without invoking a central intermediate mass black
  hole. It may be a slowly dissolving star cluster embedded in the
  counter-clockwise system\label{point:irs13};
\item the stellar disks are coeval to within 1~Myr, have an age of
  $6\pm2$~Myr and must have formed over a time period
  $\leq2$~Myr\label{point:ages};
\item the total stellar mass associated with each of the two disks does not
  exceed $10^4\sunmass$\label{point:masses};
\item the (initial) mass function of these disks is flatter than a Salpeter
  mass function by 1--1.5 dex.\label{point:IMF}
\end{enumerate}

Let us review the two star formation scenarios in view of this `report
card', for each of the two systems.

\subsubsection{The In Situ Star Formation Scenario Is in Good Agreement with the Data}

The presence of two well defined kinematical systems seems to require two
separate events of star formation. This is actually somewhat problematic
whatever the formation scenario is, since these two events must have occurred
basically at the same time. However, the two events are allowed to be
separated by $\simeq$1~Myr from each other. This timespan is sufficient for
star formation to remove most of the gas from the first disk before the second
one starts forming. The minimum time needed to form stars can be estimated as
follows.  Once the disk becomes gravitationally unstable, instabilities are
believed to grow in the disk on dynamical time scale (e.g. Toomre 1964), i.e.
$60\;\mathrm{yr}\;(R/1\arcsec)^{3/2}$. In addition to that, accretion of gas onto
proto-stars is limited by the Eddington accretion rate onto these, which sets
the stellar mass doubling time scale to about a thousand years (e.g.,
Nayakshin \& Cuadra 2005). Taken together, these two conditions constrain the
minimum duration of the star-formation episode to about $10^4$~yr.  Therefore
both gaseous disks are not required to have been in the central parsec at the
same time.  The in situ scenario passes points \ref{point:2disks} and
\ref{point:ages}.

Points \ref{point:inner edge} and \ref{point:r2} are also reasonably natural
in the context of an accretion disk.  The minimum radius where the
gravitational instability can form stars may be estimated as follows. In order
for the disk to be self-gravitating at radius $R$, the accretion disk surface
density $\Sigma_\mathrm{gas}$ must be larger than a minimum value (e.g. Fig.~2
in Collin \& Hure 1999; Levin 2003, astro-ph/0307084), which is approximately
given by $\Sigma_\mathrm{min} = 10^4 M_4 \sunmass/ (\pi R_\mathrm{a}^2)$, where
$M_4$ is the minimum unstable disk mass in units of $10^4 \sunmass$ (see top
panel in Fig.~1 in Nayakshin 2006, MNRAS, submitted, astro-ph/0512255), and
$R_\mathrm{a}$ is the radius $R$ in arcseconds. At the same time, for a given
dimensionless accretion rate $\dot{m}$, in units of the Eddington accretion
rate, and the disk viscous $\alpha$-parameter, the standard accretion disk
model (Shakura \& Sunyaev 1973) predicts that $\Sigma_\mathrm{gas} =
\Sigma_\mathrm{SS}\propto R_\mathrm{a}^{-3/5}$.  Therefore, stars may be able
to form only at disk radii greater than
\begin{equation}
R_\mathrm{min} \simeq 0.4\arcsec \alpha^{4/7} M_4^{5/7} \dot{m}^{-3/7} \;.
\end{equation}
Given the crude nature of these estimates and uncertainties in $\alpha$ and
$\dot{m}$, this estimate of $R_\mathrm{min}$ is reasonably close to the
observed inner disk cutoff radius (precise values for both $\alpha$ and
$\dot{m}$ are uncertain but should be reasonably close to unity in the case of
a massive self-gravitating disk).

The radial stellar density profile $\Sigma$ may be expected to follow the
initial gas surface density, $\Sigma_\mathrm{gas}$, if star formation
\emph{instantaneously} consumed most of the gas disk. Interestingly, the
observed $\Sigma$ varies as $R^{-2}$, as expected in a
\begin{equation}
  Q=\frac{c_\mathrm{s}\Omega}{\pi \mathcal G\Sigma}\simeq1\label{eq:toomre}
\end{equation}
marginally stable stationary self-gravitating disk (Lin \& Pringle 1987;
Collin \& Hure 1999; Thompson, Quataert, \& Murray 2005; $Q$ is the Toomre
parameter, Toomre 1964). A time-dependent self-gravitating
disk left to its own devices will also develop a steep surface density
profile, $\Sigma_\mathrm{gas}\simeq R^{-3/2}$ (Lin \& Pringle 1987), again close
to the observed steep profile, especially when compared with the standard
Shakura \& Sunyaev profile that scales as $R^{-3/5}$ at large radii (when
assuming a constant opacity, e.g. eq. 20 in Svensson \& Zdziarski 1994).
However, the time scales needed for the disk to go through a significant mass
transfer in radial direction is prohibitively long:
\begin{eqnarray}
  t_\mathrm{visc} & \simeq &
  \Omega^{-1}\left(\frac{M_\mathrm{BH}}{M_\mathrm{disk}}\right)^2  \\
  &\simeq&3\times10^8\;\mathrm{yr}\;\left(\frac R{10\arcsec}\right)^{3/2}
  \left(\frac{M_\mathrm{disk}}{10^4\sunmass}\right)^{-2}\;,\nonumber
\end{eqnarray}
where $\Omega$ is the Keplerian angular frequency.

The thickness (point \ref{point:h/R}) of the disks is also consistent with the
expectations: even though the gas (and stellar) disks should be initially
quite thin, they thicken somewhat due to relaxation ($h/R\simeq0.1$ after a
few $10^6$~yr, eq.~15 in Nayakshin \& Cuadra 2005) and get warped due to the
gravitational torque applied by each disk on the other one (Nayakshin \&
Cuadra 2005; Nayakshin et al. 2005). The low eccentricities of the orbits in
the CWS (point \ref{point:eCW}) are the natural outcome from a marginally
stable disk.

The disk mass necessary for a classical disk to fragment is
$\gtrsim10^4\sunmass$, just above the higher limit on the current stellar mass
(point \ref{point:masses}). This discrepancy is not worrisome.  A fraction of
the gas in the disks may have been expelled through stellar feed-back (either
escaping the central region, or being redistributed and later accreted by
Sgr~A*). In addition, the disk surface density was perhaps not a smooth
function of radius.  In that case, the instability criterion $Q=1$ 
may have been reached in parts of a less massive disk, which would
otherwise have been stable.

Formation of massive stars would not be unexpected in the accretion disk star
formation scenario. The initial mass of the disk fragment collapsing to form
first gravitational condensations in the disk should be of order
$M_\mathrm{frag}=\rho_\mathrm{disk} h^3$, which can be estimated for the
marginally self-gravitating disk at ($Q\simeq 1$) to be only
$\simeq0.01\sunmass$ (e.g., see the bottom left panel of Fig.~3 of Collin \&
Hure 1999. While their estimate was made for $M_\mathrm{BH}=10^6\sunmass$,
$M_\mathrm{frag}$ is a weak function of the super-massive black hole mass).
Nevertheless, since the requisite gas densities in the accretion disk are
several orders of magnitude higher than even those in molecular cores,
Bondi-Hoyle and Hill accretion rate estimates yield very high accretion rates
onto these fragments, $\dot M > 10^{-3}\sunmass$~yr$^{-1}$. With this stars
double their mass at the rate limited by the Eddington limit, which is of
order 1000 years (Goodman and Tan 2004, Nayakshin \& Cuadra 2005, Fig.~2).
Thus as little as $10^5$ years of such accretion would lead to very massive
stars. In addition, stellar mergers could contribute to the growth of the
massive stars. Finally, the first collapsed object may be more massive than
the estimate above if radiative cooling rate is only marginally sufficiently
fast to allow disk gravitational collapse (see, e.g.  second simulation in
Gammie 2001), or turbulent `pressure' support is important in preventing
smaller clouds to collapse (McKee and Tan 2004).  For these reasons, a
top-heavy mass fucntion (point~\ref{point:IMF}) may be natural in the context
of the in situ scenario.  Future detailed work is warranted to delineate the
dominant mode of massive star growth in the observed stellar disks.

Therefore, the in situ, accretion disk scenario fits perfectly for the
clockwise system. There remains however two points to be clarified for the
counter-clockwise system: the high eccentricity of the orbits (point
\ref{point:eCCW}), and the presence of a very dense star cluster
(point\ref{point:irs13}).  At first glance these two points seem to be strong
clues in the direction of the in-falling star cluster scenario. However, the
latter point is resolved if the in situ model is amended with the possibility
that gravitational collapse in the disk can lead to local cluster formation.
The likely occurrence of this process has already been pointed out by
Milosavljevic \& Loeb (2004).  The large eccentricities in the CCWS are not
incompatible either with the accretion disk scenario.  It is possible that
that the progenitor accretion disk did not circularize before forming stars,
since the disk fragmentation may have occured on a short dynamical timescale.
Alternatively, gravitational interaction between IRS~13E and the gaseous disk
may drive the former's eccentricity (Goldreich \& Sari 2003). Both of these
ideas will be investigated in greater detail in the future.
Overall, the in situ scenario seems acceptable for the CCWS, with some open
theoretical questions.

\subsubsection{The In-Spiraling Cluster Scenario Is Unlikely}

The in-spiraling cluster scenario is also able to fulfill more or less easily
most of the points of the `report-card' above. The core density of IRS~13E
($>3\times10^8\sunmass$~pc$^{-3}$) is greater than the core density required for a
cluster to sink deep into the central parsec before disruption
($10^7\sunmass$~pc$^{-3}$). Nevertheless, the in-spiraling scenario seems to
fail on point \ref{point:r2}, and fails on point \ref{point:masses} in a way
that we deem fatal.

First of all, models of such an in-falling cluster show that the cluster
should lose a lot of stars during the inspiral, leaving a stellar population
with a shallow radial profile extending over several parsecs in the radial
direction, in contradiction to point \ref{point:r2}.  G\"urkan \& Rasio (2005)
argue that this discrepancy can in principle be overcome by initial mass
segregation in the cluster. In this way, the stars that are lost at large
distance from the center are lower mass stars below our detection limit, and
all the detectable stars are brought in the central parsec. No quantitative
analysis has been done so far though, so that it is not clear whether the mass
segregation--evaporation process can bring all the OB stars into the central
0.5~pc, with a density profile as sharp as the observed one, and not leave a
telltale population of B stars at $R>0.5$~pc (which we do not observe).

Even if this strong mass segregation were possible, we would be observing all
the $>20\sunmass$ stars initially in the cluster. The total inital mass of the
cluster could not exceed $17\,000\sunmass$ for the CWS and $6,500\sunmass$ for
the CCWS, even assuming a Salpeter IMF.
The total mass required to make a cluster in-spiral from far out (a few
$10$~pc) into the central region within an O star lifetime is $>10^5\sunmass$
(Gerhard 2001; McMillan \& Portegies Zwart 2003; Kim et al.  2004; G\"urkan \&
Rasio 2005).  The inconsistency between the data and the models is a factor of
$\gtrsim6$ for the CWS, and $\gtrsim15$ for the CCWS.

Therefore, taken together, these points \emph{favor strongly the in situ
  acretion disk} model for the formation of both stellar disks in the central
parsec. We conclude that the clockwise disk almost certainly resulted from in
situ star formation in a dense gas disk. The most obvious process starting
such a disk is the infall of a large gas cloud (Morris 1993; Genzel et al.
2003), followed by dissipation of its angular momentum through shocks in a
`dispersion ring' (Sanders 1998). The in situ star formation scenario is also
plausible -- and certainly much more plausible than the infalling cluster
scenario -- for the counter-clockwise system assuming that eccentric orbits
and cluster formation can be understood within the framework of the model.

We do not have an explanation for the near-simultaneous occurrence of
two star formation events 6~Myr ago, followed by little since then,
and preceded by little for tens of Myr (Blum et al. 2003). It is
unavoidable to conclude that the epoch 4--9~Myr ago must have been
a very special one for the Galactic Center. It is interesting and
relevant to note in this context that the other two young, massive star
clusters in the central 50 pc, the Arches and Quintuplet cluster, have
comparable stellar masses ($10^4\sunmass$), stellar content (WC/WN etc.), ages
(2--7~Myr), and (flat) mass functions (Figer et al. 1999; Figer
2003; Stolte et al. 2005). We might speculate that star formation
across the Galactic Center was triggered a few Myr ago by a global
event, such as an interaction with a passing satellite galaxy that
raised the pressure in the central interstellar medium and/or lead to
increased cloud/cloud collisions.

\subsection{Origin of the central B-star cluster by scattering from a
  sea of B-stars}

We end by briefly commenting on the proposal of Alexander \& Livio (2004) that
the B-stars in the central cusp are the result of the capture of B-stars on
near-loss cone orbits, originally unbound to the central black-hole, following
a three body, direct exchange scattering process with the central massive
black hole and $\simeq10\sunmass$ stellar black holes residing in the central
cusp.

This elegant and attractive proposal requires the presence of a `reservoir' of
B-stars originally at large distances from the central hole. In the specific
model presented by Alexander \& Livio (2004) the captured fraction of
$\geq3\sunmass$ B-stars is about $10^{-4}$ for a constant star formation
scenario.  With ${\geq}$15 B-stars presently observed in the central cusp
(Eisenhauer et al. 2005), a total reservoir of about $1.5\times10^5$
$\geq3\sunmass$ B-stars (and $1.5\times10^{4}$ ${\geq}15\sunmass$ stars) are
required for the mechanism to work. The surface density in the reservoir
depends on its spatial extent. Alexander \& Livio (2004) consider stars
originating between 0.5 pc (where stars are on unbound orbits relative to the
MBH) to 2.5 pc (where the orbits are still marginally Keplerian), and find
that the required surface density over that area is about 10 B stars per
square arcsec. The deep observations reported toward the northern field
(3.3.3) limit the number of $m_K<16.5$ B-stars to about 0.4 stars per
arcsec$^2$. The discrepancy is a factor of 25. The current observations appear
to to exclude the capture of unbound stars by the mechanism proposed by
Alexander \& Livio (2004). Alternatively the reservoir may be the stellar
disks themselves. The exchange capture efficiency of such bound, short period
stars is still under investigation.  Due to crowding, it is not yet possible
to securely identify B V stars, such as populate the S-cluster in the inner
arcsec (Eisenhauer et al.  2005), throughout the disks.  While it is possible
that some of the stars we identify as OB III/V are B V stars
(Table~\ref{table:typeindisks}), deeper spectroscopy is required for a full
census of the B V content of the disks.

\section{Conclusions}

We report firm spectroscopic detections of 41 OB stars (luminosity
classes I--V) in the central parsec. The new data resolve a decade old puzzle
of the `missing O stars'. Some of these stars seem He- and N-rich (OBN stars).

We confirm the presence and define the properties of the two young star
disks first presented by Levin \& Beloborodov (2003) and Genzel et al.
(2003).

The disks rotate about the center and are at large angles with respect to each
other. They have a very well defined inner radius, a radial surface density
profile scaling as $R^{-2}$, and a moderate geometric thickness. In one of the
disks (the clockwise, `IRS~16' system) almost all stars are on close to
circular orbits.  However in the other, counter-clockwise, `IRS~13E' system,
most of the stars (including the IRS~13E cluster) orbit on eccentric orbits.
Star counts suggest that IRS~13E is a long-lived cluster of stellar mass
$\gtrsim400\sunmass$. It is the cluster with the highest known core density
after the cusp around Sgr~A* itself: $>3\times10^8\sunmass$~pc$^{-3}$.

The star disks are coeval within $\simeq$1~Myr and have formed $\simeq$6~Myr
ago. The stellar mass function is significantly flatter than Salpeter,
setting a limit to the total stellar mass in the disks of about
$1.5\times10^{4}\sunmass$.

The constraints just discussed strongly suggest that the stars in the
`IRS~16'-disk were formed in situ from a dense gaseous accretion disk.
They were not transported into the central parsec by an in-spiraling
massive star cluster. On balance the same conclusion seems plausible
(but not incontrovertibly proven) for the `IRS~13E'-disk. In that
case, the IRS~13E cluster must have formed within the disk.

\acknowledgements \emph{Acknowledgements:}We are grateful to Margaret Hanson
for giving us access to her new atlas of near-IR spectra of OB stars prior to
publication. We thank Eliot Quataert, Todd Thompson, and Norman Murray for
many useful discussions on star formation in disks. We also thank Daniel
Schaerer and William D. Vacca for making their population synthesis code
available. TA acknowledges support by Minerva grant 8484 and a New Faculty
grant by Sir H.~Djangoly, CBE, of London, UK. FM acknowledges support from the
Alexander von Humboldt foundation. Finally, we would like to thank the
anonymous referee for his useful comments.

\appendix

\section{Conventions used in this paper}
\label{app:orbelms}\label{app:conventions}

We use the usual astronomical cartesian coordinate system, in offsets from from
Sgr~A*: $x=\cos\delta\mathrm d\alpha$ increases eastwards, $y=\mathrm d\delta$
increases northwards, and $z=\mathrm dD$ increases forward along the
line-of-sight from the observer. We occasionally use spherical coordinates:
\begin{eqnarray}
  \varphi=\arctan\frac yx\;;&\theta=\arccos(z/\sqrt{x^2+y^2})\;;&R=\sqrt{x^2+y^2+z^2}\;\label{eq:spherical coordinates}\\
  x=R\sin\theta\cos\varphi\;;&y=R\sin\theta\sin\varphi\;;&z=R\cos\theta\;.\label{eq:sph. coord. inverse}
\end{eqnarray}

When discussing orbital planes, we use the orbital elements defined in Aller et
al. (1982), adapted for defining an oriented disk, which is consistent
with Eisenhauer et al. (2005). The disks are first defined by the
orientation of their line of nodes, which is the intersection of the
disks (which contain Sgr~A*) with the plane of the sky (also containing
Sgr~A*). More precisely, we refer to the ascending (=receding) half of
this line. The position angle of the line of nodes ${\Omega}$ is the
angle between the North direction and this ascending half of the line
of nodes, increasing East of North (counter-clockwise). The second
element necessary to define these oriented disks is the inclination
$i$, measured on the ascending half of the line of nodes, from
the direction of increasing ${\Omega}$ to the direction of motion on
the disk.
\begin{eqnarray}
 0\degr \leq i   <   90\degr & \Rightarrow & \mathrm{counter-clockwise\;projected\;rotation;}\nonumber\\
90\degr   <  i \leq 180\degr & \Rightarrow & \mathrm{clockwise\;projected\;rotation.\nonumber}
\end{eqnarray}

An equivalent way to define these disks is to give the three Cartesian
coordinates of their normal vector. The rotation of the stars on the disks is
then always counter-clockwise when visualized from a point towards which the
normal vector is pointing. A disk is counter-clockwise (in projection) when
the $z$ coordinate of its normal vector is negative and clockwise when it is
positive.
\begin{equation}
n_x=\sin i\cos\Omega\;;\;
n_y=-\sin i\sin\Omega\;;\;
n_z=-\cos i\;.\label{eq:orbelms}
\end{equation}
The spherical coordinates of the normal vector
relate to $i$ and $\Omega$, but are not equal to them:
\begin{equation}
\varphi_{\vect n}=-\Omega\;;
\theta_{\vect n}=180\degr-i\;.
\end{equation}

\section{Discriminating star disks}
\label{app:disks}

Genzel et al. (2003) defined the sky-projected angular momentum (or
normalized angular momentum with respect to the line of sight) as
\begin{equation}
j=\frac{J_z}{J_{z,\mathrm{max}}}=\frac{xv_y-yv_x}{(x^2+y^2)^{1/2}(v_x^2+v_y^2)^{1/2}}\;.
\label{eq:j}
\end{equation}

It  is  a simple  way  to distinguish  stars  on  projected tangential  orbits
($|j|\simeq1$) from stars on projected radial orbits ($j\simeq0$), and stars on
projected   clockwise   orbits   ($j\simeq+1$)   from   stars   on   projected
counter-clockwise orbits ($j\simeq-1$). They fit disk solutions to the data by
minimizing   as  a function  of the  normal vector  $\vect n$  the following
quantity, introduced by Levin \& Beloborodov (2003):
\begin{equation}
\chi^2=\frac{1}{N-1}\sum_{k=1}^N\frac{(\vect{n}\scalprod\vect{v}_k)^2}{(\vect{n}\scalprod\gvect{\sigma}_k)^2}\;, \label{eq:chi2}
\end{equation}
where $\vect{v}_k=(v_{x,k},v_{y,k},v_{z,k})$ is the velocity vector and
$\gvect{\sigma}_k=(\sigma_{x,k},\sigma_{y,k},\sigma_{z,k})$ the corresponding
velocity uncertainty of the $i$-th star ($Ox$ points Eastwards, $Oy$,
Northwards, and $Oz$, away from observer). The unit vector $\vect n$ describes
the orientation of the normal vector to a common plane in which all $N$ stars
are assumed to move.

There is a more straightforward way to find a disk in the data, which does not
rely on fitting. Consider the 3D space velocity of star $k$ ($k=1{\dots}N$) in
spherical coordinates (eqs. \ref{eq:spherical coordinates}, \ref{eq:sph.
  coord. inverse}),
\begin{equation}
\vect v_k=(v_{x,k},v_{y,k},v_{z,k})=||\vect v_k||(\sin \theta_k \cos \varphi_k, \sin \theta_k \sin \varphi_k, \cos \theta_k)\;.
\end{equation}

Assuming that all $N$ stars are in the plane of a common disk with normal
vector 
\begin{displaymath}
\vect n=(n_x,n_y,n_z)=(\sin i \cos \Omega, -\sin i \sin \Omega, -\cos i)
\end{displaymath}
(see eq. \ref{eq:orbelms}), then all stellar velocity vectors must obey
\begin{eqnarray}
&0=\vect{n}\scalprod\vect{v}_k=\sin i \cos\Omega \sin\theta_k\cos\varphi_k
-\sin i \sin\Omega\sin\theta_k\sin\varphi_k
-\cos i\cos\theta_k\;&\\
&\sin i\sin\theta_k\cos(\Omega+\varphi_k)=\cos i\cos\theta_k\;&\\
&\cotan\theta_k=\tan i \cos(\Omega+\varphi_k)\;.& \label{eq:cotan}
\end{eqnarray}
In the plane spanned by $\varphi$ and $\cotan\theta$, stars located in a
planar structure thus must exhibit a telltale cosine pattern.

\section{The projection effects on the determination of the eccentricity}
\label{app:MC}

For the Monte-Carlo simulations in Sect.~\ref{sect:disks:e}, for each star we
assumed normal distributions for $v_x$, $v_y$, $v_z$, $R_0$
($7.62\pm0.32$~kpc, Eisenhauer et al. 2005) and $M_\mathrm{SgrA*}$
($3.61\pm0.32\times10^6\sunmass$, Eisenhauer et al. 2005) ($x$ and $y$ have
been left out as they play a minor role in the error budget). The potential
well is assumed to be dominated by Sgr~A*, which should essentially be true
for the region where our measurements are most reliable ($p<8\arcsec$). We
have drawn $z$ as a uniform variable ($|z|<20\arcsec$).  For each one of the
$10^6$ realizations (per star), we have computed the eccentricity $e$ and the
inclination to the midplane of the system to which the star belongs
($i_\mathrm{CW}$ and $i_\mathrm{CCW}$, $i_\mathrm{(C)CW}$ for short). We then
constructed the 2D histograms of these two parameters (maps of
$P(i_\mathrm{(C)CW},e)$). In order to validate our method, we have performed the
same analysis on several sets of artificial data that assumed a single
eccentricity (including circular case) for all stars and the same geometry as
the CWS and CCWS, introducing errors typical of the real systems.  $R_0$ and
$M_\mathrm{SgrA*}$ have been varied to check their influence on the conclusions.

In these 2D histograms, strong depletions where observed close to $e=0$ and
$i_\mathrm{(C)CW}=0$.  This is simply because of the uncertainties and of the
functional dependency of $e$ and $i_\mathrm{(C)CW}$ on the 3D velocity (the true
errors on $v_x$, $v_y$, and $v_z$ are unlikely to be all smaller than their
respective $1\sigma$ error bar \emph{at the same time}). We have approximately
corrected this effect by dividing the histograms by ($1-\exp(-e/0.35)$) and
($1-\exp(-i_\mathrm{(C)CW}/20\degr)$). This makes the 2D histograms look smooth
and the values of $e$ determined for the artificial stars have the right
statistics (peaking close to the assumed eccentricity). The correction affects
only small eccentricities in practice ($e\lesssim0.2$).

\begin{figure}
  \plottwo{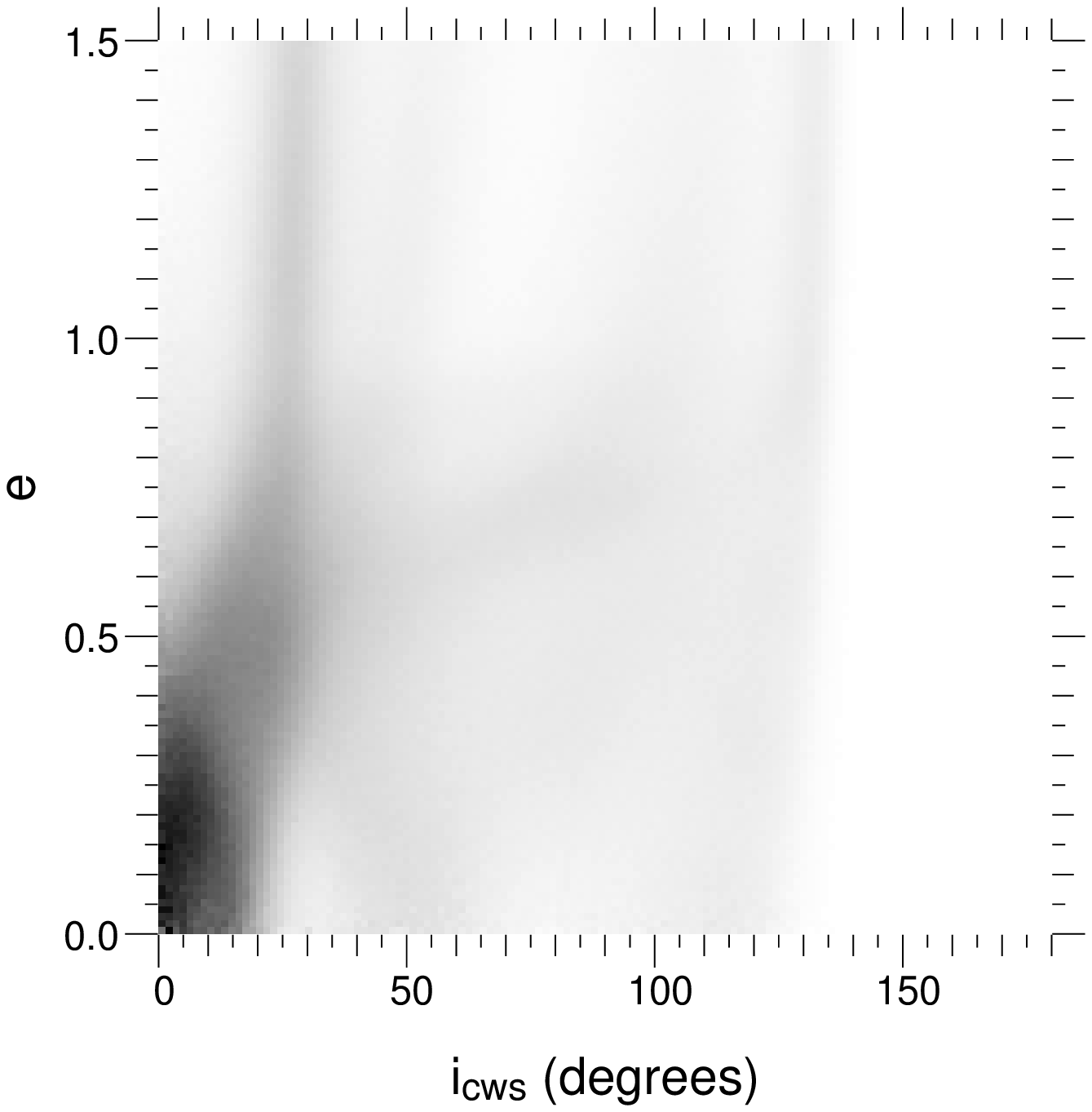}{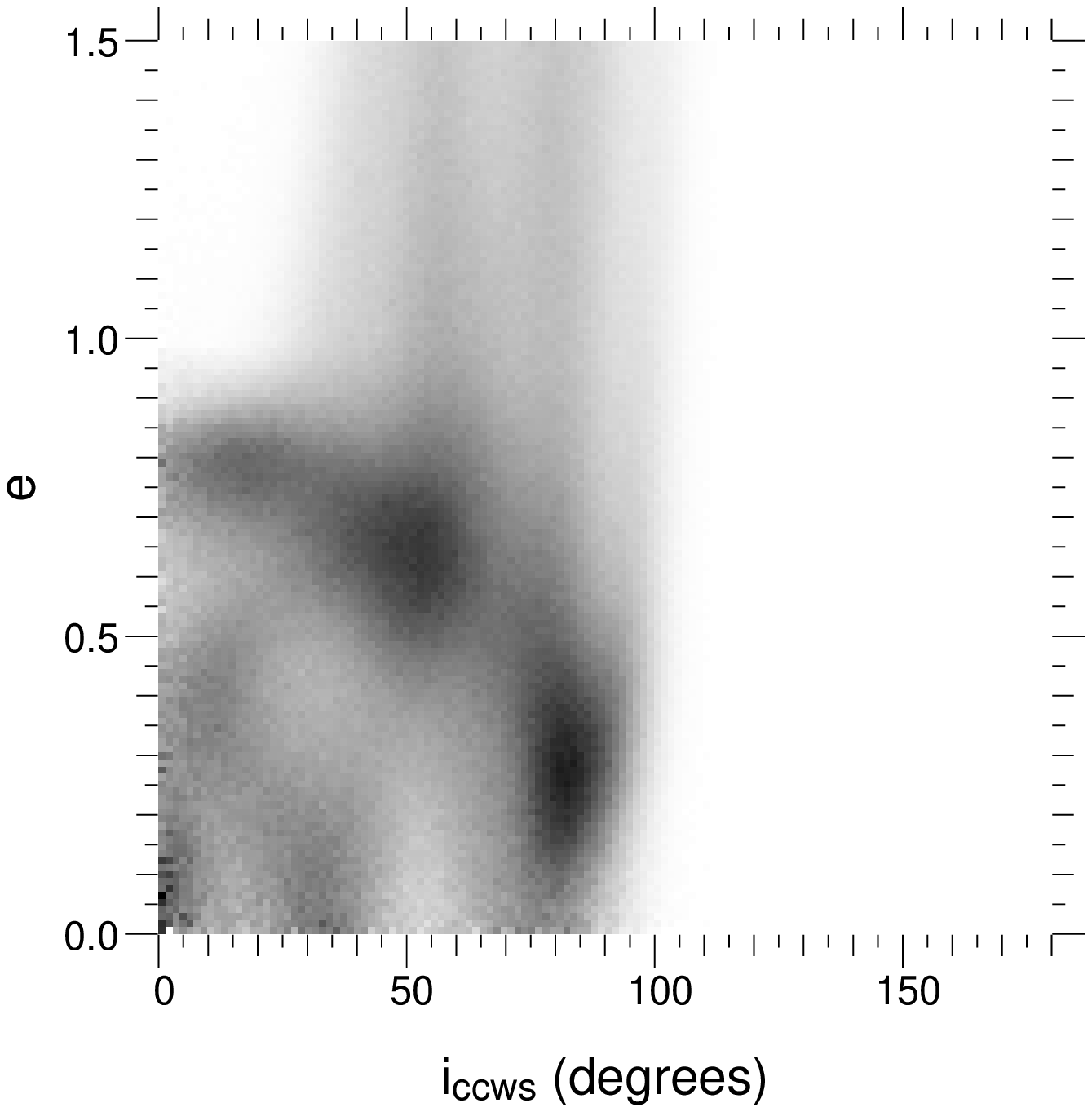}\\
  \plottwo{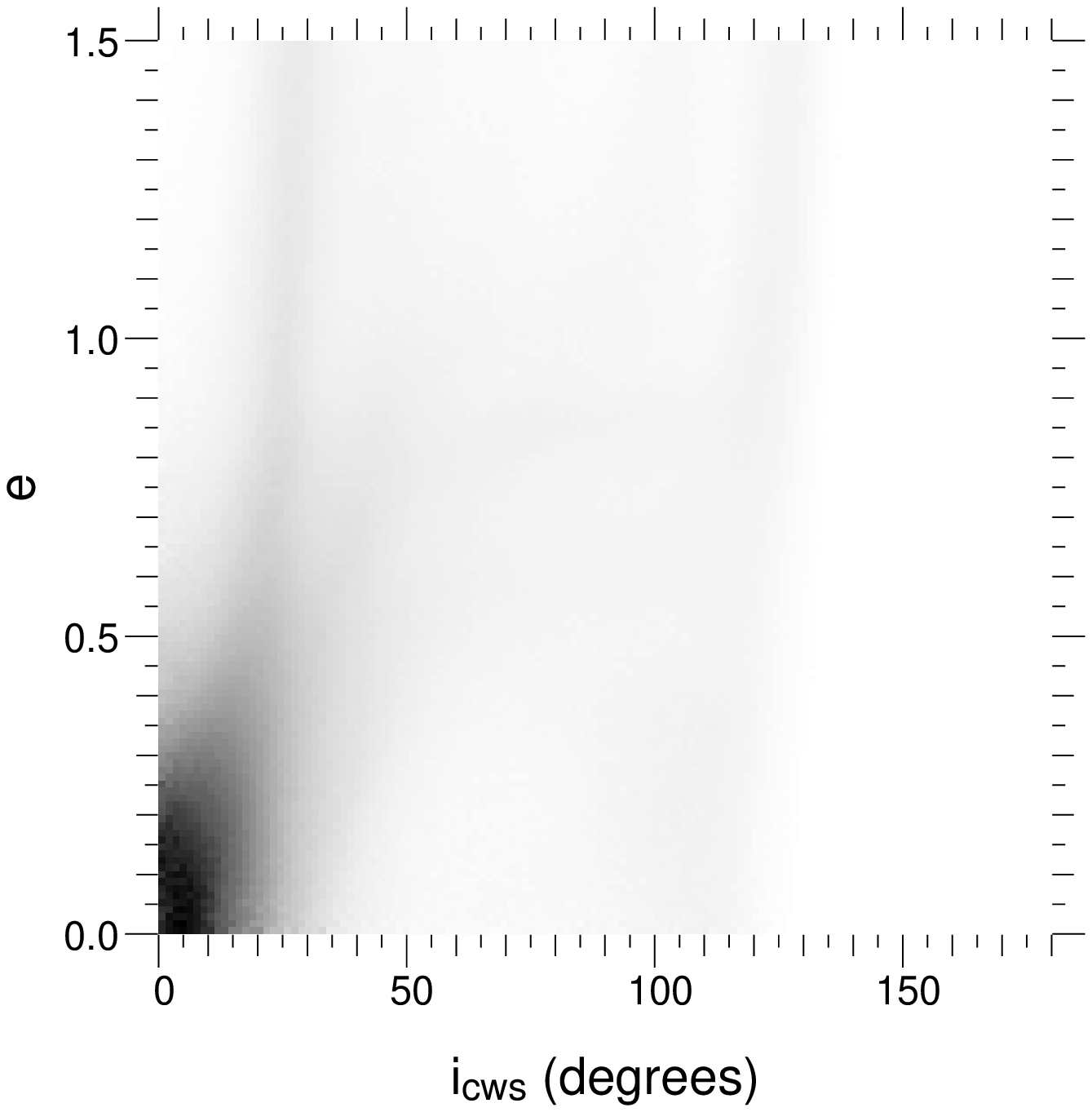}{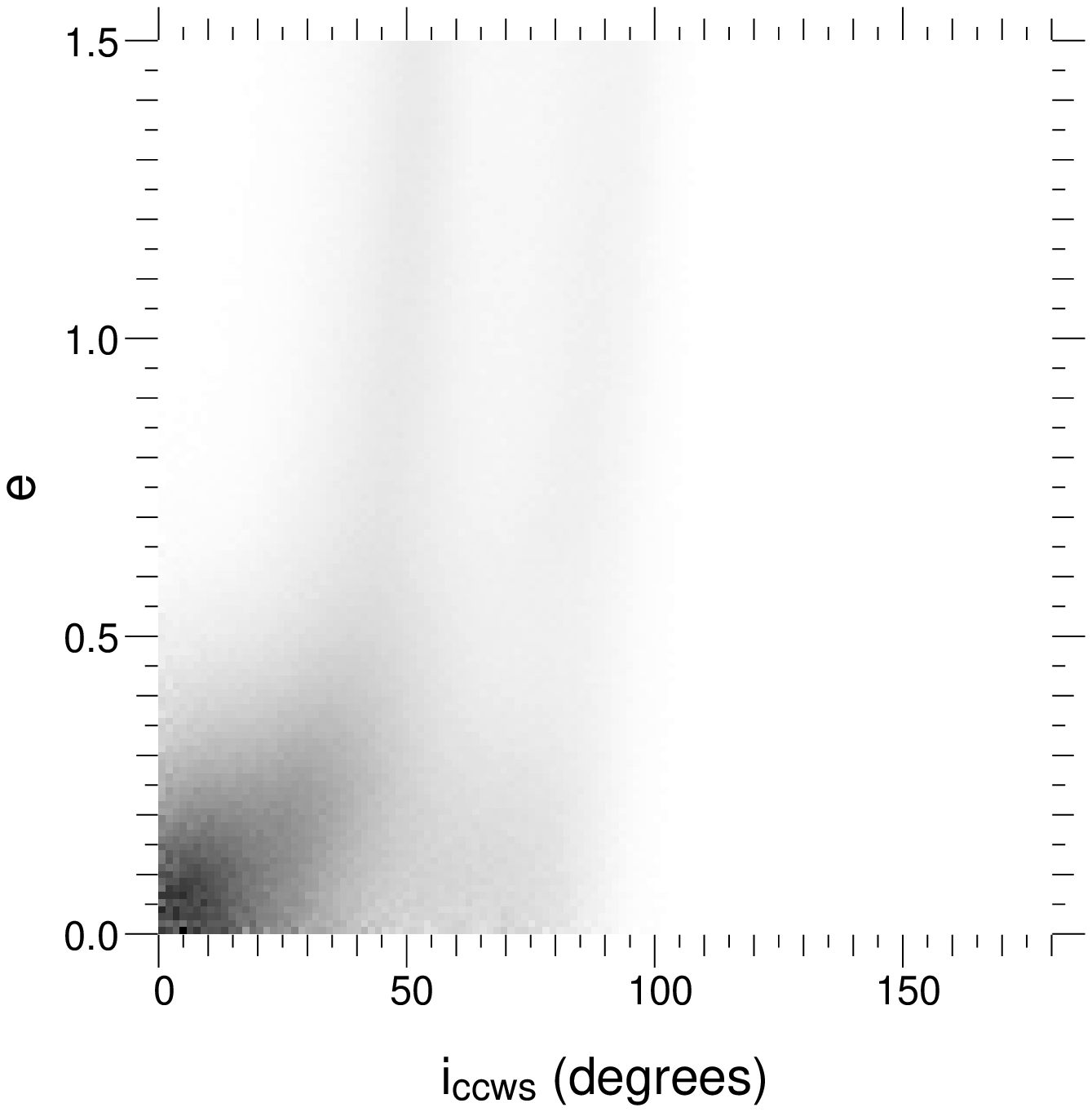}
  \caption{Co-added $e$ vs.  $i_\mathrm{(C)CWS}$ 2D histograms
    $P(i_\mathrm{(C)CW},e)$ ($z$ uniformly distributed with $|z|<20\arcsec$,
    linear gray scale) for all stars in the following systems: real CWS and
    CCWS stars (top left and right; $p<8\arcsec$), artificial CWS and CCWS
    stars on circular orbits (bottom left and right). The real and artificial CWS maps are quite
    similar whereas the CCWS ones are very different from each
    other.\label{fig:MC}}
\end{figure}
Figure~\ref{fig:MC} shows the sums of the histograms of all stars for each one
of the CWS and CCWS, for both the real data and artificial data in which all
stars are on circular motion.  These co-added $e$ vs.  $i_\mathrm{(C)CW}$
histograms yield a striking result: whereas the map for the CWS much resembles
the corresponding artificial data set that assumes circular motion, with a
maximum towards $(e=0, i_\mathrm{CWS}=0\degr)$, this is not the case for the
CCWS: for this system, the most prominent feature of the map is a (quarter of)
ring running from $(e=0.8, i_\mathrm{CCW}=0\degr)$ to $(e=0,
i_\mathrm{CCW}=80\degr)$. Such a ring-like structure on the 2D histogram of a
star is the sign that this star can either be on the disk ($i_\mathrm{(C)CW}=0$)
or on a circular orbit ($e=0$), but not both at the same time. Since indeed
most of the maps of CCWS stars exhibit this ring, and since they can not be
all at high inclination, we can already conclude that CCWS have typically high
eccentricities.

From these maps, we can go further and estimate the eccentricity for each
star. To do so, we simply compute (for each star) the distribution
\begin{equation}
P(e)=\int_\mathrm{i(C)CW}P(i_\mathrm{(C)CW},e)P(i_\mathrm{(C)CW})\mathrm di_\mathrm{(C)CW}
\end{equation}
where $P(i_\mathrm{(C)CW})$ which is known a priori: a Gaussian centered
on 0 and of width 14\degr\ and 19\degr\ respectively for the CWS and CCWS.
We then fit a Gaussian on this distribution, and take the centroid as the best
estimate, and the width of the Gaussian as the uncertainty.

It must be understood that the histograms are sometimes not nearly Gaussian.
The method gives reasonable estimates given the quoted error bars, though.

\section{Likelihood that IRS~13E is a background fluctuation}
\label{app:irs13}

In Sect.~\ref{sect:irs13:density} we use stellar counts to show that IRS~13E
is not a background density fluctuation. The first step is to determine the
background density. We have made the approximation that, given ${\Lambda}$ the
number of stars observed in the outside field of surface $A_\mathrm{out}$, the
probability $P_N$ that the actual background density of stars is indeed $N$
per such area is given by a Poisson law of parameter ${\Lambda}$.  In our
case, $\Lambda=115$ to a limiting magnitude of $m_H=20.4$ ($69$ to $m_H=19.4$)
and $A_\mathrm{out}=8.79$~arcsec$^2$.

Given $N$, the probability $P_{N\mathrm{in}|N}$ to find $N_\mathrm{in}=46$ (26)
stars in the circular aperture of surface $A_\mathrm{in}=1.45$~arcsec$^2$ is
also given by a Poisson law, of parameter $N\,A_\mathrm{in}/A_\mathrm{out}$.  The
probability, given our observations, that IRS~13E is but a spike in the
Poisson noise of the background density of stars (i.e. a chance alignment) is
given by the sum  
\begin{equation}
\int_{N=0}^{+\infty} P_N P_{N\mathrm{in}|N}\mathrm
dN\simeq\sum_{N=0}^{+\infty} P_N P_{N\mathrm{in}|N}\;.
\end{equation}

We find that the probability to find just by chance such a concentration of
star at any randomly selected line-of-sight in the central parsec is
$2\times10^{-6}$ ($2\times10^{-4}$). The over-density in IRS~13E is
significant at the $4.5\sigma$ ($3.5\sigma$) level. Of course, if one selects
a large number of lines-of-sight, the probability that one of these lines will
be associated with such a chance over-density is higher than these numbers.
The question thus arises whether the choice of the location of IRS~13E can be
considered as random with respect to the problem of chance alignment of stars.
As discussed in Sect.~\ref{sect:irs13}, the properties of IRS~13E make this
source unique in the central parsec.  However, one of the criteria for the
selection of this field was the presence of three stars.  Therefore, inasmuch
as we subtract these three stars from our count numbers, we can consider the
aperture around IRS~13E as randomly chosen. Given the number of stars that
remain after having subtracted these three, the probability that the observed
over-density associated with IRS~13E is but a chance alignment remains
$2\times10^{-5}$ ($2\times10^{-3}$).

\section{Stellar Population Synthesis}
\label{app:popsynth}

We used the synthesis code developed by Schaerer \& Vacca (1998). They
distinguished several subtypes of Wolf-Rayet (WR) stars from O (and Of) stars. They
defined any star burning H in its core and having effective temperatures
larger than 33,000 K as an O star. Here we are more generally interested in a
population of hot stars (O~I--V~+~B~I). We therefore modify their definition of
an `O' star to include the B supergiants, but not the B dwarfs, which we have
not detected. In addition to this, the effective temperature of O stars has
recently been revised downward (Martins et al. 2002, 2005; Crowther et al.
2002; Markova et al. 2004; Massey et al. 2005). In order to satisfy both
points, we have considered as an `O star' in the code every star with
$T_\mathrm{eff}>23,000$~K and $\log L/L_\sun>5.2$. We used Geneva evolutionary
tracks with rotation (Meynet \& Maeder 2003; Maeder \& Meynet 2004) for both
solar and twice solar metallicity. We chose the minimum masses for WR stars
according to the predictions of Meynet \& Maeder (2003). An initial mass
function with a minimum mass of $0.8\sunmass$ and a maximum mass of
$100\sunmass$ was used, and we varied the slope between $-2.35$ (Salpeter) and
$-1.35$ (top-heavy). We computed models for an instantaneous burst of star
formation, an extended burst (2~Myr) and a constant formation rate.

\section{Determination of $T_\mathrm{eff}$ and $M_K$}
\label{app:HR}

Accurate $T_\mathrm{eff}$s are usually derived through detailed modeling of the
spectra with atmosphere codes. This is a very long process, but first insights
can be obtained with `effective temperature scales' (c.f. Martins et al.
2005). In this way we were able to assign a spectral type to most of the stars
observed so far.  Absolute K band magnitudes can be easily computed from the
observed K magnitudes, the distance to the Galactic Center and an estimate of
the extinction. We used the currently best Galactic Center distance
($7.62\pm0.32$~kpc, Eisenhauer et al. 2005).  We applied the extinction map of
Scoville et al. (2003) to estimate the K-band extinction $A_K$. To convert the
V-band extinction $A_V$ into $A_K$, we have used the approximate relation
$A_K\simeq0.1A_V$.  The resulting absolute magnitudes (and their ${1\sigma}$
uncertainties) are given in the last two columns of Table~\ref{table:big1}.
For the 8~kpc distance adopted for the rest of this paper magnitudes would be
increased by $\simeq0.1$ mag.  We applied the effective temperature scale of
Martins et al.  (2005) for O stars.  For B stars we relied on the relations of
Schmidt-Kaler (1982). In practice, for a star classified as 09--B1~I, we have
chosen the effective temperature of a B0 supergiant as representative and the
most likely range of values was chosen by adopting $T_\mathrm{eff}$s for a O9
and a B1 supergiant as limiting values.


\clearpage
\begin{references}
\reference{} Alexander, T. \& Livio, M. 2004, ApJ, 606, L21
\reference{} Alexander, T. 2003, in `The Galactic black hole', eds.
H.Falcke \& F.W.Hehl, (Bristol: Inst. of Physics), 246
\reference{} Alexander, T. 2005, Phys. Reports, 419, 65
\reference{} Allen, D.A., Hyland, A.R. \& Hillier, D.J. 1990, MNRAS, 244, 706
\reference{} Aller, L.H. et al. 1982, in Landolt-B\"ornstein numerical data and
relationships in science and technology, New Series, volume 2,
subvolume b, Astronomy and astrophysics: Stars and star clusters,
(Berlin: Springer), 382
\reference{} Baganoff, F.K. et al. 2003, ApJ, 591, 891
\reference{} Becklin, E.E., \& Neugebauer, G. 1975, ApJ, 200, L71 \reference{} Beloborodov, A. M., \& Levin, Y. 2004, ApJ, 613, 224
\reference{} Blum, R.D., Sellgren, K., \& DePoy, D.L. 1995, ApJ, 440, L17
\reference{} Blum, R.D., Sellgren, K., \& DePoy, D.L. 1996, AJ, 112, 1988
\reference{} Blum R.D., Ramirez, S.V., Sellgren, K., \& Olsen, K. 2003, ApJ, 597, 323
\reference{} Bonnet, H. et al. 2003, Proc. SPIE, 4839, 329
\reference{} Bonnet, H. et al. 2004, The ESO Messenger, 117, 17
\reference{} Collin, S., \& Hure, J.-M. 1999, A\&A, 341, 385
\reference{} Clenet, Y. et al. 2004, A\&A, 417, L15
\reference{} Crowther, P.A., Hillier, D.J., \& Smith, L.J. 1995, A\&A, 293, 403
\reference{} Crowther, P.A., Hillier, D.J., Evans, C.J., Fullerton, A.W., DeMarco, O.,
\& Willis, A.J. 2002, ApJ, 579, 774
\reference{} Davidson, J.A., Werner, M.W., Wu, X., Lester, D.F., Harvey, P.M., Joy,
M., \& Morris, M. 1992, ApJ, 387, 189
\reference{} Davies, M.B., \& King, A. 2005, ApJ, 624, L25
\reference{} DePoy, D.L., Pepper, J., Pogge, R.W., Stutz, A., Personneault, M., \&
Sellgren, K. 2004, ApJ, 617, 1127
\reference{} Dessart, L., Crowther, P.~A., Hillier, D.~J., Willis, A.~J.,
Morris, P.~W., \& van~der~Hucht, K.~A., 2000. MNRAS, 315, 407
\reference{} Diolaiti, E., Bendinelli, O., Bonaccini, D., Close, L.,
Currie, D., \& Parmeggiani, G. 2000, A\&A Suppl., 147, 335
\reference{} Eisenhauer, F., Sch\"odel, R., Genzel, R., Ott, T., Tecza, M., Abuter,
R., Eckart, A., \& Alexander, T. 2003a, ApJ, 597, L121
\reference{} Eisenhauer, F. et al. 2003b, The ESO Messenger, 113, 17 
\reference{} Eisenhauer, F. et al. 2003c, Proc. SPIE, 4814, 1548
\reference{} Eisenhauer, F. et al. 2005, ApJ, 628, 246
\reference{} Ekers, R.D., van Gorkum, J.H., Schwarz, U.J., \& Goss, W.M. 1983, A\&A,
122, 143
\reference{} Figer, D.F., McLean, I.S., \& Najarro, F. 1997, ApJ, 486, 420
\reference{} Figer, D. F., Kim, S.S., Morris, M., Serabyn, E, Rich, R.M., \& McLean,
I.S. 1999, ApJ, 525, 750
\reference{} Figer, D. 2003, Astr. Nachr. Suppl., 324, 255
\reference{} Forrest, W.J., Shure, M.A., Pipher, J.L., \&
Woodward, C.A. 1987, in AIP Conference 155, The Galactic Center, ed. D.~Backer (New York), 153
\reference{} Genzel, R., \& Townes, C.H. 1987, ARAA, 25, 377
\reference{} Genzel, R., Thatte, N., Krabbe, A., Kroker, H., \& Tacconi-Garman, L.E.
1996, ApJ, 472, 153
\reference{} Genzel, R., Pichon, C., Eckart, A., Gerhard, O., \& Ott, T., 2000, MNRAS,
317, 348
\reference{} Genzel, R. et al. 2003, ApJ, 594, 812
\reference{} Gerhard, O. 2001, ApJ, 546, L39
\reference{} Gezari, S., Ghez, A.M., Becklin, E.E., Larkin, J., McLean, I.S., \&
Morris, M. 2002, ApJ, 576, 790
\reference{} Ghez, A.M. et al. 2003, ApJ, 586, L127
\reference{} Ghez, A.M. et al. 2004, ApJ, 601, L159
\reference{} Ghez, A.M., Salim, S., Hornstein, S.D., Tanner, A., Morris, M., Becklin,
E.E., \& Duchene, G. 2005, ApJ, 620, 744
\reference{} Goldreich, P., \& Sari, R., 2003, ApJ, 585, 1024
\reference{} Goodman, J. 2003, MNRAS, 339, 937
\reference{} Goodman, J., \& Tan, J. C. 2004, ApJ, 608, 108
\reference{} Gould, A., \& Quillen, A.C. 2003, ApJ, 592, 935
\reference{} G\"urkan, M.A., \& Rasio, F.A., 2005, ApJ, 628, 236
\reference{} G\"urkan, M.A., Freitag, M., \& Rasio, F.A., 2004, ApJ, 604, 632
\reference{} Hall, D.N.B., Kleinmann, S.G., \& Scoville, N.Z. 1982, ApJ, 260, L53
\reference{} Hansen, B. M. S., \& Milosavljevic, M., 2003, ApJ, 593, L77
\reference{} Hanson, M.M., Conti, P.S., \& Rieke, M.J. 1996, ApJ Suppl., 107, 281
\reference{} Hanson, M.M., Kudritzki, R.P., Kenworthy, M.A., Puls, J., \& Tokunaga,
A.T.2005, ApJ Suppl., 161, 154
\reference{} Herrero, A., Kudritzki, R.P., Vilchez, J.M., Kunze, D., Butler, K., \&
Haser, S. 1992, A\&A, 261, 209
\reference{} Horrobin, M. et al. 2004, Astr. Nachr., 325, 88
\reference{} Jackson, J.~M., Geis, N., Genzel, R., Harris, A.~I., Madden, S., Poglitsch,
A., Stacey, G.~J., \& Townes, C.~H. 1993, ApJ, 402, 173
\reference{} Kim, S.~S., \& Morris, M, 2003, ApJ, 597, 312
\reference{} Kim, S. S., Figer, D. F., \& Morris, M. 2004, ApJ, 617, L123
\reference{} Krabbe, A., Genzel, R., Drapatz, S., \& Rotaciuc, V. 1991, ApJ, 382, L19
\reference{} Krabbe, A. et al. 1995, ApJ, 447, L95
\reference{} Lacy, J.H., Townes, C.H., Geballe, T.R., \& Hollenbach, D.J. 1980, ApJ,
241, 132
\reference{} Langer, N. 1992, A\&A, 265, L17
\reference{} Lejeune, T., \& Schaerer, D. 2001, A\&A, 366, 538
\reference{} Levin, Y., \& Beloborodov, A.M. 2003, ApJ, 590, L33
\reference{} Levin, Y., Wu, A.S.P., \& Thommes, E.W. 2005, ApJ, 635, 341
\reference{} Lin, D.~N.~C., \& Pringle, J.~E., 1987, MNRAS, 225, 607
\reference{} Lissauer, J.J. 1987, Icarus, 69, 249
\reference{} Liszt, H.S. 2003, A\&A, 408, 1009
\reference{} Lu, J.R., Ghez, A.M., Hornstein, S.D., Morris, M., \& Becklin, E.E. 2005,
ApJ, 624, L51
\reference{} Maeder, A., \& Meynet, G. 2004, A\&A, 422, 225
\reference{} Maillard, J.-P., Paumard, T., Stolovy, S. R., \& Rigault, F. 2004, A\&A,
423, 155 
\reference{} Markova, N., Puls, J., Repolust, T., \& Markov, H. 2004, A\&A, 413, 693
\reference{} Martins, F., Schaerer, D. \& Hillier, D.J. 2002, A\&A, 382, 999
\reference{} Martins, F., Schaerer, D. \& Hillier, D.J. 2005, A\&A, 436, 1049
\reference{} Mas-Hesse, J.M., \& Kunth, D. 1991, A\&A Suppl., 88, 399
\reference{} Massey, P., Puls, J., Pauldrach, A.W.A., Bresolin, F., Kudritzki, R.P.,
\& Simon, T. 2005, ApJ, 627, 477
\reference{} McMillan, S.~L.~W., \& Portegies Zwart, S.~F., 2003, ApJ, 596, 314
\reference{} Meynet, G., \& Maeder, A. 2003, A\&A, 404, 975
\reference{} Mezger, P.G., Duschl, W., \& Zylka, R. 1996, A\&A Rev., 7, 289
\reference{} Milosavljevic, M., \& Loeb, A. 2004, ApJ, 604, L45
\reference{} Mokiem, M.R., de Koter, A., Puls, J., Herrero, A., Najarro, F., \&
Villamariz, M.R. 2005, A\&A, 441, 711
\reference{} Morisset, C., Schaerer, D, Bouret, J.C., \& Martins, F. 2004, A\&A, 415,
577
\reference{} Morris, M. 1993, ApJ, 408, 496
\reference{} Morris, M. \& Serabyn, E. 1996, ARAA, 34, 645
\reference{} Morris, P.W., Eenens, R.P.J., Hanson, M.M., Conti, P.S., \& Blum, R.D.
1996, ApJ, 470, 597
\reference{} Muno, M.P., Pfahl, E., Baganoff, F.K., Brandt, W.N., Ghez, A., Lu, J.,
\& Morris, M.R. 2005, ApJ, 622, L113
\reference{} Najarro, F., Hillier, D.J., Kudritzki, R.P., Krabbe, A., Genzel, R.,
Lutz, D., Drapatz, S., \& Geballe, T.R. 1994, A\&A, 285, 573
\reference{} Najarro, F., Krabbe, A., Genzel, R., Lutz, D., Kudritzki, R.P., \&
Hillier, D.J. 1997, A\&A, 325, 700
\reference{} Nayaksin, S., \& Cuadra, J.J. 2005, A\&A, 437, 437
\reference{} Nayakshin, S., \& Sunyaev, R. 2005, MNRAS, 364, L23
\reference{} Nayakshin, S., Dehnen, W., Cuadra, J., \& Genzel, R., MNRAS, submitted
\reference{} Ouellette, J. A., \& Pritchet, C. J. 1998, AJ, 115, 2539
\reference{} Ott, T., Eckart, A., \& Genzel, R. 1999, ApJ, 523, 248
\reference{} Paumard, T., Maillard, J.-P., Morris, M., \& Rigaut, F. 2001, A\&A,
366, 466
\reference{} Paumard, T., Genzel, R., Maillard, J.-P., Ott, T., Morris, M.~R.,
Eisenhauer, F., \& Abuter, R., 2004a, in Young Local Universe, Proceedings of
the XXXIXth Rencontres de Moriond, ed. Chalabaev, A., Fukui, T., Montmerle,
T., \& Tran-Thanh-Van, J. (Paris: \'Editions Fronti\`eres), 377,
http://www-laog.obs.ujf-grenoble.fr/ylu/ylu\_vgr/\\vgr\_index.html
\reference{} Paumard, T., Maillard, J.-P., \& Morris, M. 2004b, A\&A, 426, 81
\reference{} Portegies Zwart, S. F., \& McMillan, S. L. W., 2002, ApJ, 576, 899
\reference{} Portegies Zwart, S. F., McMillan, S. L. W., \& Gerhard, O. 2003, ApJ, 593,
352
\reference{} Reid, M.J., \& Brunthaler A. 2004, ApJ, 616, 872
\reference{} Revnivtsev, M.~G., et  al. 2004, A\&A, 425, L49
\reference{} Sanders, R.H. 1998, MNRAS, 294, 35
\reference{} Schaerer, D., \& Schmutz, W. 1994, A\&A, 288, 231
\reference{} Schaerer, D., Contini, T., Kunth, D., \& Meynet, G. 1997, ApJ, 481, L75
\reference{} Schmidt-Kaler, T., 1982, in
Landoldt-B$\ddot{\rm o}$rnstein, New Series Group, VI, Vol. 2, ed.
K. Schaifers \& H.H. Voigt (Berlin: Springer-Verlag), 1
\reference{} Sch\"odel, R. et al., 2002, Nature, 419, 694
\reference{} Sch\"odel, R., Ott, T., Genzel, R., Eckart, A., Mouawad, N., \&
Alexander, T. 2003, ApJ, 596, 1015
\reference{} Sch\"odel, R., Eckart, A., Iserlohe, C., Genzel, R., \& Ott, T. 2005,
ApJ, 625, 111
\reference{} Scoville, N.Z., Stolovy, S.R., Rieke, M., Christopher, M., \&
Yusef-Zadeh, F. 2003, ApJ, 594, 294
\reference{} Shakura, N.~I., \& Sunyaev, R.~A. 1973, A\&A, 24, 337
\reference{} Shields, J.C., \& Ferland, G.J. 1994, ApJ, 439, 236
\reference{} Smith, L.J., Norris, R.P.F., \& Crowther, P.A. 2002, MNRAS, 337, 1309
\reference{} Sternberg, A. 1998, ApJ, 506, 721
\reference{} Sternberg, A., Hoffmann, T.~L., \& Pauldrach, A.~W.~A. 2003, ApJ, 599, 1333
\reference{} Stolte, A., Brandner, W., Grebel, E.K., Lenzen, R., \& Lagrange, A.M.
2005, ApJ, 628, L113
\reference{} Svensson, R., \& Zdziarski, A. 1994, ApJ, 436, 599
\reference{} Tamblyn, P., \& Rieke, G.H. 1993, ApJ, 414, 573
\reference{} Tamblyn, P., Rieke, G.H., Hanson, M.M., Close, L.M., McCarthy, D.W., \&
Rieke, M.J. 1996, ApJ, 456, 206
\reference{} Tanner, A. et al. 2006, ApJ, in press
\reference{} Thompson, T.~A., Quataert, E., \& Murray, N. 2005, ApJ, 630, 167
\reference{} Toomre, A. 1964, AJ, 139, 1217
\reference{} Trippe, S. et al. 2006, A\&A, in press
\reference{} Trundle, C., \& Lennon, D.J. 2005, A\&A, 434, 677
\reference{} Vacca, W.D., \& Conti, P.S. 1992, ApJ, 401, 543
\reference{} Wallace, L., \& Hinkle, K. 1997, ApJ Suppl., 111, 445
\reference{} Weinberg, N. N., Miloslavljevic, M., \& Ghez, A. 2005, ApJ, 622, 878
\reference{} Zhao, J.H., \& Goss, W.M. 1998, ApJ, 499, L163
\end{references}
\end{document}